\documentclass[a4paper,11pt]{article}

\usepackage{jheppub} 
\usepackage{mathtools}
\usepackage{amsmath, amsfonts, amsthm, amssymb, graphicx, color, hyperref, slashed, braket}
\usepackage{subfigure}
\usepackage{pstool}
\usepackage{graphicx}
\usepackage{float} 
\usepackage[utf8]{inputenc}
\usepackage[normalem]{ulem}
\usepackage[english]{babel}
\usepackage{blkarray}
\usepackage{mathrsfs}
\usepackage{diagbox}
\usepackage{multirow}
\usepackage{tikz}

\allowdisplaybreaks[3]
\def\d{\mathrm{d}}
\renewcommand{\i}{\mathrm{i}}
\newcommand{\Abs}[1]{\left| #1 \right|}
\newcommand{\dif}{\mathrm{d}}
\newcommand{\eref}[1]{Eq.~\eqref{#1}}
\newcommand{\nn} {\nonumber\\}

\title{Matrix moment approach to positivity bounds and UV reconstruction from IR}

\author[a]{Shi-Lin Wan}
\author[a,b]{and Shuang-Yong Zhou}

\affiliation[a]{Interdisciplinary Center for Theoretical Study, University of Science and Technology of China, Hefei, Anhui 230026, China}
\affiliation[b]{Peng Huanwu Center for Fundamental Theory, Hefei, Anhui 230026, China}

\emailAdd{wsl9868@mail.ustc.edu.cn}
\emailAdd{zhoushy@ustc.edu.cn}

\preprint{{\footnotesize USTC-ICTS/PCFT-24-41}}

\date{\today}

\abstract{  
Positivity bounds in effective field theories (EFTs) can be extracted through the moment problem approach, utilizing well-established results from the mathematical literature. 
We generalize this formalism using the matrix moment approach to derive positivity bounds for theories with multiple field components. 
The sufficient conditions for obtaining optimal bounds are identified and applied to several example field theories, yielding results that match precisely the numerical bounds computed using other methods. 
The upper unitarity bounds can also be easily harnessed in the matrix case. Furthermore,
the moment problem formulation also provides a means to reverse engineer the UV spectrum from the EFT coefficients, often uniquely, as explicitly demonstrated in examples such as string amplitudes and the $stu$ kink theory.
}

\begin{document}
\maketitle
\flushbottom

\section{Introduction and Summary}

The space of consistent quantum field theories is strongly constrained by the fundamental principles of the S-matrix. In an effective field theory (EFT) setup, unitarity conditions in the UV can be carried down to low energies through dispersion relations, which arise from causality/analyticity and locality \cite{Adams:2006sv, deRham:2017avq, deRham:2017zjm, Arkani-Hamed:2020blm, Bellazzini:2020cot, Tolley:2020gtv, Caron-Huot:2020cmc, Sinha:2020win, Alberte:2020jsk, Chiang:2021ziz, Guerrieri:2021tak, Li:2021lpe, Bern:2021ppb, Alberte:2021dnj, Du:2021byy, Caron-Huot:2021rmr, Chiang:2022ltp} (see \cite{deRham:2022hpx} for a review). 
These constraints, often referred to as positivity bounds or causality bounds, impose stringent bounds on the EFT coefficients, especially when crossing symmetry is fully used \cite{Tolley:2020gtv, Caron-Huot:2020cmc}. 
This approach has proven fruitful when applied to the Standard Model EFT \cite{Zhang:2018shp, Bellazzini:2018paj, Remmen:2019cyz, Zhang:2020jyn, Remmen:2020vts, Yamashita:2020gtt, Bonnefoy:2020yee, Fuks:2020ujk, Gu:2020ldn, Davighi:2021osh, Li:2022rag, Davighi:2023acq, Hong:2024fbl},
chiral perturbation theory \cite{Wang:2020jxr, Fernandez:2022kzi, Albert:2023jtd, Ma:2023vgc, Li:2023qzs, Albert:2023seb}, 
gravitational EFTs \cite{Bellazzini:2015cra, Cheung:2016yqr, Tokuda:2020mlf, Caron-Huot:2022ugt, Chiang:2022jep, Henriksson:2022oeu, Hong:2023zgm, Bellazzini:2023nqj, Xu:2024iao},
string EFTs \cite{Huang:2020nqy, Chiang:2023quf, Berman:2023jys}, and cosmological models \cite{deRham:2017imi, deRham:2018qqo, Melville:2019wyy, deRham:2021fpu, Xu:2023lpq}. 
They underscore the fact that, while separation/decoupling of scales is essential for the EFT framework to function, equally important is to recognize the UV-IR mixing in defining the space of consistent EFTs. 
Alternatively, the restrictions on the EFT coefficients can be obtained by directly constructing all viable crossing-symmetric amplitudes that satisfy unitarity conditions \cite{Guerrieri:2020bto,Guerrieri:2021ivu,Haring:2022sdp,Haring:2023zwu} (see \cite{Kruczenski:2022lot} for a review).

For an EFT weakly coupled at low energies, the extraction of the positivity bounds from dispersive sum rules can be formulated as a (mathematical) moment problem \cite{Arkani-Hamed:2020blm, Bellazzini:2020cot, Chiang:2021ziz, Chiang:2022ltp, Chiang:2022jep}. The moment problem dates back as early as Stieltjes, Hausdorff and Hamburger, and has been extensively studied in the mathematical literature \cite{Schmdgen2017TheMP}, while the connections between the pion scattering and the Stieltjes moment problem, in a slightly different form, were already explored a few decades ago \cite{Common:1969nu, Yndurain:1969qm, Common:1970um, Common:1976uz}.

In the context of a single scalar EFT that is weakly coupled below the cutoff $\Lambda$, its dispersive sum rules in the forward limit can be rewritten as
\begin{align}
\label{eq:g2n0sumrules}
g^{2i, 0}=\int_{\Lambda^2}^{\infty} \frac{2 \mathrm{~d} s^{\prime}}{\pi s^{\prime 1+2 i}} \operatorname{Im} A\left(s^{\prime}, t=0\right)~~
\xrightarrow{x_1\,\equiv\, \Lambda^2/s'{}}~~ g^{2 i, 0}= \Lambda^{-4i} \int_0^1 x_1^{2i} \d \mu(x_1)
\end{align}
where $g^{2i,0}$ are the EFT coefficients when expanding the 2-to-2 scattering amplitude in terms of the Mandelstam variables $A(s,0)\sim\sum_{i} g^{2i,0}s^{2i}$ and $\d \mu$ is a positive measure due to the optical theorem $\operatorname{Im} A(s',0) \geq 0$. 
This is simply the classical Hausdorff moment problem: given a generic positive measure $\d \mu$ ({\it i.e.}, without knowledge of the specific form of the UV theory except for the unitarity condition $\operatorname{Im} A(s',0) \geq 0$), what are the allowed values for $g^{2i,0}$ according to the integral representation \eqref{eq:g2n0sumrules}? 
The solution to this problem can be given in terms of the Hankel matrices: $\mathcal{H}(g^{2 i, 0}) \succeq 0,~ \mathcal{H}(g^{2 i, 0})|_{2 i \to 2 i+2} \succeq 0, ~ \mathcal{H}(g^{2 i, 0})-\mathcal{H}(g^{2 i, 0})|_{2 i \to 2 i+2} \succeq 0$, where the Hankel matrix is defined by $\mathcal{H}^{\alpha,\beta}(g^{2 i, 0})=g^{2\alpha+2\beta+2,0}$ for $\alpha,\beta=0,1,2,...$ and the notation $\succeq 0$ denotes the matrix being positive semi-definite (PSD). 
Away from the forward limit, the EFT amplitudes can be expanded as $A(s,t) \sim \sum_{i,j} g^{i,j} (s+t/2)^i t^j$, and the dispersive sum rules can be written as a double summation over the UV mass scale $s'$ and spin $\ell$, with the latter coming from the partial wave expansion of the imaginary part of the UV amplitude. 
By introducing a second moment variable $x_2 = {\ell(\ell+1) \Lambda^2}/{s^\prime}$, these sum rules can be cast as a sum of bi-variate moments \cite{Chiang:2021ziz}: 
\begin{align}
\label{eq:gijtogij}
&g^{i,j}= \sum_{\ell} \!\int_{\Lambda^2}^{\infty} \frac{ \eta^{i,j}_{\ell(\ell+1)}\! \d s^{\prime}}{s^{\prime i+j+1}} \operatorname{Im} A^{\ell}(s') ~
\xrightarrow[x_2\equiv\frac{\Lambda^2 \ell(\ell+1)}{s'}]{x_1\equiv \frac{\Lambda^2}{s'} } ~ g^{i,j} \sim \sum_{\gamma_1,\gamma_2} \tilde{V}^{i,j}_{\gamma_1,\gamma_2}\!\! \int_{\mathcal{K}}\! x_1^{\gamma_1} x_2^{\gamma_2} \dif \mu (x_1, x_2)
\end{align}
where $\eta^{i,j}_ {\ell(\ell+1)}$ is a polynomial in $\ell(\ell+1)$ and $\mathcal{K}$ is a set of rays in the $x_1$-$x_2$ plane defined by $x_2= \ell(\ell+1) x_1$ with $\ell=0,2,4,... \, $. 
With this formulation, the positivity bounds can again be formulated as PSD conditions on (generalized) Hankel matrices. In some simple cases, the positivity bounds can even be computed analytically \cite{Chiang:2021ziz}. However, in general, to obtain optimal bounds with curved boundaries, one still needs to implement semi-definite programs (SDPs) in the final step after using the analytical results from the moment problem. 

In this paper, we shall discuss two improvements to the moment formulation above. First of all, we propose to use a mixed-variate moment formulation to speed up the computation of the final numerical bounds. 
The inefficiency of the (pure) bi-variate formulation arises from the fact that the integration over the $x_1$-$x_2$ plane is performed along discrete rays rather than over continuous 2D regions. 
As a result, much of the effort in evaluating these bounds involves eliminating redundant integration regions in the $x_1$-$x_2$ plane, which in turn leads to a proliferation of PSD conditions on Hankel matrices. 
In our mixed-variate formulation, we separate the low spin partial waves into a number of uni-variate moments $\mathfrak{a}^{i+j-2}_{\ell}$, one for each partial wave, and approximate the high spin partial waves with a bi-variate moment $\mathfrak{b}^{i-2,j}$ defined in a continuous 2D region in the $x_1$-$x_2$ plane (see Figure \ref{fig:KbKbp}):
\begin{align}
\label{eq:gijabandUch}
    g^{i,j} = \sum_{\ell=0}^{\ell_{\text{M}}-1} U^{i,j}_{\ell (\ell + 1)} \mathfrak{a}^{i+j-2}_{\ell} + \sum_{i^\prime, j^\prime} V^{i,j}_{i^\prime j^\prime} \mathfrak{b}^{i^\prime-2,j^\prime} + (\text{$u$-channel})
\end{align}
As we will see, this formulation is significantly more efficient numerically while maintaining high accuracy, especially when a large number of partial waves are involved and one seeks to impose the sufficient conditions discussed below. 

Second, we identify the sufficient conditions for obtaining the optimal positivity bounds. 
We will see that with the existing conditions on the Hankel matrices the positivity bounds are slightly different from the linear programming results of \cite{Caron-Huot:2020cmc}, and the differences increase for high-order EFT coefficients.
Mathematically, the solvability of a moment problem is equivalent to the existence of a PSD measure for the integral representation of the moment sequence.
Assuming the existence of the integral form with a PSD measure, it is easy to find some necessary conditions for the moment problem by simply constructing PSD quantities in terms of Hankel matrices (see around \eref{HposiDef}). 
By the Riesz representation as well as the Stone-Weierstrass theorem and the Positivstellensatz, the sufficient conditions can also be formulated in terms of PSD conditions on a set of Hankel matrices (see around \ref{eq:intFmu}). 
For a moment problem defined in a semialgebraic set, a sufficient set of Hankel matrices is given by the following ``power set'' 
\begin{align}
    \mathcal{P}(\hat{p}) &=\{ \hat{p}_1^{e_1}\hat{p}_2^{e_2} \cdots \hat{p}_k^{e_k} \mid e_l \text{ being either 0 or 1} \} \\
    &= \set{1, \hat{p_1}, \cdots, \hat{p}_k, \hat{p}_1 \hat{p}_2, \cdots, \hat{p}_{k-1} \hat{p}_k, \cdots, \hat{p}_1 \hat{p}_2 \cdots \hat{p}_{k-1} \hat{p}_k}
\end{align}
where $\hat{p}_i$ are the defining polynomials of the semialgebraic set $\mathcal{K}$ ({\it i.e.}, $\mathcal{K}$ is the region carved out by $\hat{p}_i\geq 0$).

Given a finite number of EFT coefficients, we are usually interested in a truncated moment problem. 
In this case, solvability additionally requires the existence of eventual flat extensions to the truncated Hankel matrices, which guarantees the existence of an underlying solvable full moment sequence. (A flat extension of a Hankel matrix is an enlarged Hankel matrix that retains the same rank as the original matrix.)  
Consequently, simply truncating the PSD conditions for the (infinite) Hankel matrices may not yield optimal positivity bounds. Instead, it may be necessary to derive the bounds using a larger set of moments than naively required.

Another main purpose of this paper is to generalize the moment problem approach to positivity bounds for theories with multiple fields or spin components. 
After all, the universe is more complex than a single scalar field! 
This generalization requires introducing {\it the  matrix moment problem}, which has also been explored but remains less developed in the mathematical literature \cite{Kimsey2022OnAS,anjos2011handbook}. 
Nonetheless, for the most part, the results of the standard moment problem can be straightforwardly extended to the matrix case, replacing numbers with matrices and generalizing the Hankel matrices. 
We will explicitly verify this with some concrete examples by comparing to bounds obtained from other methods (see Sections \ref{sec:biscalar} and \ref{sec:photonEFT}). 
The matrix moment formulation can also be easily generalized to utilize the upper bounds of partial wave unitarity (see Section \ref{sec:Lmoment}), generalizing the $L$-moment formulation of the single scalar case \cite{Chiang:2022ltp}. 
We will demonstrate the effectiveness of this approach in the context of a gravitational EFT in Section \ref{sec:gravEFT}. 

For a theory with multiple degrees of freedom (DoFs), to use all available information, one must consider all possible 2-to-2 scatterings between different particles ${\bf 12\to 34}$, and the EFT coefficients $g^{i,j}_{\bf 1234}$ now carry the information of the external particles. 
Viewing $\textbf{12}$ as a row index and $\textbf{34}$ as a column index, $g^{i,j}_{\bf 1234}$ forms a matrix for given $i$ and $j$. 
The dispersive sum rules for $g^{i,j}_{\bf 1234}$ thus become a sum of matrix moment sequences:
\begin{align}
\label{eq:gijabintro}
  \text{{\bf Single scalar:} {\sffamily moment problem}} &~~\Longrightarrow~~ \text{{\bf Multi-DoFs:} {\sffamily matrix moment problem}}
  \nonumber\\
  \text{\eref{eq:gijabandUch}}~\Longrightarrow~  g^{i,j}_{\textbf{1234}} = \sum_{\ell=0}^{\ell_{\text{M}}-1} &U^{i,j}_{\ell (\ell + 1)} \mathfrak{a}^{i+j-2}_{\ell,\textbf{1234}} + \sum_{i^\prime,j^\prime} V^{i,j}_{i^\prime j^\prime} \mathfrak{b}^{i^\prime-2,j^\prime}_{\textbf{1234}} + (\text{$u$-channel})
\end{align}
where $\mathfrak{a}^{i+j-2}_{\ell,\textbf{1234}}$ and $\mathfrak{b}^{i^\prime-2,j^\prime}_{\textbf{1234}}$ are now uni-variate and bi-variate matrix moments respectively. 
For multiple scalars, this generalization is straightforward. However, for particles with spin, the partial wave expansion is facilitated by Wigner $d$-matrices instead of the Legendre polynomials. 
Unlike the scalar case, $\eta_{\ell(\ell+1)}^{i,j;{\bf 1234}}$, the counterpart of $\eta_{\ell(\ell+1)}^{i,j}$ in \eref{eq:gijtogij}, generally contains square roots of rational functions of $\ell(\ell+1)$. 
Thus, the bi-variate part of \eref{eq:gijabintro} is no longer a standard bi-variate moment problem---it is a {\it generalized} moment problem (see Section \ref{sec:formSpin}). 
However, a simple solution to this complication is to introduce additional moment variables to convert it to a standard moment problem.  
For instance, for the spin-1 case, we will encounter the following generalized bi-variate moments
\begin{align}
    \tilde{\mathfrak{b}}_{\textbf{1234}}^{ \gamma_1,\gamma_2} &= \int_{\mathcal{K}} 
    \sqrt{x_2 (x_2 - 2 x_1)} x_1^{\gamma_1} x_2^{\gamma_2} \rho_{\textbf{1234}} (x_1, x_2) \dif x_1 \dif x_2 
\end{align}
in addition to the standard bi-variate moments. 
In this case, by introducing a new variable $x_3=\sqrt{x_2 (x_2 - 2 x_1)}$, we can convert them to tri-variate moments
\begin{align}
    \mathfrak{b}_{\textbf{1234}}^{ \gamma_1,\gamma_2, \gamma_3} &= \int_{\mathcal{K}^\prime} x_1^{\gamma_1} x_2^{\gamma_2} x_3^{\gamma_3} \rho^\prime_{\textbf{1234}} (x_1, x_2, x_3) \dif x_1 \dif x_2 \dif x_3
\end{align}
where the modified semialgebraic set $\mathcal{K}^\prime$ is $\set{(x_1,x_2,x_3) | (x_1, x_2) \in \mathcal{K}, x_3^2 = x_2^2 - 2 x_1 x_2}$.

\begin{figure}[h]
    \centering
    \includegraphics[scale=0.45]{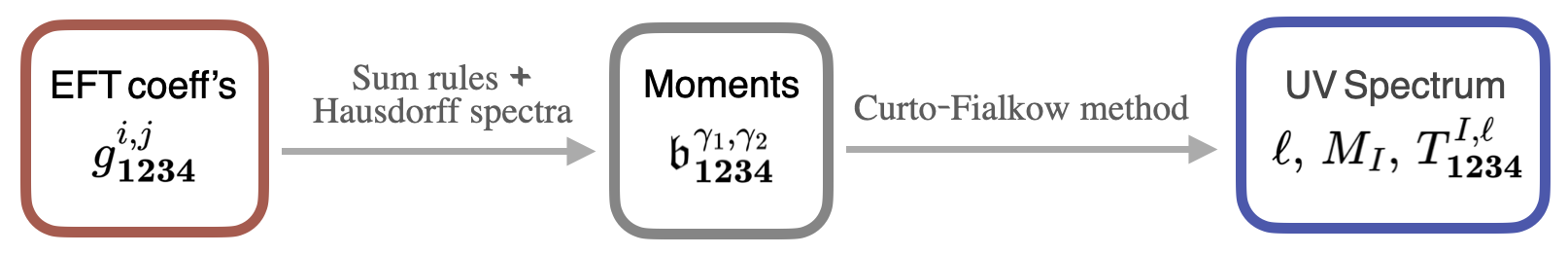}
    \caption{Inverse engineering the UV spectrum (including spin, mass and couplings) from the EFT coefficients in the amplitude via the moment problem. The conversion from the EFT coefficients to the moments cannot be done only using dispersive sum rules and it also requires computing the measures of a series of Hausdorff moment sequences.
    }
\end{figure}

In the last part of the paper, we demonstrate that the mathematical moment approach offers a systematic method for reconstructing the UV spectrum from generic EFT coefficients. 
Taking a bi-variate moment problem as an example, mathematically, a moment problem ask, for a given moment sequence $\{\mathfrak{b}_{\bf 1234}^{\gamma_1,\gamma_2}\}$, whether there exists a measure $\mu_{\bf 1234}$ such that the integral $\mathfrak{b}_{\bf 1234}^{\gamma_1, \gamma_2}=\int_{\mathcal{K}} x_1^{\gamma_1} x_2^{\gamma_2} \mathrm{d} \mu_{\bf 1234}$ holds. 
As long as the moment problem is solvable, it is possible to find a representing measure for the moment sequence.
Particularly, the Curto-Fialkow method \cite{curto1998flat,Kimsey2022OnAS} allows
one to construct an atomic representing measure (see around \eref{eq:HWequation0}). 
An atomic measure is a simple measure that only consists of a sum of products of delta functions: $\mathrm{d} \mu_{\bf 1234}=\sum_n T_{\bf 1234}^{n} \delta(x_1-\omega_{1 n}) \delta(x_2-\omega_{2 n}) \mathrm{d} x_1 \mathrm{d} x_2$, where $\omega_{1n}, \omega_{2n}, T_{\bf 1234}^n$ are constants. 
For our physical problem, the measure corresponds to the UV spectrum, and the atomic measure gives rise to isolated UV particle masses $M_{I}$ and spins $\ell$ along with their couplings $\mathcal{T}^{I, \ell}_{\textbf{1234}}$:
\begin{align}
\mathrm{d} \mu_{\bf 1234}&= \sum_{I,\ell} T^{I,\ell}_{\textbf{1234}} \delta (s^\prime - M_I^2) \delta ( J^2 - \ell (\ell + 1)) \d s^\prime \d  J^2
\\
&~~~~~~~~~~~~~~~~~~~~~\big\Downarrow\nonumber
\\
 A_{\textbf{1234}}^{\rm (UV)} (s,t) & =  \text{ poles } + 
        \sum_{I,\ell} \frac{\mathcal{T}^{I, \ell}_{\textbf{1234}} d^\ell_{h_{12},h_{34}}(\cos\theta)}{M_I^2-s-i\epsilon}
        + {\rm (crossing~terms)},
\end{align}
For a truncated moment problem, it is essential for this method to obtain flat extended Hankel matrices, which may require augmenting the truncated moment sequence with some fictitious moments. 
We will illustrate this method with explicit examples. 
This UV reconstruction is always unique in the forward limit, as in that limit it reduces a Hausdorff problem, which satisfies Carleman's criterion \cite{Schmdgen2017TheMP}. 
However, away from the forward limit, the uniqueness of the reconstructed UV spectrum is not guaranteed unless the moments satisfy some determinacy criteria. 
We will demonstrate the UV reconstruction with a few examples, 
including the $stu$ kink theory, the Veneziano and Virasoro amplitudes. In the explicit examples we examined, which are by no means general, the obtained UV spectra do reliably converge to our intended UV theory (see Figure \ref{fig:SingleFullMPExample1}, Figure \ref{fig:DoubleFullMPExample1}, and Figure \ref{fig:Vis_Spec}). For an EFT living on the boundary of the positivity bounds, its UV spectrum may also be extracted by saturating the PSD conditions used to derive the positivity bounds \cite{Chiang:2022jep, Albert:2023seb, Komargodski:2016auf, Simmons-Duffin:2016wlq}.

A subtlety in inverse engineering the UV spectrum is that, because of the presence of the $u$-channel in the dispersive sum rules (such as \eref{eq:gijabandUch}), one can not directly obtain the moments $\{\mathfrak{b}_{\bf 1234}^{\gamma_1,\gamma_2}\}$ from the EFT coefficients via the sum rules. 
Roughly half of the moments are undetermined. 
This issue can be overcome by utilizing the atomic measures constructed from moments that can be directly or indirectly inferred from the EFT coefficients (see Section \ref{sec:EFT_gives_Moments}). 
This method uniquely converts the EFT coefficients to the moments, as the Hausdorff moment problems involved in this process have unique atomic measures. Then these moments can be used to inverse engineer the UV spectrum. Alternatively, one can numerically compute the moments from the sum rules and the PSD conditions of the Hankel matrices. We will demonstrate these conversions with explicit examples.

\section{Sum rules as matrix moment problem}
\label{sec:SumRules}

In Section \ref{sec:DispSumrules}, we first derive the dispersive sum rules for scattering amplitudes in a theory with multiple degrees of freedom in the IR. 
For simplicity, we assume a large hierarchy between the UV and IR scales, allowing us to take the massless limit for all IR particles. 
In Section \ref{sec:momentFormulation}, we then reformulate the problem of extracting bounds from these sum rules as a mixed-variate moment problem, treating scalar particles and those with spin separately for technical reasons.

\subsection{Dispersive sum rules}
\label{sec:DispSumrules}

Consider a 2-to-2 scattering amplitude between the IR particles $A_{\textbf{1234}} (s,t)$ in 4D, where $\textbf{1},\textbf{2},\textbf{3}$ and $\textbf{4}$ represent the four scattering particles and $s,t,u$ are the standard Mandelstam variables. We will work with partial waves, particularly utilizing (part of) the partial wave unitarity, so we expand the amplitude as follows
\begin{align}
    \begin{gathered}
        A_{\textbf{1234}} (s,t) = 16 \pi \sum_{\ell} \left(2\ell+1\right) \ d_{h_{12} h_{43}}^{\ell} \left( \theta \right) A^{\ell}_{\textbf{1234}} (s), \\ ~~h_{ij} = h_{i} - h_{j}, ~~\cos\theta = 1 + \frac{2t}{s}
    \end{gathered}
\end{align}
where $h_i$ is the helicity of particle $i$, $\theta$ is the scattering angle between the particle $1$ and $3$, and $d_{h_{12} h_{43}}^{\ell}$ is the Wigner (small) $d$-matrix. 
$d_{h_{12} h_{43}}^{\ell}$ can be expressed in terms of hypergeometric function ${}_2 F_1 (a,b;c;x)$ (see for example \cite{biedenharn_louck_1984})
\begin{align}
    \begin{aligned}
        d_{h_{12} h_{43}}^{\ell}(\cos\theta) &= \dfrac{(-1)^{\lambda}}{a!} \sqrt{ \dfrac{(J+a+b)!(J+a)!}{(J)!(J+b)!}} \left( \sin \dfrac{\theta}{2} \right)^a \left( \cos \dfrac{\theta}{2} \right)^b \\
        &~~~ \cdot {}_2 F_1 \left( -J, 1+a+b+J; 1 + a; \sin^2 \frac{\theta}{2} \right)
    \end{aligned}
\end{align}
with $a \geq 0$, $b \geq 0$ and
\begin{gather}
\label{eq:Jablambda}
    (J, a, b, \lambda) = 
    \begin{cases}
        (\ell - |h_{43}|,~ |h_{43}| + h_{12},~ |h_{43}| - h_{12},~|h_{43}| +h_{12} ), & \text{if } - h_{43} \geq |h_{12}|; \\
        (\ell - |h_{43}|,~ |h_{43}| - h_{12},~ |h_{43}| + h_{12},~ \phantom{|h_{43}|} 0 \phantom{+h_{12}} ), &\text{if } + h_{43} \geq |h_{12}|; \\
        (\ell - |h_{12}|,~ |h_{12}| + h_{43},~ |h_{12}| - h_{43},~ \phantom{|h_{12}|} 0 \phantom{-h_{43}} ), &\text{if } - h_{12} \geq |h_{43}|; \\
        (\ell - |h_{12}|,~ |h_{12}| - h_{43},~ |h_{12}| + h_{43},~ |h_{12}| - h_{43} ),  &\text{if } + h_{12} \geq |h_{43}|.
    \end{cases}
\end{gather}
The four cases discussed above are interconnected through the symmetry of the Wigner $d$-matrices: $d^{\ell}_{-h_{43}, -h_{12}}  = d^{\ell}_{h_{12}, h_{43}}$ and $d_{h_{43} h_{12}}^{\ell} = (-1)^{h_{43}-h_{12}} d_{h_{12} h_{43}}^{\ell}$.
Since the $S$-matrix is block-diagonal for different partial waves, the unitarity of the full amplitudes imposes the following partial wave unitarity conditions
\begin{align}
\label{eq:AbsAellUnitarity}
    \operatorname{Abs} A_{\textbf{1234}}^{\ell} = \sum_{X} A^{\ell}_{\textbf{12} \rightarrow X} ( A^{\ell}_{\bar{\textbf{3}} \bar{\textbf{4}} \rightarrow X} )^*
\end{align}
where we have defined the absorptive part of the partial wave
\begin{align}
    \operatorname{Abs} A_{\textbf{1234}}^{\ell} \equiv \dfrac{1}{2 \i} \left( A_{\textbf{1234}}^{\ell} (s + \i \epsilon) - ( A_{\bar{\textbf{3}} \bar{\textbf{4}} \bar{\textbf{1}} \bar{\textbf{2}}}^{\ell} (s + \i \epsilon) )^* \right) .
\end{align}
Here, for example, $\bar{\textbf{3}}$ is a shorthand notation for particle $3$ with helicity $-h_3$, and $A^{\ell}_{\textbf{12} \rightarrow X}$ is the partial wave amplitude from $\textbf{12}$ to state $X$,
where $X$ denotes all intermediate states in a complete basis of the Hilbert space.
We will use the fact that $\operatorname{Abs} A_{\textbf{1234}}^{\ell}$ is PSD, if $\textbf{1}\textbf{2}$ is viewed as the row index and $\textbf{3}\textbf{4}$ as the column index of a matrix, and as will be discussed in Section \ref{sec:Lmoment}, the moment approach can also be easily extended to leverage the fact that $\tilde{I}-{\text{Abs}} A^\ell$ is PSD, where $\tilde{I}$ is a diagonal matrix. 
This PSD condition, referred to as the ``upper bound", can be inferred by isolating the elastic part on the right hand side of \eref{eq:AbsAellUnitarity}. 
Thanks to the Hermitian analyticity of the $S$-matrix $(A_{\bar{\textbf{3}}\bar{\textbf{4}}\bar{\textbf{1}}\bar{\textbf{2}}}^{\ell} (s + \i \epsilon) )^* = A_{\textbf{1234}}^{\ell} (s - \i \epsilon)$, the absorptive part can be written as the discontinuity $\operatorname{Abs} A_{\textbf{1234}}^{\ell} = \operatorname{Disc} A_{\textbf{1234}}^{\ell}/(2\i) \equiv  \left( A_{\textbf{1234}}^{\ell} (s + \i \epsilon) - A_{\textbf{1234}}^{\ell} (s - \i \epsilon) \right)/(2\i)$.
If the theory is time reversal invariant, we additionally have $( A_{\bar{\textbf{3}} \bar{\textbf{4}} \bar{\textbf{1}} \bar{\textbf{2}}}^{\ell} (s + \i \epsilon) )^* = (A_{\textbf{1}\textbf{2}\textbf{3}\textbf{4}}^{\ell} (s + \i \epsilon) )^*$, which leads to $\operatorname{Abs} A^{\ell}_{\textbf{1}\textbf{2}\textbf{3}\textbf{4}} (s) = \operatorname{Im} A^{\ell}_{\textbf{1}\textbf{2}\textbf{3}\textbf{4}} (s+ \i \epsilon)$.

The dispersive sum rules arise fundamentally due to the analytic nature of scattering amplitudes. 
The amplitudes are widely conjectured to be analytic apart from the singularities already known in the perturbation theory, but rigorous results about it are scarce \cite{Eden:1966dnq, Martin:1969ina}. 
We shall consider an EFT scenario where the theory is weakly coupled in the low energy region such that we can neglect the loop corrections to the amplitudes below the cutoff $\Lambda$, which means that $A_{\textbf{1234}} (s,t)$ only have simple poles below the cutoff, arising from the exchange diagrams of the light particles. 
The analyticity we assume in this paper is that for fixed $t$ below the cutoff, the amplitudes $A_{\textbf{1234}} (s,t)$ are analytic in the whole complex $s$ plane except for some poles and branch cuts on the real axis inferred from unitarity and the bound states in the theory, and the amplitudes are crossing symmetric. 
Another essential ingredient for deriving the dispersive sum rules is the asymptotic behavior of the amplitudes, which will allow us to get rid of the contour integral at infinity.
We shall assume that, for fixed $t$, the UV amplitudes are bounded by polynomials of $s$ with degree $N_s$ for large $|s|$:
\begin{align}
\label{eq:Asinfinity0}
    \lim_{\Abs{s} \rightarrow \infty} \frac{A_{\textbf{1234}}(s,t)}{s^{N_{s}}} = 0.
\end{align}
For fields with spin less than 2, we can take $N_{s}=2$, which can be rigorously proven for a theory with a mass gap
\cite{Jin:1964zza}; but for the spin-2 case, we will take $N_{s}=3$ to avoid the complications of the $t$-channel pole \cite{Alberte:2020jsk, Tokuda:2020mlf, Caron-Huot:2021rmr}.
The $N_{s}=2$ dispersive sum rules exist for the spin-2 case, but their moment problem formulation is challenging.

\begin{figure}[h]
    \centering
    \includegraphics[scale=0.45]{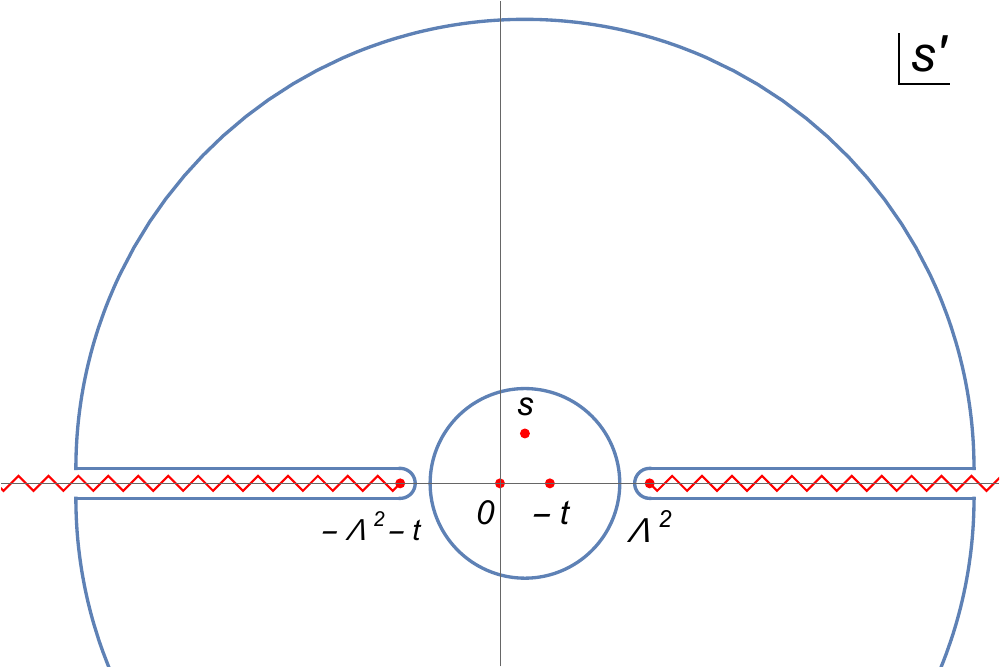}
    \caption{
    Analytic structure of the $2$-to-$2$ scattering amplitude on the complex $s'$-plane, along with the contour for deriving the fixed-$t$ dispersive sum rules.
    }
    \label{fig:contour}
\end{figure}
With these ingredients established, the fixed-$t$ dispersive sum rules can be easily obtained by using Cauchy's integral formula on complex $s$ plane for the integration contour of Figure \ref{fig:contour}:
\begin{align}
    \begin{aligned}
    \label{eq:A0orignal}
    &~~~~ A_{\textbf{1234}} (s, t) - \sum_{\text{EFT poles }} \operatorname*{Res}_{s^\prime \rightarrow s_{\text{poles}}} A_{\textbf{1234}}(s^\prime, t) \\
    &= \int^{+\infty}_{\Lambda^2} \dfrac{\dif s^\prime}{\pi} \dfrac{\operatorname{Abs} A_{\textbf{1}\textbf{2}\textbf{3}\textbf{4}} (s^\prime, t)}{s^\prime - s} + \int_{\Lambda^2}^{+\infty} \dfrac{\dif s^\prime}{\pi} \dfrac{\operatorname{Abs} A_{\textbf{1432}} (s^\prime, t)}{s^\prime - u} + \int_{C^{\pm}_{\infty}} \dfrac{\dif s^\prime}{2 \pi \i} \dfrac{A_{\textbf{1}\textbf{2}\textbf{3}\textbf{4}}(s^\prime,t)}{s^\prime - s}
    \end{aligned} 
\end{align}
where we have used $su$-crossing symmetry for the second integral on the right hand side and the summation over the EFT poles is for all the poles within the EFT. 
\eref{eq:A0orignal}, as it stands, is not very useful, as the terms on the right hand side can be divergent. 
These terms would be convergent if the integrands go like $1/s'{}^{N_s}$ at large $|s'|$ (cf. \eref{eq:Asinfinity0}).
To remedy the divergence, we can make use of the following algebraic identity
\begin{align}
    \dfrac{1}{s^\prime - s} &= \dfrac{(s - s_0)^{N_{s}}}{(s^\prime - s_0)^{N_{s}}} \dfrac{1}{s^\prime - s} +  \sum_{i=0}^{N_{s}-1} \dfrac{N_{s}!}{i! (N_{s}-i)!} \dfrac{(s-s_0)^{i} (s^\prime - s)^{N_{s}-i-1}}{(s^\prime-s_0)^{N_{s}}} 
\end{align}
where the arbitrarily chosen $s_0$ is called a subtraction point. 
We will choose $s_0=-t/2$.
Then, introducing an auxiliary variable 
\begin{align}
v = s+ \frac{t}{2}, 
\end{align}
\eref{eq:A0orignal} can be rewritten as $N_s$-th subtracted dispersion relations
\begin{align}
\label{eq:centralDisp}
    & A_{\textbf{1234}} (s, t) - \sum_{\text{EFT poles }} \operatorname*{Res}_{s^\prime \rightarrow s_{\text{poles}}} A_{\textbf{1234}} (s^\prime, t) = \\
    &~~\sum_{i=0}^{N_{s}-1} b^{i,(N_{s})}_{\textbf{1234}} (t) s^i + \int^{+\infty}_{\Lambda^2} \dfrac{\dif s^\prime}{\pi} \dfrac{v^{N_{s}}}{(s^\prime + t/2)^{N_{s}}} \left( \dfrac{\operatorname{Abs} A_{\textbf{1234}} (s^\prime, t)}{s^\prime + t/2 - v} + \dfrac{(-1)^{N_{s}} \operatorname{Abs} A_{\textbf{1432}} (s^\prime, t)}{s^\prime + t/2 + v} \right)\nonumber
\end{align}
We have collected the integrals whose integrands involve polynomials in $s$ of degree less than $N_{s}$ as $\sum_{i=0}^{N_{s}-1} b^{i,(N_{s})}_{\textbf{1234}} (t) s^i$, and the remaining contributions from the infinite upper semicircle and lower semicircle vanish thanks to \eref{eq:Asinfinity0}. The usefulness of these dispersion relations arises from the fact that we can approximate the left hand side with the EFT, and the positivity bounds arise from the fact that, despite the unknown details of the UV amplitudes on the right hand side, they must comply with partial wave unitarity conditions.

To draw connections with the mathematical moment problem, we shall further extract, from \eref{eq:centralDisp}, dispersive sum rules that directly link the Wilson coefficients with the UV partial amplitudes. With sufficient subtractions, the right hand side of \eref{eq:centralDisp} is analytic both around $s = 0$ and $t = 0$, so we can Taylor-expand both sides around $s = 0$ and $t = 0$ and match the coefficients of $v^{i} t^{j}$. For the right hand side, we also perform partial wave expansion for $\operatorname{Abs} A_{\textbf{1}\textbf{2}\textbf{3}\textbf{4}} (s, t)$ and further expand the Wigner $d$-matrices. The results are
\begin{align}
    \begin{aligned} 
    \label{eq:gijrelation0}
        g^{i,j}_{\textbf{1234}} &= \sum_{k=0}^{j} \int^{+\infty}_{\Lambda^2} \sum_{\ell}  8 (2 \ell + 1 )  \dfrac{ \mathcal{D}_{\ell, h_{12}, h_{43}}^{k} \mathcal{E}_{i+1}^{j-k} }{(s^\prime)^{i+j+1}} \operatorname{Abs} A_{\textbf{1234}}^{\ell} (s^\prime) \dif s^\prime \\
        &~~~~~ + \sum_{k=0}^{j} (-1)^i  \int^{+\infty}_{\Lambda^2} \sum_{\ell} 8 (2 \ell + 1 )  \dfrac{\mathcal{D}_{\ell, h_{1\bar{4}}, h_{\bar{2}3}}^{k} \mathcal{E}_{i+1}^{j-k} }{(s^\prime)^{i+j+1}} \operatorname{Abs} A^\ell_{\textbf{1432}} (s^\prime) \dif s^\prime
    \end{aligned}
\end{align}
where $g^{i,j}_{\textbf{1234}}$ is defined via
\begin{align}
    &\sum_{i \geq N, j \geq 0} g^{i,j}_{\textbf{1234}} v^i t^j \equiv A_{\textbf{1234}} (s, t) - \sum_{\text{EFT poles }} \operatorname*{Res}_{s^\prime \rightarrow s_{\text{poles}}} A_{\textbf{1234}}(s^\prime, t) - \sum_{i=0}^{N_{s}-1} b^{i,(N_{s})}_{\textbf{1234}} (t) s^i
\end{align}
For tree-level EFT amplitudes, the amplitude coefficients/EFT coefficients, $g^{i,j}_{\textbf{1234}}$, can be easily linked to the Wilson coefficients, 
so we will be directly concerned with the amplitude coefficients in this paper, which have the benefit of being free of the ambiguity of field redefinition.
$\mathcal{D}_{\ell, h_{12}, h_{43}}^{k}$ and $\mathcal{E}_{i+1}^{k}$ are Taylor coefficients defined by the following series
\begin{align}
    d_{h_{12} h_{43}}^{\ell} (\operatorname{arccos} (1 + 2 z) ) &\equiv  \sum_{k=0}^{\infty} \mathcal{D}_{\ell, h_{12}, h_{43}}^{k} z^k, &
    \dfrac{1}{(1 + z/2)^{i+1}} &\equiv  \sum_{k=0}^{\infty} \mathcal{E}_{i+1}^{k} z^k.
\end{align}
Explicitly, for $h_{12} \geq |h_{43}|$, for example, we have
\begin{align}
    J = \ell - h_{12}, a = h_{12} - h_{43}, b = h_{12} + h_{43}, \lambda = h_{12} - h_{43}
\end{align}
and $\mathcal{D}_{\ell, h_{12}, h_{43}}^{k}$ is given by 
\begin{align}
    \begin{aligned}
        \mathcal{D}_{\ell, h_{12}, h_{43}}^{k} &= \dfrac{2^k}{k!} \left( \dfrac{\dif^{k}}{\dif z^{k}} d_{h_{12} h_{43}}^{\ell} (\arccos{z}) \right)_{z \rightarrow 1} \nonumber \\
        &= \sum_{j=0}^{k} \dfrac{2^k}{j! (k-j)!} \left[ \dfrac{\dif^{k-j}}{\dif x^{k-j}} \left( \dfrac{1+z}{2} \right)^{\frac{h_{12}+h_{43}}{2}} \left(\dfrac{1-z}{2} \right)^{\frac{h_{12}-h_{43}}{2}} \right]_{z \rightarrow 1} \mathcal{F}_{\ell, h_{12}, h_{43}}^{j}  \\
    \end{aligned}
\end{align}
where we have defined
\begin{align}
\mathcal{F}_{\ell, h_{12}, h_{43}}^{j} &= \dfrac{(-1)^{ h_{12} - h_{43} }}{( h_{12} - h_{43} )!}  \sqrt{\frac{(\ell + h_{12})! (\ell - h_{43})!}{(\ell - h_{12})! (\ell + h_{43})!}}  \nonumber\\
&~~~~~~~~~~ \cdot \left[ \dfrac{\dif^j}{\dif z^j} {}_2 F_1 \left( h_{12} - \ell, 1 + h_{12} + \ell; 1 + h_{12} - h_{43}; \dfrac{1-z}{2} \right) \right]_{z \rightarrow 1} \ .
\end{align}
Here, the factorials should be understood as the corresponding Gamma functions, i.e., $n!\equiv \Gamma (n+1)$, for the cases of half integer spins. To formulate \eref{eq:gijrelation0} as a moment problem, we want to further simplify the expression for $ \mathcal{D}_{\ell, h_{12}, h_{43}}^{k}$. To this end, note that ${}_2 F_1 (a,b;c;x)$ can be written as 
\begin{align}
    \sum_{n=0}^{\infty} \left( \prod_{i=1}^{n} \dfrac{(a+i)(b+i)}{(c+i)} \right) \dfrac{x^n}{n!}.
\end{align}
where we have defined $\prod_{i=1}^{n=0} {(a+i)(b+i)}/{(c+i)} \equiv 1$. 
With this expression, it is easy to see that
\begin{align}
    \begin{gathered}
    \label{eq:Fsimple}
    \mathcal{F}_{\ell, h_{12}, h_{43}}^{j} \!=\! 
    \begin{cases}
         \dfrac{(-1)^{h_{12} - h_{43}}}{2^j (h_{12} - h_{43} + j)!} \sqrt{
            \dfrac{1
            }
            {
                \mathcal{G} (\ell,h_{12}) \mathcal{G} (\ell,h_{43})
            }
         }\mathcal{G} (\ell,h_{12}+j), & \ell \geq h_{12} + j; \\
         0, & \ell < h_{12} + j.
    \end{cases}
    \end{gathered}
\end{align}
where we have introduced
$\mathcal{G} (\ell,h) \equiv  \Gamma(\ell+h+1)/\Gamma(\ell-h+1) \equiv \prod_{i=1}^{|h|} (\ell (\ell + 1) - i (i - 1) )^{\operatorname*{sign} (h)}$ with $\operatorname*{sign} (h)$ denoting the sign of $h$.
We see that all the $\ell$ dependence in $\mathcal{F}_{\ell, h_{12}, h_{43}}^{j}$, and thus $\mathcal{D}_{\ell, h_{12}, h_{43}}^{k}$, is through the combination $\ell(\ell+1)$, the eigenvalue of the spin Casimir operator.
Similarly, we can also get explicit expressions for $\mathcal{F}_{\ell, h_{12}, h_{43}}^{j}$ for all other three cases of \eref{eq:Jablambda}, where all the $\ell$ dependence is also through the combination $\ell(\ell+1)$. Because of this $\ell(\ell+1)$ dependence, we shall later use $\ell(\ell+1)$, instead of $\ell$, as a moment variable.
As will be discussed in more detail shortly, we see from \eref{eq:Fsimple} that for generic spins the sum rules will involve square roots of rational functions, in contrast to the scalar case where $\mathcal{F}_{\ell, h_{12}, h_{43}}^{j}$ is simply a polynomial in $\ell(\ell+1)$. 
Fortunately, the moment problem defined on a semialgebraic set has been well established, which can accommodate semialgebraic functions, including square roots of rational functions \cite{Lasserre2012}. 

The above sum rules already contain the information of the $su$ crossing symmetry of the amplitudes. 
However, the amplitudes have more crossing symmetry that is not contained in the fixed-$t$ dispersion relations. 
These extra crossing symmetry can be incorporated by imposing the $st$ crossing symmetry \cite{Tolley:2020gtv,Caron-Huot:2020cmc,Du:2021byy}
\begin{align}
\label{eq:AstSymmetry}
A_{\textbf{1234}}(s,t)=A_{\textbf{1324}}(t,s)
\end{align}
To translate this into constraints on the EFT coefficients, one can first subtract the pole terms from both sides, and then Taylor-expand both sides and match the coefficients in front of $s^i t^j$: $a^{i,j}_{\textbf{1234}} = a^{j,i}_{\textbf{1234}}$. 
Expressing these $a^{i,j}_{\textbf{1324}}$ in terms of $g^{i,j}_{\textbf{1234}}$, we obtain
\begin{align}
    \label{eq:st_sym_1}
     \sum_{k=0}^{j} \dfrac{(i+k)!}{2^k i!k!} g^{i+k,j-k}_{\textbf{1234}} = \sum_{k=0}^{i} \dfrac{(j+k)!}{2^k j!k!} g^{j+k,i-k}_{\textbf{1324}} . 
\end{align}
Because of the $N_s$-th subtraction, we will make use of the null constraints with only $g^{i\geq N_s,j\geq 0}_{\textbf{1234}}$. 
Note that for various reasons, some of the lower-order EFT coefficients may vanish, in which case certain lower-order null constraints can also be used.

Alternatively, one can substitute \eref{eq:gijrelation0} into \eref{eq:st_sym_1} and get a series of integral equations. 
Each of these integral equations also contains a sum over all the different partial waves/spin, and dictates that a certain spectral combination of the UV spins need to cancel out identically, which strongly constrains how the high spins can affect the EFT coefficients. 
In principle, fully combining the $su$ crossing symmetry with the $st$ symmetry will lead to the full crossing symmetry. 
However, practically, one often needs to resort to truncation in evaluating a specific problem, in which case it is sometimes beneficial to also impose the $tu$ crossing symmetry. For example, $\ell$ can be only even integers for the single scalar case due to $tu$ crossing symmetry.

\subsection{Matrix moment problem formulation}
\label{sec:momentFormulation}

Now, we are ready to formulate the problem of extracting the bounds on $g^{i,j}_{\textbf{1234}}$ as a matrix moment problem. For the single scalar case, this has been done in Refs.~\cite{ Arkani-Hamed:2020blm, Bellazzini:2020cot, Chiang:2021ziz}, in which case it reduces to a ``standard''/non-matrix moment problem. 
For a single field with spin, which already contains multiple degrees of freedom, or a multi-field case, the matrix moment generalization is needed. 

Apart from the matrix generalization, our moment problem formulation will be slightly different from the bi-variate formulation of Ref.~\cite{Chiang:2021ziz}. For the most part, we will use a uni-variate ({\it i.e.}, the UV scale $1/s^\prime$) moment formulation, and only adopt a multi-variate formulation for the high partial waves, which is used to accelerate the numerical convergence. 
If the uni-variate method is evaluated to a high partial wave order, the multi-variate part will be not needed, but that will be computationally more costly, although still faster than a purely multi-variate formulation.

Note that for the multi-variate moment formulation, the cases with spin (in the IR) are slightly differently from the multi-scalar case. 
This is because, as can be seen from \eref{eq:Fsimple}, for the multi-scalar case $\mathcal{D}_{\ell, h_{12}=0, h_{43}=0}^{k}$ is a $k$-th order polynomial function of $\ell(\ell+1)$, while for the cases with spinning particles $\mathcal{D}_{\ell, h_{12}, h_{43}}^{k}$ may contain square roots of rational functions of $\ell(\ell+1)$. 
Thus, for particles with spin, its multi-variate formulation needs more than two moment variables. 
In the following, we shall first show the formulation of the moment problem for a multi-scalar theory and then discuss the differences for the fields with spin.

\subsubsection{Multi-scalar}
\label{sec:multiScalar}

For the multi-scalar case, we can write the right hand side of \eref{eq:gijrelation0} as a linear combination of uni-variate moments and bi-variate moments. To see this, let us define moment variables
\begin{align}
\label{eq:x1x2def}
    x_1 = \dfrac{\Lambda^2}{s^\prime},~~~ x_2 = \dfrac{\ell(\ell+1) \Lambda^2}{s^\prime}
\end{align}
and the PSD spectral functions for the low and high partial wave parts are respectively 
\begin{align}
\label{sigmaspe}
    \sigma_{\textbf{1234}}^{\ell} (x_1) &= 16 (2 \ell + 1) x_1^{N_s-1} \operatorname*{Abs} A^{\ell}_{\textbf{1234}} (\Lambda^2 / x_1 ) \\
\label{rhospe}
    \rho_{\textbf{1234}} (x_1, x_2) &= \sum_{\ell = \ell_{\text{M}}}^{\infty} \sigma_{\textbf{1234}}^{\ell} (x_1) \delta \Big(x_2 - \ell (\ell + 1) x_1 \Big).
\end{align}
where $N_s$ is the subtraction order of the relevant dispersion relation. (For a large $\ell$, $x_2$ is essentially the square of the impact parameter $b=\ell/\sqrt{s'}$ with $\ell$ being the angular momentum.)
With these definitions, we can split the right hand side of \eref{eq:gijrelation0} into two parts for each of the $s$ and $u$ channels:
\begin{align}
\label{gij1234Exp}
    g^{i,j}_{\textbf{1234}} = \sum_{\ell=0}^{\ell_{\text{M}}-1} U^{i,j}_{\ell (\ell + 1), \textbf{1234}} \mathfrak{a}^{\gamma}_{\ell,\textbf{1234}} + \sum_{\gamma_2^\prime} V^{(\gamma)}_{\gamma_2,\gamma_2^\prime} \mathfrak{b}_{\textbf{1234}}^{\gamma_1,\gamma_2^\prime} + (\text{$u$-channel})
\end{align}
with 
\begin{align}
\label{amomentPro}
    \mathfrak{a}^{\gamma}_{\ell,\textbf{1234}} &= \int_{\mathcal{K}_{\mathfrak{a}}} x_1^\gamma \sigma_{\textbf{1234}}^{\ell} (x_1) \dif x_1 
    \\
\label{bmomentPro}
    \mathfrak{b}_{\textbf{1234}}^{\gamma_1,\gamma_2} &= \int_{\mathcal{K}_{\mathfrak{b}}} \ x_1^{\gamma_1} x_2^{\gamma_2} \rho_{\textbf{1234}} (x_1, x_2) \dif x_1 \dif x_2 
\end{align}
where we have defined $\gamma = i + j - N_s$,
$\gamma_1 = i - N_s$, $\gamma_2 = j$ and semialgebraic sets\,\footnote{In a moment problem, the domain of integration $\mathcal{K}$ can generally be a semialgebraic set, which is a finite union of subsets of $\mathbb{R}^n$ delineated by a finite system of polynomial inequalities.}
\begin{align}
    \mathcal{K}_{\mathfrak{a}} &= \{ x_1 | x_1 \geq 0, 1 - x_1 \geq 0  \} \\
    \mathcal{K}_{\mathfrak{b}} &= \{ (x_1, x_2) |  x_1 \geq 0, 1 - x_1 \geq 0, x_2 \geq 0, x_2 - \ell_{\infty} (\ell_{\infty} + 1) x_1 \leq 0,
    \\
    & ~~\big( x_2 - \ell (\ell + 1) x_1 \big) \big( x_2 - (\ell + 1) (\ell + 2) x_1 \big) \geq 0, ~\ell =\ell_{\text{M}},\ell_{\text{M}}+1, \ell_{\text{M}}+2, \cdots, \ell_{\infty}-1 \}\nonumber
\end{align}
$\ell_{\infty}$ is a large integer regulator introduced to make $\mathcal{K}_{\mathfrak{b}}$ strictly semialgebraic, but the numerical results are insensitive to it as long as $\ell_{\infty}$ is sufficiently large. 
$U^{i,j}_{\ell (\ell + 1)}$ and $V^{(\gamma)}_{\gamma_2,\gamma_2^\prime}$ can be easily inferred from \eref{eq:gijrelation0}. From the partial wave unitarity conditions, we can infer that $\sigma_{\textbf{1234}}^{\ell} (x_1)$ and $\rho_{\textbf{1234}} (x_1, x_2)$ are positive semi-definite matrices, if $\textbf{12}$ ($\textbf{34}$) is viewed as the row (column) index, so $\sigma_{\textbf{1234}}^{\ell} (x_1) \dif x_1$ and $\rho_{\textbf{1234}} (x_1, x_2) \dif x_1 \dif x_2$ are matrix-valued positive measures. This means that \eref{amomentPro} and \eref{bmomentPro} define a uni-variate and bi-variate matrix moment sequence respectively. 

We could have formulated \eref{eq:gijrelation0} as a pure uni-variate moment problem ($\ell_{\text{M}}=\infty$ and without $\mathfrak{b}_{\textbf{1234}}^{\gamma_1,\gamma_2}$) or a pure bi-variate moment problem ($\ell_{\text{M}}=0$ and without $\mathfrak{a}_{\ell,\textbf{1234}}^{\gamma}$), but neither of them is numerically as efficient as the mixed formulation (\ref{gij1234Exp}). 
In particular, in the pure uni-variate formulation, it is awkward to include the large/infinite $\ell$ contributions, as $V^{(\gamma)}_{\gamma_2,\gamma_2^\prime}$ can become very large/go to infinity as $\ell\to \infty$.

\begin{figure}
    \centering
    \includegraphics[width=0.3\linewidth]{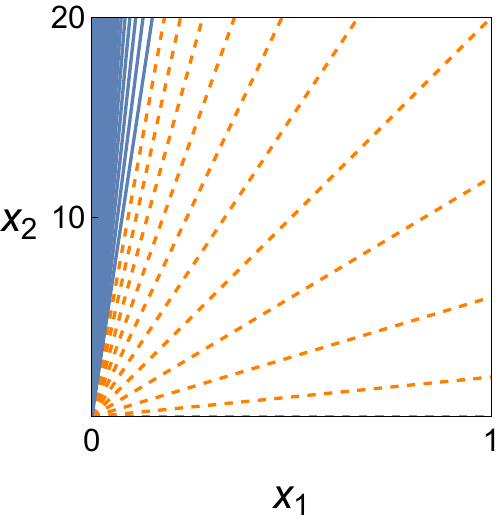}
~~~~~~~\raisebox{68pt}{$\Longrightarrow$}~~~~~
 \includegraphics[width=0.3\linewidth]{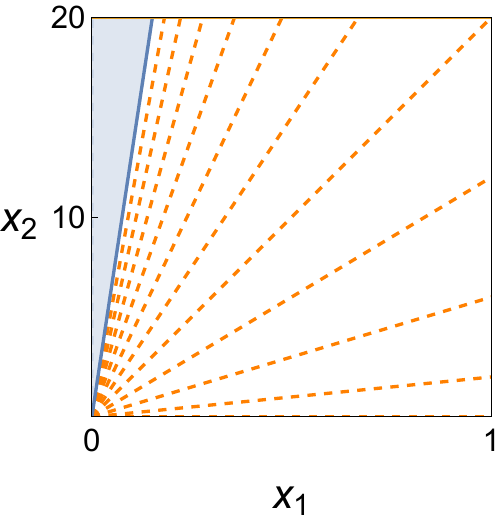}
    \caption{Set $\mathcal{K}_{\mathfrak{b}}$ (blue lines) and its approximate set $\mathcal{K}'_{\mathfrak{b}}$ (blue region) as the integration regions of the moment problems. The orange dashed lines are not part of $\mathcal{K}_{\mathfrak{b}}$ and $\mathcal{K}'_{\mathfrak{b}}$. Instead, each dashed line is formulated as a uni-variate moment problem $\mathfrak{a}_{\ell,\textbf{1234}}^{\gamma_1,\gamma_2}$. This mixed-variate formulation improves numerical efficiency. $\ell_{\rm M}=10$ is chosen for these plots.}
    \label{fig:KbKbp}
\end{figure}

The rationale behind the mixed formulation is that we shall compute the dominant contributions to $g^{i,j}_{\textbf{1234}}$ from the $\mathfrak{a}^{\gamma}_{\ell,\textbf{1234}}$ moments with just the first few leading $\ell$, and then the $\mathfrak{b}_{\textbf{1234}}^{\gamma_1,\gamma_2}$ moments allow us to obtain the subleading corrections from the large $\ell$. In the partial wave expansion, it is natural to expect the results are dominated by the low spins \cite{Bern:2021ppb}. 
This is also implied by the null constraints, which explains the effectiveness of formulating the leading spins as uni-variate moment problems. 
$\mathcal{K}_{\mathfrak{b}}$ is a set of line segments. 
For a relative large $\ell_{\text{M}}$, we relax $\mathcal{K}_{\mathfrak{b}}$ to a continuous region $\mathcal{K}'_{\mathfrak{b}}$:
\begin{align}
\label{eq:Kbpdef}
\mathcal{K}^{\prime}_{\mathfrak{b}} = \{ (x_1, x_2) | 0 \leq x_1 \leq 1, x_2 - \ell_{\text{M}} (\ell_{\text{M}} + 1) x_1 \geq 0, x_2 - \ell_{\infty} (\ell_{\infty} + 1) x_1 \leq 0 \}.
\end{align}
See Figure \ref{fig:KbKbp} for a pictorial description of the set $\mathcal{K}_{\mathfrak{b}}$ and the approximate set $\mathcal{K}'_{\mathfrak{b}}$. 

With the formulation (\ref{gij1234Exp}), we can solve the matrix moment problems of 
$\mathfrak{a}^{\gamma}_{\ell,\textbf{1234}}$ and $\mathfrak{b}_{\textbf{1234}}^{\gamma_1,\gamma_2}$ separately, extract the conditions for each, and then combine them to infer the bounds on $g^{i,j}_{\textbf{1234}}$. 
Specifically, note that, denoting
$\mathfrak{a}^{\gamma}_{\ell,\textbf{1234}}$ and $\mathfrak{b}_{\textbf{1234}}^{\gamma_1,\gamma_2}$ collectively as
\begin{align}
\mathfrak{m}_{\textbf{1234}}^{\gamma_1,\gamma_2}= \mathfrak{a}^{\gamma}_{\ell,\textbf{1234}} \text{ or }\mathfrak{b}_{\textbf{1234}}^{\gamma_1,\gamma_2} ,
\end{align}
the solvability condition of the matrix moment problem for $\mathfrak{m}^{\gamma_1 \gamma_2}_{\textbf{1234}}$ is just a set of explicit conditions on $\mathfrak{m}^{\gamma_1 \gamma_2}_{\textbf{1234}}$.
These conditions can be written as linear matrix inequalities of the form: $\mathfrak{m}^T \mathcal{M} \geq 0$, where $\mathfrak{m}$ is a vector whose components are the relevant elements of $\mathfrak{m}^{\gamma_1 \gamma_2}_{\textbf{1234}}$ and $\mathcal{M}$ is a vector whose components are constant matrices.
To compute the bounds on a specific $g^{i,j}_{\textbf{1234}}$, we express it as a linear combination of $\mathfrak{m}_{\textbf{1234}}^{\gamma_1,\gamma_2}$ via \eref{gij1234Exp} at first, and then optimize it as the objective in a garden-variety SDP.
In this process, we can also include the null constraints (\ref{eq:st_sym_1}) to get two-sided bounds. 
These null constraints can be implemented by direct substitution or by imposing them as part of the constraints in the SDP. This final-step SDP can be easily implemented in, say, {\tt Mathematica} with minimal coding, if extremely high accuracy is not required. 

\subsubsection{Fields with spin}
\label{sec:formSpin}

As mentioned above, the main difference for fields with spin is that $\mathcal{D}_{\ell, h_{12}, h_{43}}^{k}$ is now generally a function of $\ell(\ell+1)$ that involves square roots of rational functions of $\ell(\ell+1)$, not merely a polynomial in $\ell(\ell+1)$ of degree $k$. 
For a fast and efficient determination of the bounds, we still want to adopt the mixed-variate formulation. 
That is, for the low partial waves, we continue use the uni-variate moment formulation, for which the non-polynomial nature of $\mathcal{D}_{\ell, h_{12}, h_{43}}^{k}$ does not create any difficulty at all, as we formulate each partial wave as a uni-variate moment problem. 
However, for the high partial wave part, the bi-variate formulation with Eqs.~(\ref{eq:x1x2def})-(\ref{sigmaspe}) fails to yield \eref{gij1234Exp}, due to the presence of square roots of rational functions of $\ell(\ell+1)$. 
Instead, the equivalence of \eref{gij1234Exp} for fields with spin becomes
\begin{align}
\label{gij1234Expspin}
    &g^{i,j}_{\textbf{1234}} = \sum_{\ell=0}^{\ell_{\text{M}}-1} U^{i,j}_{\ell (\ell + 1)} \mathfrak{a}^{\gamma}_{\ell,\textbf{1234}} + \sum_{k, \gamma_2^\prime} V^{(k,\gamma)}_{\gamma_2,\gamma_2^\prime} \mathfrak{b}_{\textbf{1234}}^{k,\gamma_1,\gamma_2^\prime} + \text{($u$ channel)}
    \\
    &\mathfrak{b}_{\textbf{1234}}^{k, \gamma_1,\gamma_2} = \int_{\mathcal{K}_{\mathfrak{b}}} \  \mathscr{F}_k(x_1, x_2) x_1^{\gamma_1} x_2^{\gamma_2} \rho_{\textbf{1234}} (x_1, x_2) \dif x_1 \dif x_2
\end{align}
where $\mathscr{F}_k$ takes the form of $\mathscr{F}_k=\sqrt{\cdots}$, with $\cdots$ potentially also containing square roots.  
Thus, we get a so-called {\it generalized moment problem} \cite{anjos2011handbook} for the the high partial wave part. 
For our specific problem, $\mathscr{F}_k$ are positive semialgebraic functions\,\footnote{For a semialgebraic set $\mathcal{K}$, a semialgebraic function $\mathscr{F}(x_1, \cdots, x_n)$ defined on $\mathcal{K}$ is a function such that $\{ (x_1, x_2, \cdots, x_n, \mathscr{F} ) \mid (x_1, x_2, \cdots, x_n) \in \mathcal{K} \}$ is also a semialgebraic set. 
Functions that are rational and include radical roots are semialgebraic functions.} on $\mathcal{K}_\mathfrak{b}$. 
In this case, we can easily transform a generalized moment problem into a standard moment problem by introducing extra moment variables.

Let us demonstrate this with the example of massless spin-$1$ particle scattering. In this case, we will have the following three bi-variate moment sequences:
\begin{align}
    \mathfrak{b}_{\textbf{1234}}^{ \gamma_1,\gamma_2} &= \int_{\mathcal{K}_{\mathfrak{b}}} \ x_1^{\gamma_2} x_2^{\gamma_2} \rho_{\textbf{1234}} (x_1, x_2) \dif x_1 \dif x_2 \\
    \hat{\mathfrak{b}}_{\textbf{1234}}^{ \gamma_1,\gamma_2} &= \int_{\mathcal{K}_{\mathfrak{b}}} \  \sqrt{ x_2 (x_2 - 2 x_1)} 
    x_1^{\gamma_2} x_2^{\gamma_2} \rho_{\textbf{1234}} (x_1, x_2) \dif x_1 \dif x_2 \\
    \tilde{\mathfrak{b}}_{\textbf{1234}}^{ \gamma_1,\gamma_2} &= \int_{\mathcal{K}_{\mathfrak{b}}} \ \frac{1}{\sqrt{x_2 (x_2 - 2 x_1)}} x_1^{\gamma_2} x_2^{\gamma_2} \rho_{\textbf{1234}} (x_1, x_2) \dif x_1 \dif x_2
\end{align}
As $1/\sqrt{x_2 (x_2 - 2 x_1)}$ is positive, we can first absorb it into the positive spectral function $\rho_{\textbf{1234}}$:
\begin{align}
\rho^{\prime}_{\textbf{1234}} (x_1, x_2) = \frac{1}{\sqrt{x_2 (x_2 - 2 x_1)}} \rho_{\textbf{1234}} (x_1, x_2)
\end{align}
and the moment problems become
\begin{align}
\label{bmomentPro1}
    \mathfrak{b}_{\textbf{1234}}^{ \gamma_1,\gamma_2} &= \int_{\mathcal{K}_{\mathfrak{b}}} \ 
    \sqrt{x_2 (x_2 - 2 x_1) } x_1^{\gamma_1} x_2^{\gamma_2} \rho^\prime_{\textbf{1234}} (x_1, x_2) \dif x_1 \dif x_2 \\
\label{bmomentPro2}
    \hat{\mathfrak{b}}_{\textbf{1234}}^{ \gamma_1,\gamma_2} &= \int_{\mathcal{K}_{\mathfrak{b}}} x_2 (x_2 - 2 x_1) 
    x_1^{\gamma_1} x_2^{\gamma_2} \rho^\prime_{\textbf{1234}} (x_1, x_2) \dif x_1 \dif x_2 
    \\
\label{bmomentPro3}
    \tilde{\mathfrak{b}}_{\textbf{1234}}^{ \gamma_1,\gamma_2} &= \int_{\mathcal{K}_{\mathfrak{b}}} \ x_1^{\gamma_1} x_2^{\gamma_2} \rho^\prime_{\textbf{1234}} (x_1, x_2) \dif x_1 \dif x_2
\end{align}
(For the case of multiple different square roots in the denominator, one can simply absorb a large common factor into $\rho'_{\textbf{1234}} (x_1, x_2)$ to get rid of the square roots in the denominator.) The moment problems (\ref{bmomentPro2}) and (\ref{bmomentPro3}) are bi-variate, the same as the multi-scalar case. For \eref{bmomentPro2}, we introduce a new moment variable
\begin{align}
x_3= \sqrt{x_2 (x_2 - 2 x_1)}
\end{align}
along with
\begin{align}
\rho^{\prime\prime}_{\textbf{1234}} (x_1, x_2, x_3) = \rho^{\prime}_{\textbf{1234}}(x_1, x_2 )\delta (x_3 - \sqrt{x_2 (x_2 - 2 x_1)})
\end{align}
which allows us to write \eref{bmomentPro2} as one tri-variate problem 
\begin{align}
    \mathfrak{b}_{\textbf{1234}}^{ \gamma_1,\gamma_2, \gamma_3} &= \int_{\mathcal{K}^{(3)}_{\mathfrak{b}}} x_1^{\gamma_2} x_2^{\gamma_2} x_3^{\gamma_3} \rho^{\prime\prime}_{\textbf{1234}} (x_1, x_2, x_3) \dif x_1 \dif x_2 \dif x_3
\end{align}
where now the semialgebraic set becomes
\begin{align}
    \mathcal{K}^{(3)}_\mathfrak{b} &= \{(x_1, x_2, x_3) | (x_1, x_2)\in \mathcal{K}_\mathfrak{b}  ,~
 x_3^2 - x_2 (x_2 - 2 x_1) = 0 \}.
\end{align}
With one extra moment variable, the solution of the moment problem will be expressed with Hankel matrices (which will be introduced in the next section) with very high dimensions. 

For scatterings between massless spin-$s$ particles, the relevant new moment variables needed are listed in Table \ref{tab:my_label}. We see that in those cases only one extra variable is needed. For scatterings with mixed spins or between massive particles, more moment variables are needed. For example, for scatterings betwen massive spin-2 particles, we would need the following new moment variables $x_3=\sqrt{x_1 x_2}$ $x_4=\sqrt{x_1 (x_2 - 2 x_1)}$, $x_5=\sqrt{x_1 (x_2 - 6 x_1)}$, $x_6=\sqrt{x_1 (x_2 - 12 x_1)}$ or a subset of them depending on the specific scattering process.

\begin{table}[h]
    \centering
    \begin{tabular}{|c|c|c|c|c|c|}
        \hline
       extra moment variable & spin-$0$ & spin-$1$ & spin-$2$ & spin-$\frac{1}{2}$ \\
        \hline
        $x_3=\sqrt{x_1 x_2}$  & & & & $\checkmark$ \\
        \hline
        $x_3=\sqrt{x_2 (x_2 - 2 x_1)}$ & & $\checkmark$ & & \\
        \hline
        $x_3=\sqrt{x_2 (x_2 - 2 x_1) (x_2 - 6 x_1)  (x_2 - 12 x_1)}$ & & & $\checkmark$ & \\
        \hline
    \end{tabular}
    \label{tab:my_label}
    \caption{Additional moment variable needed for a scattering amplitude between massless particles with (equal) spin in 4D.}
\end{table}

\section{Solvability of matrix moment problem}
\label{sec:MP}

In the previous section, we have formulated the problem of finding positivity bounds as a set of matrix moment problems plus a standard SDP, without actually specifying how to solve the matrix moment problems. 
In this section, we will demonstrate how to solve these matrix moment problems using existing mathematical results. 
Since a uni-variate moment problem (low $\ell$ part) is a special case of a bi-variate problem (high $\ell$ part of the scalar case) and the problems with more moment variables (high $\ell$ part of the spinning field case) are similar to the bi-variate problem, we shall discuss the results primarily for the bi-variate case: $\mathfrak{m}^{\gamma_1 \gamma_2}_{\textbf{ij}} \equiv \mathfrak{m}^{\gamma_1 \gamma_2}_{\textbf{1234}}$. 
The solvability (necessary and sufficient) conditions for the truncated moment problem, including the flat extension condition, will be established in this section.

\subsection{Full matrix moment problem}
\label{sec:FullMP}

Let us first consider the following (infinite) bi-variate matrix $\mathcal{K}$-moment problem
    \begin{align}
    \label{momentpro0}
        \mathfrak{m}_{\mathbf{ij}}^{\gamma_1\gamma_2} = \int_{\mathcal{K}} x_1^{\gamma_1} x_2^{\gamma_2} \dif \mu_{\mathbf{ij}} ( x_1, x_2) \equiv \int_{\mathcal{K}} x_1^{\gamma_1} x_2^{\gamma_2} ~ \rho_{\mathbf{ij}} (x_1, x_2) \dif x_1 \dif x_2
    \end{align}
where $(\gamma_1, \gamma_2) \in \mathbb{N}_0^2$ is a pair of integers, 
$\textbf{i}=\textbf{12}$ ($\textbf{j}=\textbf{34}$) enumerates the ingoing (outgoing) asymptotic state in a 2-to-2 scattering and ${\cal K}$ is a semialgebraic set (see, {\it e.g.}, \eref{amomentPro}). 
Since $\rho_{\mathbf{ij}} (x_1, x_2)$ is a PSD matrix-valued function, we have introduced a PSD matrix-valued measure $\mu_{\mathbf{ij}}$. 
In this subsection, we will show how to solve this matrix moment problem in principle, demonstrating the solvability condition. 
In practice, however, one often can only solve a truncated moment problem, where $\gamma_1$ and $\gamma_2$ are allowed to take some finite values, say, $\gamma_1+\gamma_2\leq G$, which will be discussed in the next subsection.

A full matrix moment problem is solved if a possible representing measure $\mu_{\mathbf{ij}}$ can be found for the infinite $\mathcal{K}$-moment sequence $\{\mathfrak{m}_{\mathbf{ij}}^{\gamma_1\gamma_2}\}$ enumerated by $\mathbf{i}, \mathbf{j}, \gamma_1, \gamma_2$. Therefore, {\it any conditions on the existence of a representing measure $\mu_{\mathbf{ij}}$ give rise to necessary conditions for the sovable moment problem on the moment sequence $\{\mathfrak{m}_{\mathbf{ij}}^{\gamma_1\gamma_2}\}$.} 

So let us first assume the existence of PSD measure $\mu_{\mathbf{ij}}$ in \eref{momentpro0} and derive some necessary conditions for $\mathfrak{m}_{\mathbf{ij}}^{\gamma_1\gamma_2}$. To that end, note that, thanks to $\mu_{\mathbf{ij}}$ being PSD, the following quantity is obviously PSD:
\begin{align}
    \label{HposiDef}
    \begin{aligned}
        &\sum_{ \mathbf{i}\mathbf{j} } \int_{\mathcal{K}} \left[ \sqrt{p (x_1, x_2)} \left( \sum_{\alpha_1 \alpha_2} c_{\alpha_1 \alpha_2}^{\mathbf{i}} x_1^{\alpha_1} x_2^{\alpha_2}  \right) \right] \left[ \sqrt{p (x_1, x_2)} \left( \sum_{\beta_1 \beta_2} c_{\beta_1 \beta_2}^{\mathbf{j}}  x_1^{\beta_1} x_2^{\beta_2} \right) \right]  \d \mu_{\mathbf{ij}} \\
        &= \sum_{\alpha_1 \alpha_2 \beta_1 \beta_2 \mathbf{i} \mathbf{j}} c_{\alpha_1 \alpha_2}^{\mathbf{i}} c_{\beta_1 \beta_2}^{\mathbf{j}} \int_{\mathcal{K}} p (x_1, x_2) x_1^{\alpha_1+\beta_1} x_2^{\alpha_2+\beta_2} \d \mu_{\mathbf{ij}}  \\
        &\equiv \sum_{\alpha_1 \alpha_2 \beta_1 \beta_2 \mathbf{i} \mathbf{j}} c_{\alpha_1\alpha_2}^{\mathbf{i}} ( \mathcal{H}_p^{\mathfrak{m}} )^{\alpha_1 \alpha_2,\beta_1 \beta_2}_{\mathbf{ij}}  c_{\beta_1 \beta_2}^{\mathbf{j}} \geq 0.
    \end{aligned}
\end{align}
where $c_{\alpha_1, \alpha_2}^{\mathbf{i}}$ are a set of real constants, $p (x_1, x_2)$ is a positive polynomial of $x_1$ and $x_2$ over the set $\mathcal{K}$, and we have defined the following Hankel matrix
\begin{align}
\label{GHankelpro}
    \Big( \mathcal{H}_{p(x_1,x_2)}^{\mathfrak{m}} \Big)^{\alpha_1 \alpha_2,\beta_1 \beta_2}_{\mathbf{ij}}
    \equiv \int_{\mathcal{K}} p (x_1, x_2) \, x_1^{\alpha_1+\beta_1} x_2^{\alpha_2+\beta_2} \d \mu_{\mathbf{ij}} (x_1, x_2)
\end{align}
If we take $(\alpha_1, \alpha_2, \mathbf{i})$ as the row index and $(\beta_1, \beta_2, \mathbf{j})$ as the column index,  \eref{HposiDef} implies that $(\mathcal{H}_p^{\mathfrak{m}})^{\alpha_1\alpha_2, \beta_1 \beta_2}_{\mathbf{ij}}$ must be PSD for any positive polynomial $p(x_1,x_2)$ on $\mathcal{K}$:
\begin{align}
\label{eq:HnecesCon}
\mathcal{H}_{p}^{\mathfrak{m}} \succeq 0 ,~~~\forall ~p(x_1,x_2) \geq 0 ~{\rm on}~\mathcal{K}
\end{align}
We emphasize that these conditions follow from assuming the existence of a PSD measure $\mu_{\mathbf{ij}}$, and they are necessary conditions for the existence of $\mu_{\mathbf{ij}}$, and thus also necessary conditions for the matrix moment problem (\ref{momentpro0}) to have a solution.
Since the entries of $(\mathcal{H}_p^{\mathfrak{m}})^{\alpha_1\alpha_2, \beta_1 \beta_2}_{\mathbf{ij}}$ can be written as linear combinations of $\mathfrak{m}_{\mathbf{ij}}^{\gamma_1\gamma_2}$, these PSD conditions impose conditions on $(\mathcal{H}_p^{\mathfrak{m}})^{\alpha_1\alpha_2, \beta_1 \beta_2}_{\mathbf{ij}}$, and thus on the EFT coefficients. 

Evidently, the condition \eqref{eq:HnecesCon} is impractical as it necessitates consideration of all positive polynomial $p(x_1,x_2)$ on $\mathcal{K}$. 
Furthermore, so far, we have only established \eref{eq:HnecesCon} as necessary conditions. 
The next step is to show that it is possible to find a sufficient set of polynomials $\{p(x_1,x_2)\}$ such that the PSD of the corresponding $(\mathcal{H}_p^{\mathfrak{m}})^{\alpha_1\alpha_2, \beta_1 \beta_2}_{\mathbf{ij}}$ gives rise to the solution of the full matrix moment problem. 

Before proceeding, to grasp a sense of these conditions, let us look at two examples that are related to the physical problems \cite{ Arkani-Hamed:2020blm, Bellazzini:2020cot, Chiang:2021ziz}. 
First, consider a simple, non-matrix-valued, uni-variate moment problem, in which case $\mathfrak{m}^{\gamma_1}$ is simply a real number for a given $\gamma_1$. 
If we are interested in the set $\mathcal{K}= \set{x_1|x_1 \geq 0, 1 - x_1 \geq 0} = [0,1]$, that is, we are concerned with the Hausdorff moment problem, which is related to the forward scattering in a single scalar theory, we can choose $p(x_1)=1,x_1,1 - x_1$ to get the solution for the moment problem. 
That is, we can simply require the PSD of the following matrices:
\begin{align}
    (\mathcal{H}_1^{\mathfrak{m}})^{\alpha_1,\beta_1} &= \int_{\mathcal{K}} x_1^{\alpha_1+\beta_1} \ \d \mu, \\
    (\mathcal{H}_{x_1}^{\mathfrak{m}})^{\alpha_1,\beta_1} &= \int_{\mathcal{K}} x_1^{\alpha_1+\beta_1+1} \ \d \mu, \\
    (\mathcal{H}_{1-x_1}^{\mathfrak{m}})^{\alpha_1,\beta_1} &= \int_{\mathcal{K}} (x_1^{\alpha_1+\beta_1} - x_1^{\alpha_1+\beta_1+1}) \d \mu,
\end{align}
These three matrices are sufficient because for any positive polynomial $p (x_1)$ on $\mathcal{K}$ can be expanded as $p (x_1) = q_0^2 (x_1) + x_1 q_1^2 (x_1) + (1 - x_1) q_2^2 (x_1)$, where $q_0, q_1, q_2$ are polynomials\,\footnote{In fact, $q_0^2 (x_1)$ can be omitted for this case because $1=x_1+(1-x_1)$.}.

Another example is the bi-variate formulation of a scalar field \cite{Chiang:2021ziz}. In this case, the second moment variable is nontrivial. With the continueous approximation, the semialgebraic set $\mathcal{K}$ in this case is given by the 2D regions carved out by the following inequalities
\begin{align}
        x_1 \geq 0,~ 1 - x_1 \geq 0,~ x_2 \geq 0,~ x_2 - \ell_{\text{M}} (\ell_{\text{M}} + 1) \geq 0, - x_2 + \ell_{\infty} (\ell_{\infty} + 1) \geq 0.
\end{align}
By the construction (\ref{HposiDef}), this set $\mathcal{K}$ immediately implies that we can obtain the following positivity bounds for the single scalar field theory 
\begin{align}
    \begin{gathered}
    \label{HSS0}
        \mathcal{H}_{p}^{\mathfrak{m}} \succeq 0, ~~~p \in \set{ x_1,  1 - x_1, x_2 - \ell_{\text{M}} (\ell_{\text{M}} + 1) , - x_2 + \ell_{\infty} (\ell_{\infty} + 1) }
    \end{gathered}
\end{align}
More explicitly, this means that the following matrices are PSD:
\begin{gather}
    \mathcal{H}_{x_1}^{\mathfrak{m}} = 
    \begin{pmatrix}
        \mathfrak{m}^{1,0} & \mathfrak{m}^{2,0} & \mathfrak{m}^{1,1} & \cdots \\
        \mathfrak{m}^{2,0} & \mathfrak{m}^{3,0} & \mathfrak{m}^{2,1} & \cdots \\
        \mathfrak{m}^{1,1} & \mathfrak{m}^{2,1} & \mathfrak{m}^{1,2} & \cdots \\
        \vdots   & \vdots  & \vdots & \ddots
    \end{pmatrix}, \\
    \mathcal{H}_{1 - x_1}^{\mathfrak{m}} = 
    \begin{pmatrix}
        \mathfrak{m}^{0,0} - \mathfrak{m}^{1,0} & \mathfrak{m}^{1,0} - \mathfrak{m}^{2,0} & \mathfrak{m}^{0,1} - \mathfrak{m}^{1,1} & \cdots \\
        \mathfrak{m}^{1,0} - \mathfrak{m}^{2,0} & \mathfrak{m}^{2,0} - \mathfrak{m}^{3,0} & \mathfrak{m}^{1,1} - \mathfrak{m}^{2,1} & \cdots \\
        \mathfrak{m}^{0,1} - \mathfrak{m}^{1,1} & \mathfrak{m}^{1,1} - \mathfrak{m}^{2,1} & \mathfrak{m}^{0,2} - \mathfrak{m}^{1,2} & \cdots \\
        \vdots   & \vdots  & \vdots  & \ddots
    \end{pmatrix}, \\
    \mathcal{H}_{x_2 - \ell_{\text{M}} (\ell_{\text{M}} + 1) x_1 }^{\mathfrak{m}} =\hspace{200pt}\qquad\\
    \begin{pmatrix}
        \mathfrak{m}^{0,1} - \ell_{\text{M}} (\ell_{\text{M}} + 1) \mathfrak{m}^{1,0} & \mathfrak{m}^{1,1} - \ell_{\text{M}} (\ell_{\text{M}} + 1) \mathfrak{m}^{2,0} & \mathfrak{m}^{0,2} - \ell_{\text{M}} (\ell_{\text{M}} + 1) \mathfrak{m}^{1,1} & \cdots \\
        \mathfrak{m}^{1,1} - \ell_{\text{M}} (\ell_{\text{M}} + 1) \mathfrak{m}^{2,0} & \mathfrak{m}^{2,1} - \ell_{\text{M}} (\ell_{\text{M}} + 1) \mathfrak{m}^{3,0} & \mathfrak{m}^{1,2} - \ell_{\text{M}} (\ell_{\text{M}} + 1) \mathfrak{m}^{2,1} & \cdots \\
        \mathfrak{m}^{0,2} - \ell_{\text{M}} (\ell_{\text{M}} + 1) \mathfrak{m}^{1,1} & \mathfrak{m}^{1,2} - \ell_{\text{M}} (\ell_{\text{M}} + 1) \mathfrak{m}^{2,1} & \mathfrak{m}^{0,3} - \ell_{\text{M}} (\ell_{\text{M}} + 1) \mathfrak{m}^{1,2} & \cdots \\
        \vdots   & \vdots  & \vdots  & \ddots
    \end{pmatrix}, \nonumber \\
    \mathcal{H}_{\ell_{\infty} (\ell_{\infty} + 1) x_1 - x_2}^{\mathfrak{m}} = \hspace{200pt} \qquad \\
    \begin{pmatrix}
        \ell_{\infty} (\ell_{\infty} + 1) \mathfrak{m}^{1,0} - \mathfrak{m}^{0,1} & \ell_{\infty} (\ell_{\infty} + 1) \mathfrak{m}^{2,0} - \mathfrak{m}^{1,1} & \ell_{\infty} (\ell_{\infty} + 1) \mathfrak{m}^{1,1} - \mathfrak{m}^{0,2} & \cdots \\
        \ell_{\infty} (\ell_{\infty} + 1) \mathfrak{m}^{2,0} -\mathfrak{m}^{1,1} & \ell_{\infty} (\ell_{\infty} + 1) \mathfrak{m}^{3,0} - \mathfrak{m}^{2,1} & \ell_{\infty} (\ell_{\infty} + 1) \mathfrak{m}^{2,1} - \mathfrak{m}^{1,2} & \cdots \\
        \ell_{\infty} (\ell_{\infty} + 1) \mathfrak{m}^{1,1} -\mathfrak{m}^{0,2} & \ell_{\infty} (\ell_{\infty} + 1) \mathfrak{m}^{2,1} - \mathfrak{m}^{1,2} & \ell_{\infty} (\ell_{\infty} + 1) \mathfrak{m}^{1,2} - \mathfrak{m}^{0,3} & \cdots \\
        \vdots   & \vdots  & \vdots  & \ddots
    \end{pmatrix}, \nonumber 
\end{gather}
As mentioned, these are necessary conditions, and we will see in Section \ref{sec:singleScalar}, they are indeed not sufficient.

Let us return to the general case and establish the sufficient conditions. 
To solve a matrix moment problem, we must ensure the existence of the measure $\mu_{\mathbf{ij}}$ for given matrix moments $\mathfrak{m}_{\mathbf{ij}}^{ \gamma_1\gamma_2}$. To this end, we invoke a matrix version of the Riesz representation theorem (Theorem 3 of \cite{Cimpri2013}), which asserts that the existence of the measure $\mu_{\mathbf{ij}} (x_1, x_2)$ is equivalent to the requirement that for all PSD, continuous matrices $\mathcal{F}^{\mathbf{ij}} (x_1, x_2)$ on $\mathcal{K}$, the following conditions are satisfied
\begin{align}
\label{eq:intFmu}
\int_{\mathcal{K}} \mathcal{F}^{\mathbf{ij}} (x_1, x_2) \d \mu_{\mathbf{ij}} \geq 0 .
\end{align}
By the Stone-Weierstrass theorem, any continuous function defined on a compact Hausdorff space can be uniformly approximated by a polynomial function, so the condition (\ref{eq:intFmu}) can be relaxed to apply to all possible positive polynomials if the semialgebraic set $\mathcal{K}$ is compact. 
This is the case for the uni-variate moment problems of $\mathfrak{a}^{\gamma}_{\ell,\textbf{ij}}$ and also for the multi-variate problems once a large order partial wave truncation is in place. 

Then, by a special case of a matrix-valued Positivstellensatz theorem (Theorem 2.5 of \cite{guo2023momentsos} as well as Corollary 1 of \cite{HOL2005451}),  
we can write $\mathcal{F}^{\mathbf{ij}}$ as 
\begin{align}
\label{eq:mcFdef1}
    \mathcal{F}^{\mathbf{ij}} (x_1, x_2) =  \sum_{l=0}^{l_{\text{max}}} p_{l} (x_1, x_2) q_l^{\mathbf{i}} (x_1, x_2) q_l^{\mathbf{j}} (x_1, x_2),
\end{align}
where $q_l^{\mathbf{i}}$ are polynomials of $x_1$ and $x_2$, $p_l$ are positive polynomials of $x_1$ and $x_2$, and we have defined $p_0 = 1$. 
For a given $\mathcal{K}$, we can choose a fixed, finite set of $p_{l} (x_1, x_2)$, but $q_l^{\mathbf{i}}$ are generally different for a different $\mathcal{F}^{\mathbf{ij}}$. 
If we parameterize $q_l^{\mathbf{i}}$ as
\begin{align}
q_l^{\mathbf{i}} = \sum_{\alpha_1 \alpha_2} (c_l)_{\alpha_1 \alpha_2}^{\mathbf{i}} x_1^{\alpha_1} x_2^{\alpha_2}
\end{align}
and substitute \eref{eq:mcFdef1} into \eref{eq:intFmu}, we obtain
\begin{align}
\sum_{l=0}^{l_{\rm max}}\sum_{\alpha_1 \alpha_2 \beta_1 \beta_2 \mathbf{i} \mathbf{j}} (c_l)_{\alpha_1\alpha_2}^{\mathbf{i}} (\mathcal{H}_{p_l}^{\mathfrak{m}})^{\alpha_1 \alpha_2,\beta_1 \beta_2}_{\mathbf{ij}}  (c_l)_{\beta_1 \beta_2}^{\mathbf{j}} \geq 0
\end{align}
Therefore, the sufficient conditions for ensuring the existence of $\mu_{\mathbf{ij}} (x_1, x_2)$, {\it i.e.}, to solve the matrix moment problem, are that
\begin{align}
     \mathcal{H}_{ p_l }^{\mathfrak{m}} \succeq 0, ~~\text{for a sufficient set of}~p_l
\end{align}

Since we have established the existence of a sufficient set of $p_l(x_1, x_2)$, the next step is to select an appropriate such set. For this, a couple of possible options have been established in the mathematical literature for functions and matrix-valued functions $\mathcal{F}^{\mathbf{ij}} (x_1, x_2)$.
One option is to choose the ``power set'' of the defining polynomials of the semialgebraic set
(see \cite{Schmdgen2017TheMP} for the function case and \cite{cimpric2010strictpositivstellensatzematrixpolynomials} for the matrix-valued case). Suppose the semialgebraic set $\mathcal{K}$ is defined by
\begin{align}
\label{Kgeneralpdef}
   \mathcal{K} = \{ (x_1, x_2) \mid \hat{p}_1 (x_1, x_2) \geq 0 , ~\hat{p}_2 (x_1, x_2) \geq 0, \cdots, \hat{p}_k (x_1, x_2) \geq 0 \} .
\end{align}
where $\hat{p}_l (x_1, x_2)$ are the defining polynomials. Then, a simple choice of a sufficient set of $p_l$ polynomials is given by linearly multiplying $\hat{p}_l (x_1,x_2)$ together:
\begin{align}
         \mathcal{P}(\hat{p})&=\{ \hat{p}_1^{e_1}\hat{p}_2^{e_2} \cdots \hat{p}_k^{e_k} \mid e_l \text{ being either 0 or 1} \} \\
        &=\set{1, \underbrace{\hat{p}_1, \cdots, \hat{p}_k}_{k}, \underbrace{\hat{p}_1 \hat{p}_2, \cdots, \hat{p}_{k-1} \hat{p}_{k}}_{k(k-1)/2}, \cdots, \underbrace{\hat{p}_1 \cdots \hat{p}_{k-1}, \cdots, \hat{p}_{2} \cdots \hat{p}_k}_{k}, \hat{p}_1 \cdots \hat{p}_{k}}. \label{eq:OrderedPowerSetOfP}
\end{align}
Actually, at least for the non-matrix case, it has been proven (Theorem 12.27 of \cite{Schmdgen2017TheMP})
that, counting after the element $1$, the first $2^{k-1}$ elements of the set (\ref{eq:OrderedPowerSetOfP}) are already sufficient:
\begin{align}
\label{mcPp}
   \mathcal{P}_{\frac{1}{2}}(\hat{p})= \{ \underbrace{\hat{p}_1, \cdots, \hat{p}_k,\hat{p}_1 \hat{p}_2, \cdots, \hat{p}_{k-1} \hat{p}_{k}, \cdots}_{2^{k-1}} \}.
\end{align}

Therefore, the solvability ({\it i.e.}, necessary and sufficient) condition of the full matrix moment problem for a compact $\mathcal{K}$ is
\begin{align}
\label{eq:Hsolvability}
        \mathcal{H}_{p_l}^{\mathfrak{m}} \succeq 0,~~ \forall p_l \in  \mathcal{P}(\hat{p}) .
\end{align}
At this point, as a solution to the full moment problem, the Hankel matrices above are infinite dimensional. 

\subsection{Truncated moment problem and flat extension}
\label{sec:TruncatedMP}

In the previous subsection, we have obtained the solution to the full matrix moment problem with truncated partial waves but without truncating the moment sequence.
In practice, however, we may only wish to know the bounds on the first few EFT coefficients, {\it i.e.,} the first few moments $\mathfrak{m}^{\gamma_1\gamma_2}_{\mathbf{ij}}$ with $\gamma_1$ and $\gamma_2$ taking the first few values, agnostic about the higher order moments. After all, numerically, we only have finite resources to compute matrices with finite dimensions. 
That is, we are interested in solving a truncated moment problem. 
Generally, Stochel's theorems \cite{Stochel_2001} allow us to, in some sense, approximate the full moment problem by solving the truncated moment problem with some specific truncations. 
An important concept here is the flat extension of the truncated moment problem, which allows one to find the optimized solution for the given information. 

Suppose we only consider the a finite number of moments $\mathfrak{m}^{\gamma_1 \gamma_2}_{\mathbf{ij}}$ specified by the set
\begin{align}
\Theta (G) = \set{ (\gamma_1, \gamma_2)\in \mathbb{N}_0^2 \mid \gamma_1 + \gamma_2 \leq G} ,
\end{align}
where $G$ is a non-negative integer. 
In general, if we simply discard all the rows and columns of the $\mathcal{H}_{p_l}^{\mathfrak{m}}$ matrices containing the excluded $\mathfrak{m}^{\gamma_1\gamma_2}_{\mathbf{ij}}$ and use the condition \eqref{eq:Hsolvability} for the naively trunacted Hankel matrices,
we get some truncated conditions on the EFT coefficients 
\begin{align}
    \mathcal{H}_{p_l}^{\mathfrak{m}} (G) \succeq 0,
    ~~ \forall p_l \in   \mathcal{P}(\hat{p}) .
\end{align}
These conditions are weaker than the PSD conditions of the full moment problem, because $\mathcal{H}_{p_l}^{\mathfrak{m}} (G)$ is a principal submatrix of $\mathcal{H}_{p_l}^{\mathfrak{m}} (+\infty)$, and the PSD of a matrix requires all of its principal submatrices to be PSD. 

If we extend the set from $\Theta (G)$ to $\Theta (G+2)$, we obtain a slightly larger matrix $\mathcal{H}_{p_l}^{\mathfrak{m}} (G + 2)$, which is referred to as an \textit{extension} of $\mathcal{H}_{p_l}^{\mathfrak{m}} (G)$. 
For this given $\Theta(G)$-truncation,
a \textit{flat extension} of $\mathcal{H}_{p_l}^{\mathfrak{m}} (G)$ is said to exist if one can choose the extra moments in $\Theta (G+2) - \Theta (G)$ such that 
\begin{align}
\label{eq:rankHG2}
\operatorname*{rank} \left(\mathcal{H}_{p_l}^{\mathfrak{m}} (G+2) \right) = \operatorname*{rank} \left(\mathcal{H}_{p_l}^{\mathfrak{m}} (G) \right).
\end{align}
In this case, $(\mathcal{H}_{p_l}^{\mathfrak{m}} (G+2))^{\alpha_1\alpha_2,\beta_1\beta_2}_{\mathbf{ij}}$ is said to be \textit{flat}. 
If \eref{eq:rankHG2} is not satisfied, one verifies a similar equation comparing the ranks of $G+4$ and $G+2$, and so on, until an \textit{eventual flat extension} is found, which satisfies
\begin{align}
\operatorname*{rank} \left(\mathcal{H}_{p_l}^{\mathfrak{m}} (G+2G_l'+2) \right) = \operatorname*{rank} \left(\mathcal{H}_{p_l}^{\mathfrak{m}} (G+2G_l') \right) 
\end{align}
where $G'_l$ is a non-negative integer. 
A truncated moment problem has a solution if and only if all $\mathcal{H}_{p_l}^{\mathfrak{m}} (G)$ are PSD and every $\mathcal{H}_{p_l}^{\mathfrak{m}} (G)$ has an eventual flat extension.
The ``only if'' part of the above statement is established in Ref.~\cite{Kimsey2022OnAS} for the matrix-valued generalization of the Bayer-Teichmann-Tchakaloff theorem \cite{Kimsey2014AnOG}.
The ``if'' part can be proven in a constructive way, that is, one can always construct a special representing measure for a flat Hankel matrix, which will be discussed in detail in Section \ref{sec:AllSpec}.
Furthermore, it is unnecessary to verify whether all of  $\mathcal{H}_{p_l}^{\mathfrak{m}} (G)$ have flat extensions. 
It is suffice to only verify whether $\mathcal{H}_1^{\mathfrak{m}} (G)$ has a flat extension, at least in the case of the non-matrix-valued moment problem \cite{curto2000truncated}. 
This is because, if $\mathcal{H}_1^{\mathfrak{m}} (G)$ has a flat extension, all the other Hankel matrices will eventually have a flat extension, thereby guaranteeing the solvability of the moment problem.  
Our empirical evidence from numerical computations suggests that the same conclusion may also apply to the matrix moment problem. 

Put another way, without verifying the existence of a flat extension, even if a truncated moment sequence satisfies the PSD conditions of the naively truncated Hankel matrices, there is no guarantee that this truncated moment sequence can truly have a moment integral representation (with a positive measure). 
On the other hand, the PSD conditions of flat Hankel matrices are reliable because a truncated moment sequence satisfying these conditions must have a moment integral representation. In terms of the original physical problem, only those EFT coefficients within the boundaries defined by the PSD conditions of the flat Hankel matrices satisfy the dispersion relations.

Now, we introduce Stochel’s theorem \cite{Stochel_2001}, which establishes the relation between a truncated and a full moment problem defined on a closed semialgebraic set $\mathcal{K}$. (The theorem is rigorously proven for a ``standard'' moment problem, but, as with many other results, it might be generalized to the matrix-valued case.) The theorem states that 
if for every $G \geq 0$, the $\Theta(G)$-truncated moment problem is solvable, then the full moment problem is solvable. 
As will be shown in Section \ref{sec:AllSpec}, one can actually construct a specific representing measure for every solvable $\Theta(G)$-truncated moment problem. 
If the full moment problem is determinate on $\mathbb{R}^n$, the measure of the truncated problem converges to that of the full problem as $G \rightarrow \infty$. We will come back to this discussion about determinacy in Section \ref{sec:Determinacy}.

In summary, it is usually sufficient to work with a truncated moment problem, as we typically must do when extracting positivity bounds numerically.
For our physical problem, there are some null constraints, which are linear equations involving a certain number of moments. 
To calculate the positivity bounds for a given set of null constraints, the lowest order of the $\Theta(G)$-truncation, denoted as $G_1$, should be chosen such that all moments appearing in this set of null constraints are included. 
Concretely, the steps to compute the positivity bounds are outlined as follows.
\begin{itemize}
\item  For a given set of null constraints, we begin by imposing the the lowest order PSD constraints
\begin{align}
    \mathcal{H}_{p_l}^{\mathfrak{m}} (G_1) \succeq 0,
    ~~ \forall p_l \in  \mathcal{P}(\hat{p}),
\end{align}
together with the chosen null constraints, and optimize some linear combinations of the EFT coefficients that we wish to bound. 
Here, the polynomials $\hat{p}_l (x_1, x_2)$, $l=1,2,...,k$, are the ones that define the semialgebraic set $\mathcal{K}$, and $\mathcal{P}(\hat{p})$ is the ``power set'' of $\{\hat{p}_l\}$.
\item One can verify the flatness of $\mathcal{H}^{\mathfrak{m}}_{p_l}(G_1) $
\begin{align}
\operatorname*{rank} \left(\mathcal{H}_{p_l}^{\mathfrak{m}} (G_1) \right) = \operatorname*{rank} \left(\mathcal{H}_{p_l}^{\mathfrak{m}} (G_1-2) \right).
\end{align}
If they are flat, one has obtained the positivity bounds. 
Often, it is sufficient to check $\operatorname*{rank} \left(\mathcal{H}_{1}^{\mathfrak{m}} (G_1) \right) = \operatorname*{rank} \left(\mathcal{H}_{1}^{\mathfrak{m}} (G_1-2) \right)$.
\item Sometimes, they are not flat, in which case we shall choose a bigger $G$, starting from $\Theta(G_1+2)$, $\Theta(G_1+4)$ and so on, until eventual flat Hankel matrices are obtained. The optimal bounds are then obtained by instead imposing the stronger constraints from the eventually flat Hankel matrices. 
\end{itemize}
To compute the bounds for higher order EFT coefficients or with more null constraints, a higher order $G$-truncation is required. 
Essentially, we often should use higher order Hankel matrices in order to obtain the optimal constraint for a truncated moment problem.

In our numerical computations later, we will find that for a sufficiently large $\ell_{\rm M}$ ($\sim 20$), replacing $\mathcal{K}_{\mathfrak{b}}$ with $\mathcal{K}'_{\mathfrak{b}}$ (see Section \ref{sec:momentFormulation}) generally does not affect the numerical accuracy (at least within the errors of $10^{-5}$). Apart from $\mathcal{K}'_{\mathfrak{b}}$ being a good approximate set for a large $\ell_{\rm M}$, this also reflects the fact that we are evaluating a truncated moment problem, which involves only a finite number of moments for a given set of null constraints. 
With the conditions on the Hankel matrices established, we optimize the moments to find the positivity bounds. 
To a given order in the moment truncation, there are only a finite number of moments in the linear matrix inequalities of the SDP. 
As it happens, adding more high $\ell$ uni-variate moments is often redundant and does not improve the accuracy for a given truncation.

\subsection{$L$-moment problem for multiple DoFs}
\label{sec:Lmoment}

In the previous discussions, we only made use of the positivity part of the UV partial wave unitarity conditions \eqref{eq:AbsAellUnitarity}: ${\rm Abs} A_\ell\succeq 0$. 
While the full unitarity conditions are straightforward to implement in the primal bootstrap approach 
\cite{Paulos:2017fhb,Homrich:2019cbt}, it is more complicated for a dual approach with dispersion relations \cite{Guerrieri:2021tak}. 
Nevertheless, the ``upper bound'' of the unitarity conditions, which also manifest as positivity conditions $\tilde{I}-{\rm Abs} A_\ell\succeq 0$, can be also be easily incorporated in the moment problem approach. 
The (non-matrix) uni-variate moment problem where the measure spectrum $\rho (x)$ is bounded both from above and below, $0 \leq \rho (x) \leq L$ is sometimes referred to as the $L$-moment problem \cite{Akhiezer1962SomeQI}. 
By taking the Minkowski sum for various partial waves, each of which is formulated as an $L$-moment problem, and tapping into the results of the $L$-moment problem, Refs.~\cite{Chiang:2022jep,Chiang:2022ltp} were able to analytically derive positivity bounds for the single degree of freedom, taking into account the upper bounds of unitarity. 
However, this approach does not yet account for the inclusion of multiple null constraints or the extension to multiple degrees of freedom. 
Other ways to utilize the upper bounds of unitarity and beyond for positivity bounds include the approaches of \cite{Chiang:2022jep, Chiang:2022ltp, Chen:2023bhu, Hong:2024fbl, Caron-Huot:2022ugt}. 
Here we propose a straightforward way to implement the upper bounds of unitarity within the matrix moment approach that can easily accommodate multiple null constraints.

Note that the PSD of the measure in a moment problem is equivalent to the PSD of the corresponding Hankel matrices. Therefore, the (reduced) unitarity conditions $\tilde{I}_{\textbf{ij}} \succeq \rho_{\textbf{ij}} (x) \succeq 0$ can be reformulated as the PSD of the Hankel matrices generated by two measures  $\rho_{\textbf{ij}} (x) \succeq 0$ and $\tilde{\rho}_{\textbf{ij}}\equiv \tilde{I}_{\textbf{ij}} - \rho_{\textbf{ij}} (x) \succeq 0$, where $\tilde{I}_{\textbf{ij}}$ is a diagonal matrix whose entries are either 1 or 2, depending on whether the scattering is between identical particles. 
In the previous subsections, we already make use of $\rho_{\textbf{ij}} (x) \succeq 0$. 
To take into account $\tilde\rho_{\textbf{ij}} (x) \succeq 0$,
for a bi-variate moment problem for example, we define
\begin{align}
    \mathfrak{I}^{\gamma_1 \gamma_2} = \int_{\mathcal{K}} x_1^{\gamma_1} x_2^{\gamma_2} \tilde{I}_{\textbf{ij}} \dif x_1 \dif x_2, 
    ~~~~\text{so we have}~~
    \mathfrak{I}^{\gamma_1 \gamma_2} - \mathfrak{m}^{\gamma_1 \gamma_2} = \int_{\mathcal{K}} x_1^{\gamma_1} x_2^{\gamma_2} \tilde{\rho}_{\textbf{ij}} \dif x_1 \dif x_2,
\end{align}
Then, the {\it extra} Hankel matrix conditions we need to impose are
\begin{align}
\mathcal{H}_{p_l}^{\mathfrak{I} - \mathfrak{m}}(G) \succeq 0,
    ~~~ \forall p_l \in  \mathcal{P}(\hat{p}).
\end{align}
where $\mathcal{H}_{p_l}^{\mathfrak{I} - \mathfrak{m}}$ denotes a Hankel matrix constructed from the moment sequence $\mathfrak{I} - \mathfrak{m}$.

\begin{figure}[h]
    \centering
    \includegraphics[scale=0.4]{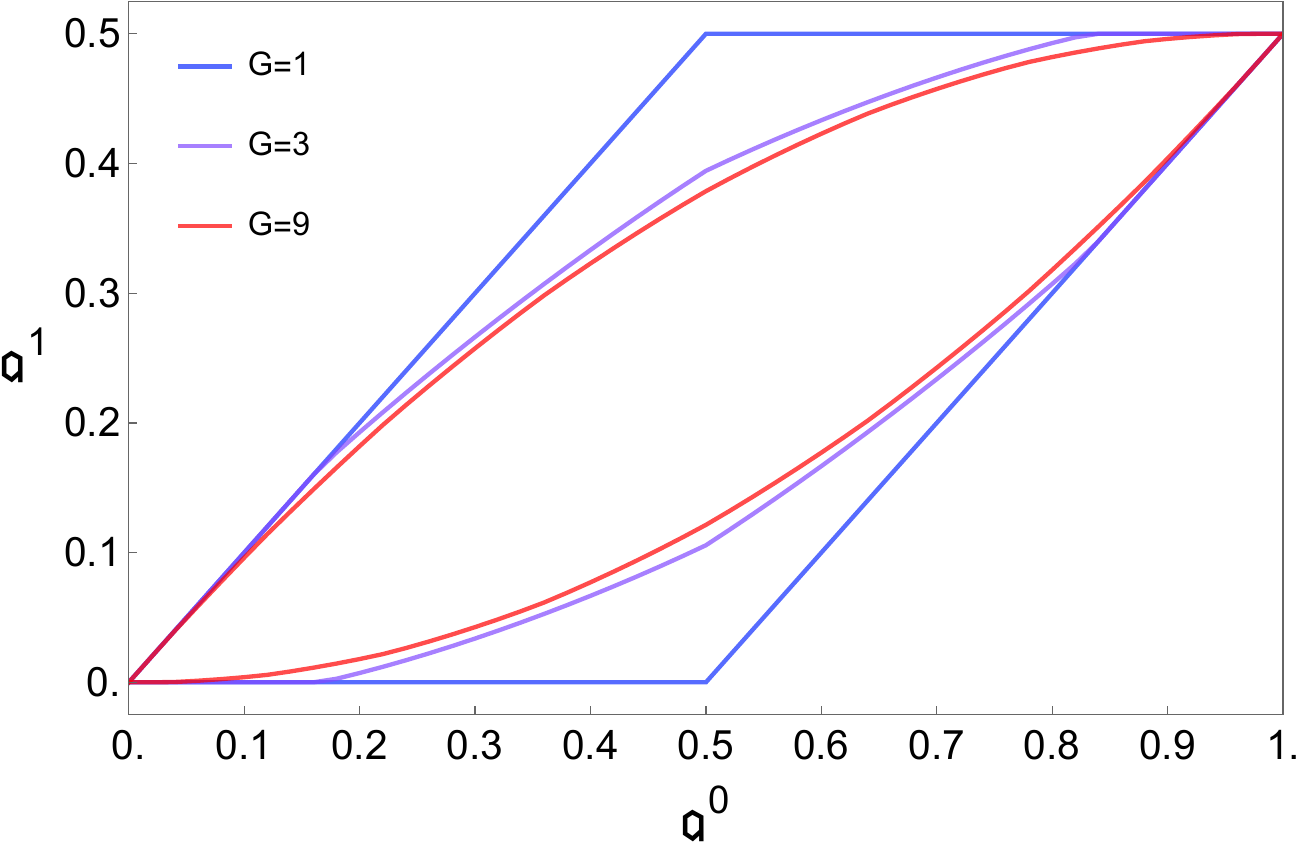}
    \caption{Numerical bounds for various truncation orders $G$ for the simple example \eqref{alowerupperrho}.}
    \label{fig:LMP_ToyExam}
\end{figure}

For a toy model example, consider the Hausdorff moment problem modified with the upper bound for $\rho(x_1)$. 
That is, we look at the moment sequence given by
\begin{align}
\label{alowerupperrho}
    \mathfrak{a}^{\gamma_1} = \int_{0}^{1} x_1^{\gamma_1} \rho(x_1) \dif x_1, \qquad 0 \leq \rho(x_1) \leq 1.
\end{align}
To incorporate the upper bound $\rho \leq 1$ in addition to the conditions from the Hausdorff moment problem, we simply need to impose PSD conditions for the following {\it extra} truncated Hankel matrices:
\begin{align}
    \mathcal{H}^{\mathfrak{I} - \mathfrak{a}}_{x_1} \succeq 0,~~~~ \mathcal{H}^{\mathfrak{I} - \mathfrak{a}}_{1-x_1} \succeq 0 .
\end{align}
Putting all the PSD conditions together, $\mathfrak{a}^{0}$ and $\mathfrak{a}^{1}$ are bounded in a closed region, and we see that the convergence is rather quick as the truncation order $G$ increases; see Figure \ref{fig:LMP_ToyExam}. The case of $G=9$ is visually indistinguishable from the analytical result of \cite{Chiang:2022ltp, Chiang:2022jep} for this toy example. However, it is also easy to apply this method to numerically compute the upper bounds for multi-field theories with spin, including multiple null constraints, as will be briefly discussed in Section \ref{sec:EFTBounds} for a spin-2 theory.

\section{Bounds on EFT models}
\label{sec:EFTBounds}

In this section, we demonstrate how to use our matrix moment formulation to derive bounds on the EFT coefficients for several example EFTs. 
Most (but not all) of these positivity bounds for these models have been previously calculated using other methods. 
Therefore, our goal here is not to obtain comprehensive bounds for these models, but rather to illustrate how to compute the bounds using our approach, showcasing both consistency with and improvements over previous results.

\subsection{Single scalar}
\label{sec:singleScalar}

Let us first consider the case of a single scalar, in which case the matrix moments reduce to the normal moments and some necessary conditions of the moment problem have been used to bound general EFT coefficients via a pure bi-variate formulation \cite{Chiang:2021ziz} (see \cite{Arkani-Hamed:2020blm,Bellazzini:2020cot} for earlier work). 
Here we revisit this problem with our mixed-variate moment formulation, which is numerically more efficient. 
With the continuous large $\ell$ approximation, we implement the sufficient conditions for both the uni-variate and bi-variate parts of the moment problem. 
We find that these sufficient conditions generally give rise to small improvements to the previous results from the necessary conditions in the moment approach, and the bounds derived from these sufficient conditions agree to a high degree with the results obtained using the method of \cite{Caron-Huot:2020cmc}.

Let us use this single scalar case to reiterate the essential steps for obtaining the positivity bounds in the moment problem approach, ignoring for now the complications arising from multiple degrees of freedom and spin.
For this case, we have $\textbf{1}=\textbf{2}=\textbf{3}=\textbf{4}$, so we shall omit the subscripts of the external particles. We parametrize the scalar EFT amplitude as follows
\begin{align}
    A (s, t) = \sum_{i \geq 0, j \geq 0} g^{i,j} \left(s + \dfrac{t}{2}\right)^i t^j.
\end{align}
Following the prescription of Section \ref{sec:DispSumrules}, we can derive the dispersive sum rules and the corresponding null constraints:
\begin{align}
g^{i,j}=\sum_k\int\d s' \sum_\ell (\cdots)_{i,j},~~~~~
\sum_{k=0}^{j} \dfrac{(i+k)!}{2^k i!k!} g^{i+k,j-k} = \sum_{k=0}^{i} \dfrac{(j+k)!}{2^k j!k!} g^{j+k,i-k}
\end{align}
Thanks to the positivity part of partial wave unitarity, which imposes the PSD on the spectral functions, these sum rules are then transformed to a linear combination of moment problems, as prescribed in Section \ref{sec:multiScalar},
\begin{align}
    g^{i,j} = \sum_{\ell=0}^{\ell_{\text{M}}-1} U^{i,j}_{\ell (\ell + 1)} \mathfrak{a}^{\gamma}_{\ell} + \sum_{\gamma_2^\prime} V^{(\gamma)}_{\gamma_2,\gamma_2^\prime} \mathfrak{b}^{\gamma_1,\gamma_2^\prime} + (\text{$u$-channel})
\end{align}
where $\gamma = i + j - N_s$,
$\gamma_1 = i - N_s$, $\gamma_2 = j$ and the uni-variate moments $\mathfrak{a}_\ell^{\gamma}$ and the bi-variate moments $\mathfrak{b}^{\gamma_1,\gamma_2}$ along with their integration regions $\mathcal{K}_{\mathfrak{a}}$ and $\mathcal{K}_{\mathfrak{b}}^{\prime\prime}$ are given by
\begin{align}
    \mathfrak{a}_\ell^{\gamma} &= \int_{\mathcal{K}_{\mathfrak{a}}} x_1^{\gamma} \dif \mu_\ell (x_1),~~~~
    \mathfrak{b}^{\gamma_1,\gamma_2} = \int_{\mathcal{K}_{\mathfrak{b}}^{\prime\prime}} x_1^{\gamma_1} x_2^{\gamma_2} \dif \mu (x_1, x_2) 
    \\
    \mathcal{K}_{\mathfrak{a}} &= \{ x_1 | x_1 \geq 0, 1 - x_1 \geq 0 \} 
    \nn
    \mathcal{K}_{\mathfrak{b}}^{\prime\prime} &= \{ (x_1, x_2) | x_1 \geq  0,1- x_1 \geq 0, x_2 - \ell_{\text{M}}(\ell_{\text{M}}+1) x_1\geq 0  \}
\end{align}
We have replaced $\mathcal{K}_{\mathfrak{b}}^{\prime}$ of \eref{eq:Kbpdef} with the $\mathcal{K}_{\mathfrak{b}}^{\prime\prime}$ above, as numerically it does not affect the results for a sufficiently large $\ell_\infty$. (In comparison, the pure bi-variate moment approach corresponds to the choice of $\ell_{\rm M}=0$, that is, there being no $\mathfrak{a}_\ell^{\gamma}$ part.) The solutions of the $\mathfrak{a}_\ell^{\gamma}$ and $\mathfrak{b}^{\gamma_1,\gamma_2}$ moment problems are given respectively by 
\begin{align}
\label{scalarHaCon}
    \mathcal{H}^{\mathfrak{a}_\ell}_{p_l} (G^\mathfrak{a}_{\ell}) &\succeq 0,
    ~~ \forall p_l \in  \set{x_1,1-x_1}
    \\
\label{scalarHbCon}
   \mathcal{H}^{\mathfrak{b}}_{p_l} (G^\mathfrak{b}) &\succeq 0,
    ~~ \forall p_l \in \set{x_1, 1-x_1, x_1(1-x_2),x_2-x_1 \ell_{\text{M}} (\ell_{\text{M}} + 1) }.
\end{align}
where the orders of the moment truncations  $G^\mathfrak{a}_{\ell}$ and $G^\mathfrak{b}$ will be specified momentarily. 

Our goal is to infer bounds on the ratios of the EFT coefficients $\tilde{g}^{i,j}=g^{i,j}/g^{2,0}$ from the Hankel matrix constraints as well as the null constraints, where $g^{2,0}$ is positive as can be easily seen from its dispersive sum rule. 
In a given optimization run, we select an optimization objective $\sum_{i,j}\alpha_{ij} \tilde{g}^{i,j}$, which entails choosing a direction $\alpha_{ij}$ within the space spanned by $\tilde{g}^{i,j}$ to probe the boundary of the positivity bounds. 
Additionally, we also choose the number of null constraints we want to impose. 
Since each null constraint encompasses several EFT coefficients, the minimal moment truncations $G^\mathfrak{a}_{\ell}$ and $G^\mathfrak{b}$ are such that the moments $\mathfrak{a}_\ell^{\gamma}$ and $\mathfrak{b}^{\gamma_1,\gamma_2}$ associated with these EFT coefficients are just fully included in the Hankel matrices. 
For these minimal moment truncations, we can verify whether the corresponding Hankel matrices are flat or not. 
In principle, the flatness of all these Hankel matrices needs to be verified. 
However, in practice, it is often sufficient to check $\mathcal{H}^{\alpha, \beta}_{1,\mathbf{ij}} (G^\mathfrak{a}_{\ell})$ and $\mathcal{H}^{\alpha_1 \alpha_2,\beta_1 \beta_2}_{1,\mathbf{ij}} (G^\mathfrak{b})$. 
If these Hankel matrices are flat, it is unnecessary to consider higher-order moment truncations.
Conversely, if they are not flat, higher-order moment truncations should be pursued until flat Hankel matrices are found. The eventual flat Hankel matrices are the ones to impose PSD conditions in order to yield optimal positivity bounds.

As an explicit example, suppose we aim to obtain the lower bound of $\tilde{g}^{3,1}$, making use the null constraints involving $\tilde{g}^{i,j}$ with $i+j \leq 6$. 
We shall execute the following SDP:
\begin{align}
    \begin{array}{ll}
    \text {\bf minimize } & g^{3,1} = \sum_{\ell} \Big( \dfrac{3}{2} - \ell (\ell + 1) \Big) \mathfrak{a}^{1}_{\ell} + \dfrac{3}{2} \mathfrak{b}^{1,0} - \mathfrak{b}^{0,1}, \\
    \text {\bf such that } & g^{2,0} \equiv \sum_{\ell} \mathfrak{a}^{0}_{\ell} +\mathfrak{b}^{0,0} = 1 \\
    & \sum_{\ell} \Big( \ell^2 (\ell+1)^2 - 8 \ell (\ell + 1) \Big) \mathfrak{a}_{\ell}^{2} + \mathfrak{b}^{0,2} - 8 \mathfrak{b}^{1,1} = 0, \\
    & \sum_{\ell} \Big( 2 \ell^3 (\ell+1)^3 - 43 \ell^2 (\ell + 1)^2 + 150 \ell (\ell + 1) \Big) \mathfrak{a}^{3}_{\ell} \\
    &\phantom{\sum_{\ell}} \qquad + 2\mathfrak{b}^{0,3} - 43 \mathfrak{b}^{1,2} + 150 \mathfrak{b}^{2,1} = 0, \\
    & \sum_{\ell} \Big( \ell^4 (\ell+1)^4 - 44 \ell^3 (\ell + 1)^3 + 588 \ell^2 (\ell+1)^2 - 2448 \ell (\ell + 1) \Big) \mathfrak{a}^{4}_{\ell} \\
    &\phantom{\sum_{\ell}} \qquad + \mathfrak{b}^{0,4} - 44 \mathfrak{b}^{1,3} + 588 \mathfrak{b}^{2,2} - 2448 \mathfrak{b}^{3,1} = 0, \\
     &\mathcal{H}^{\mathfrak{a}_\ell}_{x_1} (5) \succeq 0, ~~~\mathcal{H}^{\mathfrak{a}_\ell}_{1-x_1} (5) \succeq 0, ~~~~~ \ell = 0, 2, 4, ..., \ell_{\text{M}},\\
     &\mathcal{H}^{\mathfrak{b}}_{x_1} (5) \succeq 0, ~\mathcal{H}^{\mathfrak{b}}_{1-x_1} (5) \succeq 0, ~
     \mathcal{H}^\mathfrak{b}_{x_1 (1 - x_1)} (5) \succeq 0, ~\mathcal{H}^\mathfrak{b}_{x_2-x_1 \ell_{\text{M}} (\ell_{\text{M}} + 1)} (5) \succeq 0 .
    \end{array}
\end{align}
where $\mathfrak{a}^{\gamma}_\ell$ and $\mathfrak{b}^{\gamma_1,\gamma_2}$ are decision variables and we have chosen the normalization $g^{2,0}=1$. For $\ell_{\text{M}}=8$, the result of this program gives: 
\begin{align}
\label{g31accurate}
\operatorname*{min} (\tilde{g}_{3,1}) = 5.18309 .
\end{align}
which is in perfect agreement with the result of \cite{Caron-Huot:2020cmc}. 
(The result remains unchanged if the pure bi-variate formulation is applied using these sufficient base polynomials.) 
Indeed, after running the program, one can get the values for the decision variables $\mathfrak{a}_\ell^{\gamma}$ and $\mathfrak{b}^{\gamma_1,\gamma_2}$, and substituting them back into the Hankel matrices $\mathcal{H}^{\mathfrak{a}_\ell}_{1}$, one can check that these Hankel matrices are already flat. The flatness of the Hankel matrices can also be seen from the fact that one can obtain an {\it atomic representing measure} from these matrices, which will be discussed in Section \ref{sec:AllSpec}. Once flatness is achieved, the results will agree perfectly with those obtained using the method of Ref.~\cite{Caron-Huot:2020cmc}.

Typically, for given orders of moment and null constraint truncations, there exists a maximum $\ell_{\text{M}}$, beyond which increasing $\ell_{\text{M}}$ will not affect the numerical results.
This is because we are optimizing over a finite number of constraints, and a large $\ell_{\text{M}}$ merely leads to redundant decision variables. 

\begin{figure}[h]
    \centering
    \includegraphics[width=0.8\textwidth]{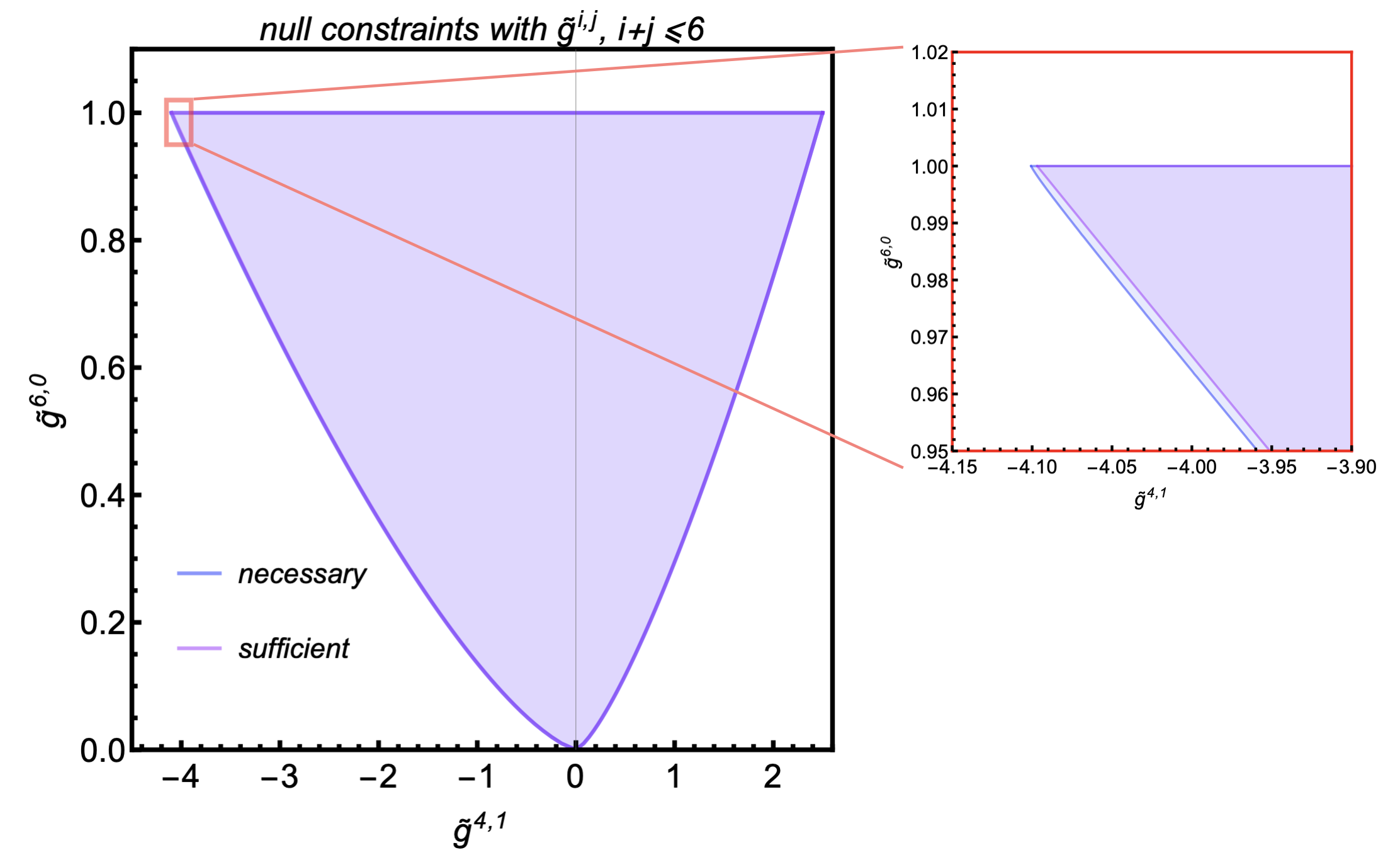}
    \caption{Comparison between the positivity bounds on $\tilde{g}^{6,0}=g^{6,0}/g^{2,0}$ and $\tilde{g}^{4,1}=g^{4,1}/g^{2,0}$ from sufficient (Eqs.~(\ref{scalarHaCon} and \ref{scalarHaCon})) and necessary (see around \eref{scalarBiVList}) Hankel matrix conditions in the single scalar theory. The sufficient conditions agree to a high degree with the results using the method of \cite{Caron-Huot:2020cmc}. }
    \label{fig:g60g41}
\end{figure}

For comparison, we also compute the bounds using the necessary set of base polynomials for the Hankel matrices used in \cite{Chiang:2021ziz} in the pure bi-variate formulation:
\begin{align}
\label{scalarBiVList}
        x_2, x_1, 1 - x_1, (x_2 - \ell (\ell + 1) x_1) (x_2 - (\ell + 2) (\ell + 3) x_1),~ \ell = 0, 2, 4, \cdots, \ell_{M}
\end{align}
These are also the defining polynomials for the integration region of the corresponding moment problem, {\it i.e.}, they are the defining polynomials of set $\mathcal{K}_{\mathfrak{b}}$ with $\ell_{\text{M}}=0$ and without the $x_2-\ell_\infty(\ell_\infty+1)x_1\leq 0$ truncation. 
(If the $x_2-\ell_\infty(\ell_\infty+1)x_1\leq 0$ truncation is included, a sufficient set of base polynomials in this formulation is treating these polynomials as $\hat{p}_l$ and get the ``power set'' or the first half of the ``power set'', according to the prescription of Eqs.~(\ref{Kgeneralpdef}-\ref{mcPp}.) 
Keeping all the other setups the same, it would give rise to 
\begin{align}
\operatorname*{min} (\tilde{g}_{3,1}) = 5.18321
\end{align}
which slightly deviates from the minimum in \eqref{g31accurate}. 
Larger discrepancies can arise for the higher order EFT coefficients. 
For example, for the lower bound on $\tilde{g}_{6,0}$, the differences can reach a percentage level for a large $\tilde{g}_{4,1}$. See Figure \ref{fig:g60g41} for a comparison of the effects of the necessary and sufficient sets of base polynomials on the positivity bounds of $\tilde{g}_{6,0}$ and $\tilde{g}_{4,1}$. 
Notably, these discrepancies decrease close to the origin of the $\tilde{g}_{6,0}$ and $\tilde{g}_{4,1}$ plane. Therefore, for the single scalar example, the necessary conditions of \cite{Chiang:2021ziz}, although incomplete, are already quite accurate. 

In the pure bi-variate formulation, the number of Hankel matrices, {\it i.e.}, the cardinality of the set of the sufficient base polynomials grows exponentially with $\ell_M$. 
In the mixed-variate formulation, on the other hand, the number of Hankel matrices increases linearly with $\ell_{\text{M}}$. The reason is that in the pure bi-variate formulation, considerable effort is spent eliminating regions on the $x_1$-$x_2$ plane that do not correspond to the true integration domain of the underlying moment problem. In contrast, the mixed-variate formulation, for the most part, directly uses the true integration domain.

\subsection{Double $\mathbb{Z}_2$ bi-scalar}
\label{sec:biscalar}

Now, we move to one of the simplest cases of multiple degrees of freedom where the matrix moment formulation is needed: a bi-scalar with double $\mathbb{Z}_2$ symmetry:  $\phi_1 \leftrightarrow \phi_2, \phi_1 \leftrightarrow - \phi_1$. We follow the steps described in Section \ref{sec:momentFormulation}. 
In particular, we use the bi-variate moment formulation for the large $\ell$ region. The main difference from the single scalar case is that now many quantities have additional indices to label the particle degrees of freedom. 
This mostly means that the SDP involves a greater number of decision variables.
For the Hankel matrices, these particle labels become parts of their row and column indices, so the dimensions of the matrices are significantly enlarged, but the sufficient set of base polynomials is still given by those of Eqs.~(\ref{scalarHaCon}) and (\ref{scalarHbCon}). 
For example, for a generic bi-field case, the Hankel matrices $\mathcal{H}_{1}^{\mathfrak{a}_{\ell}} (2)$ and $\mathcal{H}_{1}^{\mathfrak{b}} (2)$ are respectively enlarged to be 
\begin{gather}
{ \small
    \begin{pmatrix}
        \mathfrak{a}^{0}_{\ell,1111} & \mathfrak{a}^{0}_{\ell,1112} & \mathfrak{a}^{0}_{\ell,1121} \mathfrak{a}^{0}_{\ell,1122} & \mathfrak{a}^{1}_{\ell,1111} & \mathfrak{a}^{1}_{\ell,1112} & \mathfrak{a}^{1}_{\ell,1121} \mathfrak{a}^{1}_{\ell,1122} \\
        \mathfrak{a}^{0}_{\ell,1211} & \mathfrak{a}^{0}_{\ell,1212} & \mathfrak{a}^{0}_{\ell,1221} \mathfrak{a}^{0}_{\ell,1222} & \mathfrak{a}^{1}_{\ell,1211} & \mathfrak{a}^{1}_{\ell,1212} & \mathfrak{a}^{1}_{\ell,1221} \mathfrak{a}^{1}_{\ell,1222} & \\
        \mathfrak{a}^{0}_{\ell,2111} & \mathfrak{a}^{0}_{\ell,2112} & \mathfrak{a}^{0}_{\ell,2121} \mathfrak{a}^{0}_{\ell,2122} & \mathfrak{a}^{1}_{\ell,2111} & \mathfrak{a}^{1}_{\ell,2112} & \mathfrak{a}^{1}_{\ell,2121} \mathfrak{a}^{1}_{\ell,2122} & \\
        \mathfrak{a}^{0}_{\ell,2211} & \mathfrak{a}^{0}_{\ell,2212} & \mathfrak{a}^{0}_{\ell,2221} \mathfrak{a}^{0}_{\ell,2222} & \mathfrak{a}^{1}_{\ell,2211} & \mathfrak{a}^{1}_{\ell,2212} & \mathfrak{a}^{1}_{\ell,2221} \mathfrak{a}^{1}_{\ell,2222} & \\
        \mathfrak{a}^{1}_{\ell,1111} & \mathfrak{a}^{1}_{\ell,1112} & \mathfrak{a}^{1}_{\ell,1121} \mathfrak{a}^{1}_{\ell,1122} & \mathfrak{a}^{2}_{\ell,1111} & \mathfrak{a}^{2}_{\ell,1112} & \mathfrak{a}^{2}_{\ell,1121} \mathfrak{a}^{2}_{\ell,1122} & \\
        \mathfrak{a}^{1}_{\ell,1211} & \mathfrak{a}^{1}_{\ell,1212} & \mathfrak{a}^{1}_{\ell,1221} \mathfrak{a}^{1}_{\ell,1222} & \mathfrak{a}^{2}_{\ell,1211} & \mathfrak{a}^{2}_{\ell,1212} & \mathfrak{a}^{2}_{\ell,1221} \mathfrak{a}^{2}_{\ell,1222} \\
        \mathfrak{a}^{1}_{\ell,2111} & \mathfrak{a}^{1}_{\ell,2112} & \mathfrak{a}^{1}_{\ell,2121} \mathfrak{a}^{1}_{\ell,2122} & \mathfrak{a}^{2}_{\ell,2111} & \mathfrak{a}^{2}_{\ell,2112} & \mathfrak{a}^{2}_{\ell,2121} \mathfrak{a}^{2}_{\ell,2122} \\
        \mathfrak{a}^{1}_{\ell,2211} & \mathfrak{a}^{1}_{\ell,2212} & \mathfrak{a}^{1}_{\ell,2221} \mathfrak{a}^{1}_{\ell,2222} & \mathfrak{a}^{2}_{\ell,2211} & \mathfrak{a}^{2}_{\ell,2212} & \mathfrak{a}^{2}_{\ell,2221} \mathfrak{a}^{2}_{\ell,2222} 
    \end{pmatrix}
}, \\ 
{ \small
    \begin{pmatrix}
        \mathfrak{b}^{0,0}_{1111} & \mathfrak{b}^{0,0}_{1112} & \mathfrak{b}^{0,0}_{1121} \mathfrak{b}^{0,0}_{1122} & \mathfrak{b}^{1,0}_{1111} & \mathfrak{b}^{1,0}_{1112} & \mathfrak{b}^{1,0}_{1121} \mathfrak{b}^{1,0}_{1122} & \mathfrak{b}^{0,1}_{1111} & \mathfrak{b}^{0,1}_{1112} & \mathfrak{b}^{0,1}_{1121} \mathfrak{b}^{0,1}_{1122}  \\
        \mathfrak{b}^{0,0}_{1211} & \mathfrak{b}^{0,0}_{1212} & \mathfrak{b}^{0,0}_{1221} \mathfrak{b}^{0,0}_{1222} & \mathfrak{b}^{1,0}_{1211} & \mathfrak{b}^{1,0}_{1212} & \mathfrak{b}^{1,0}_{1221} \mathfrak{b}^{1,0}_{1222} & \mathfrak{b}^{0,1}_{1211} & \mathfrak{b}^{0,1}_{1212} & \mathfrak{b}^{0,1}_{1221} \mathfrak{b}^{0,1}_{1222} \\
        \mathfrak{b}^{0,0}_{2111} & \mathfrak{b}^{0,0}_{2112} & \mathfrak{b}^{0,0}_{2121} \mathfrak{b}^{0,0}_{2122} & \mathfrak{b}^{1,0}_{2111} & \mathfrak{b}^{1,0}_{2112} & \mathfrak{b}^{1,0}_{2121} \mathfrak{b}^{1,0}_{2122} & \mathfrak{b}^{0,1}_{2111} & \mathfrak{b}^{0,1}_{2112} & \mathfrak{b}^{0,1}_{2121} \mathfrak{b}^{0,1}_{2122} \\
        \mathfrak{b}^{0,0}_{2211} & \mathfrak{b}^{0,0}_{2212} & \mathfrak{b}^{0,0}_{2221} \mathfrak{b}^{0,0}_{2222} & \mathfrak{b}^{1,0}_{2211} & \mathfrak{b}^{1,0}_{2212} & \mathfrak{b}^{1,0}_{2221} \mathfrak{b}^{1,0}_{2222} & \mathfrak{b}^{0,1}_{2211} & \mathfrak{b}^{0,1}_{2212} & \mathfrak{b}^{0,1}_{2221} \mathfrak{b}^{0,1}_{2222} \\
        \mathfrak{b}^{1,0}_{1111} & \mathfrak{b}^{1,0}_{1112} & \mathfrak{b}^{1,0}_{1121} \mathfrak{b}^{1,0}_{1122} & \mathfrak{b}^{2,0}_{1111} & \mathfrak{b}^{2,0}_{1112} & \mathfrak{b}^{2,0}_{1121} \mathfrak{b}^{2,0}_{1122} & \mathfrak{b}^{1,1}_{1111} & \mathfrak{b}^{1,1}_{1112} & \mathfrak{b}^{1,1}_{1121} \mathfrak{b}^{1,1}_{1122}  \\
        \mathfrak{b}^{1,0}_{1211} & \mathfrak{b}^{1,0}_{1212} & \mathfrak{b}^{1,0}_{1221} \mathfrak{b}^{1,0}_{1222} & \mathfrak{b}^{2,0}_{1211} & \mathfrak{b}^{2,0}_{1212} & \mathfrak{b}^{2,0}_{1221} \mathfrak{b}^{2,0}_{1222} & \mathfrak{b}^{1,1}_{1211} & \mathfrak{b}^{1,1}_{1212} & \mathfrak{b}^{1,1}_{1221} \mathfrak{b}^{1,1}_{1222} \\
        \mathfrak{b}^{1,0}_{2111} & \mathfrak{b}^{1,0}_{2112} & \mathfrak{b}^{1,0}_{2121} \mathfrak{b}^{1,0}_{2122} & \mathfrak{b}^{2,0}_{2111} & \mathfrak{b}^{2,0}_{2112} & \mathfrak{b}^{2,0}_{2121} \mathfrak{b}^{2,0}_{2122} & \mathfrak{b}^{1,1}_{2111} & \mathfrak{b}^{1,1}_{2112} & \mathfrak{b}^{1,1}_{2121} \mathfrak{b}^{1,1}_{2122} \\
        \mathfrak{b}^{1,0}_{2211} & \mathfrak{b}^{1,0}_{2212} & \mathfrak{b}^{1,0}_{2221} \mathfrak{b}^{1,0}_{2222} & \mathfrak{b}^{2,0}_{2211} & \mathfrak{b}^{2,0}_{2212} & \mathfrak{b}^{2,0}_{2221} \mathfrak{b}^{2,0}_{2222} & \mathfrak{b}^{1,1}_{2211} & \mathfrak{b}^{1,1}_{2212} & \mathfrak{b}^{1,1}_{2221} \mathfrak{b}^{1,1}_{2222} \\
        \mathfrak{b}^{0,0}_{1111} & \mathfrak{b}^{0,0}_{1112} & \mathfrak{b}^{0,0}_{1121} \mathfrak{b}^{0,0}_{1122} & \mathfrak{b}^{1,0}_{1111} & \mathfrak{b}^{1,0}_{1112} & \mathfrak{b}^{1,0}_{1121} \mathfrak{b}^{1,0}_{1122} & \mathfrak{b}^{0,1}_{1111} & \mathfrak{b}^{0,1}_{1112} & \mathfrak{b}^{0,1}_{1121} \mathfrak{b}^{0,1}_{1122}  \\
        \mathfrak{b}^{0,1}_{1211} & \mathfrak{b}^{0,1}_{1212} & \mathfrak{b}^{0,1}_{1221} \mathfrak{b}^{0,1}_{1222} & \mathfrak{b}^{1,1}_{1211} & \mathfrak{b}^{1,1}_{1212} & \mathfrak{b}^{1,1}_{1221} \mathfrak{b}^{1,1}_{1222} & \mathfrak{b}^{0,2}_{1211} & \mathfrak{b}^{0,2}_{1212} & \mathfrak{b}^{0,2}_{1221} \mathfrak{b}^{0,2}_{1222} \\
        \mathfrak{b}^{0,1}_{2111} & \mathfrak{b}^{0,1}_{2112} & \mathfrak{b}^{0,1}_{2121} \mathfrak{b}^{0,1}_{2122} & \mathfrak{b}^{1,1}_{2111} & \mathfrak{b}^{1,1}_{2112} & \mathfrak{b}^{1,1}_{2121} \mathfrak{b}^{1,1}_{2122} & \mathfrak{b}^{0,2}_{2111} & \mathfrak{b}^{0,2}_{2112} & \mathfrak{b}^{0,2}_{2121} \mathfrak{b}^{0,2}_{2122} \\
        \mathfrak{b}^{0,1}_{2211} & \mathfrak{b}^{0,1}_{2212} & \mathfrak{b}^{0,1}_{2221} \mathfrak{b}^{0,1}_{2222} & \mathfrak{b}^{1,1}_{2211} & \mathfrak{b}^{1,1}_{2212} & \mathfrak{b}^{1,1}_{2221} \mathfrak{b}^{1,1}_{2222} & \mathfrak{b}^{0,2}_{2211} & \mathfrak{b}^{0,2}_{2212} & \mathfrak{b}^{0,2}_{2221} \mathfrak{b}^{0,2}_{2222}
    \end{pmatrix}.
}
\end{gather}
where $\textbf{1}$, $\textbf{2}$, $\textbf{3}$ and $\textbf{4}$ take the values of $1$ and $2$, denoting the two types of scalars. With the double $\mathbb{Z}_2$ symmetry, some of the components are identical to each other, significantly reducing the number of independent coefficients in the EFT.

\begin{figure}[h]
    \centering
    \includegraphics[scale=0.36]{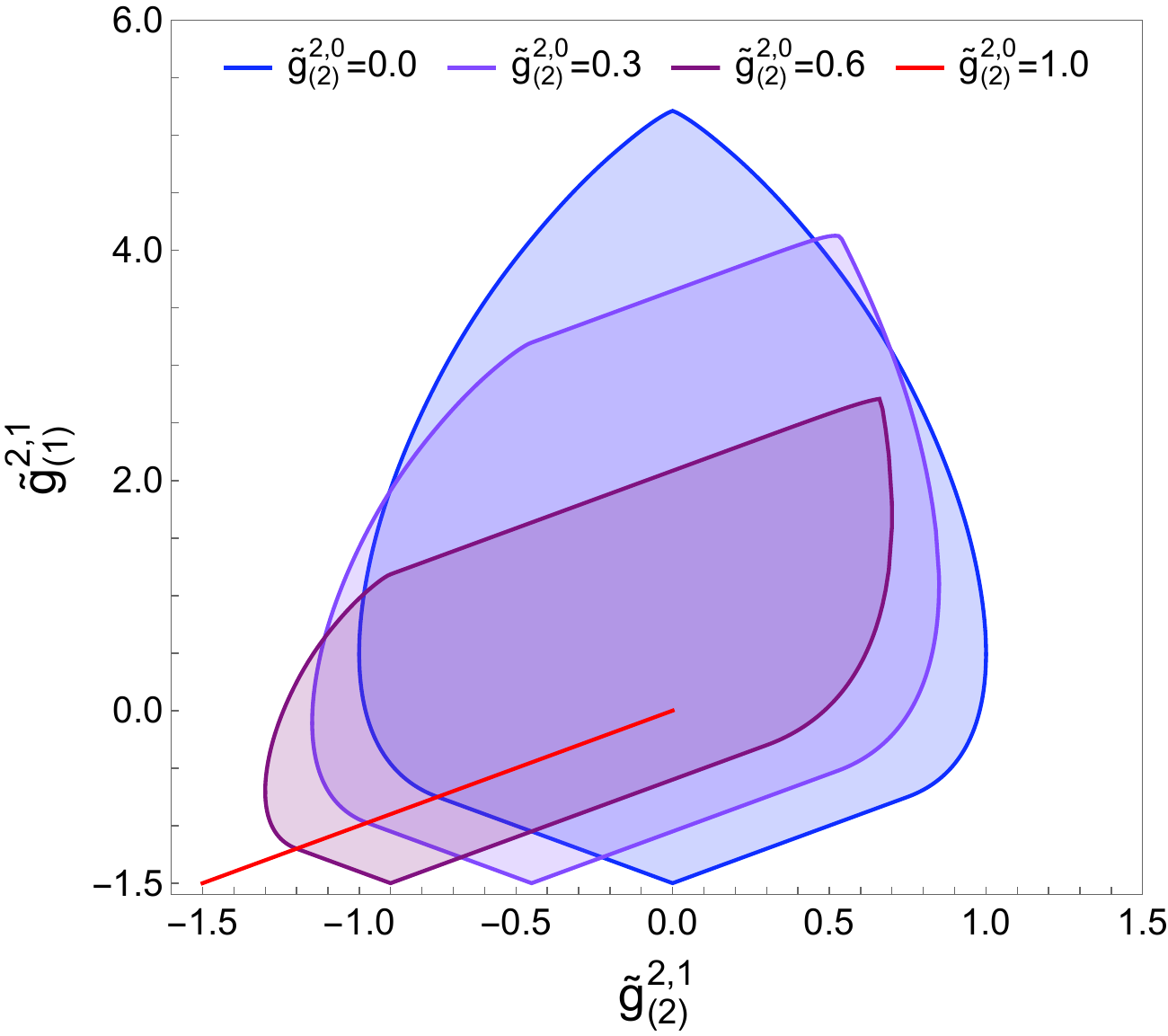}
    \caption{Positivity bounds on $\tilde{g}^{2,1}_{(1)}$ and $\tilde{g}^{2,1}_{(2)}$ with all $\tilde{g}^{i,j}_{(3)} = 0$ and the $i+j \leq 7$ null constraints in double $\mathbb{Z}_2$ bi-scalar theory. 
    We have used notations similar to Ref.~\cite{Du:2021byy}: $g^{2,0}_{(1)} = 2 g^{2,0}_{1111}$, $\tilde{g}^{i,j}_{(1)} = 2 g^{i,j}_{1122} / g^{2,0}_{(1)}$, $\tilde{g}^{i,j}_{(2)} = 4 g^{2,1}_{1111} / g^{2,0}_{(1)}$ and $\tilde{g}^{i,j}_{(3)} = 4 g^{i,j}_{1122} / g^{2,0}_{(1)}$.}
    \label{fig:DoubleZ2Scalar}
\end{figure}

The positivity bounds on this model have been computed in Ref.~\cite{Du:2021byy}. Here, we reproduce some of the bounds using the matrix moment formulation, which serves as a verification of the formalism, in particular, the choice of the base polynomials for the Hankel matrices. 
We parameterize the IR amplitude as
\begin{align}
    A_{\textbf{1234}} (s, t) = \text{ poles } + g^0_{\textbf{1234}} (t) + s \ g^1_{\textbf{1234}} (t) + \sum_{i \geq 2,j \geq 0} g_{\textbf{1234}}^{i,j}  (s + t/2)^i t^j
\end{align}
Let us compute, for example, the bounds on $g^{i,j}_{1122} / g^{2,0}_{1111}$ and $g^{i,j}_{1111} / g^{2,0}_{1111}$ for various $g^{2,0}_{1122} / g^{2,0}_{1111}$; see Figure \ref{fig:DoubleZ2Scalar}. 
These bounds can be easily computed with the {\tt SemidefniteOptimazition} function in {\bf Mathematica}, and the results are consistent with those of \cite{Du:2021byy}.

\subsection{Photon EFT}
\label{sec:photonEFT}

The spin effects are another way to introduce multiple degrees of freedom in a field theory. Let us first consider the case of an Abelian spin-1 EFT whose leading terms in the Lagrangian are given by
\begin{align}
    \mathcal{L} = - \dfrac{1}{4} F_{\mu\nu} F^{\mu\nu} + a_1 (F_{\mu\nu} F^{\mu\nu})^2 + a_2 (F_{\mu\nu} \tilde{F}^{\mu\nu})^2 + \cdots \ ,
\end{align}
where ${F}^{\mu\nu}$ is the field strength and its dual is defined as $\tilde{F}^{\mu\nu} =  \epsilon_{\mu\nu\rho\sigma} F^{\rho\sigma}/2$. 
For this model, the particle labels $\textbf{1}$, $\textbf{2}$, $\textbf{3}$ and $\textbf{4}$ in the 2-to-2 scattering amplitudes take the values of the two helicity states $h=+1,-1$. In the context of QED, when the electron is considered as a heavy particle and integrated out, we will get the famous Euler-Heisenberg EFT where the EFT
coefficients are suppressed by powers of the electron mass.
Additionally, dark photons have been popular in modeling the dark sector of the universe \cite{Fabbrichesi:2020wbt}.  
The positivity bounds of this theory have been comprehensively investigated by \cite{Henriksson:2021ymi, Henriksson:2022oeu}. 
Here we use the moment approach to reproduce some of the previous results.

For an Abelian spin-1 EFT with conserved parity and time reversal invariance, the amplitudes have the following symmetries
\begin{align}
    A_{\textbf{1234}} = A_{\bar{\textbf{1}}\bar{\textbf{2}}\bar{\textbf{3}}\bar{\textbf{4}}}, \quad
    A_{\textbf{1234}} = A_{\bar{\textbf{3}}\bar{\textbf{4}}\bar{\textbf{1}}\bar{\textbf{2}}},
     \quad
    A_{\textbf{1234}} = A_{\textbf{2143}},
\end{align}
where $\bar{\textbf{1}}$, for example, denotes the opposite helicity state of $\textbf{1}$.
Combined with crossing symmetry, we can parametrize the amplitudes as follows
\begin{gather}
    \begin{gathered}
        \begin{pmatrix}
        A_{++--} & A_{++-+} & A_{+++-} & A_{++++} \\
        A_{+---} & A_{+--+} & A_{+-+-} & A_{+-++} \\
        A_{-+--} & A_{-+-+} & A_{-++-} & A_{-+++} \\
        A_{----} & A_{---+} & A_{--+-} & A_{--++}
    \end{pmatrix} 
   \\
   =
    \begin{pmatrix}
       s^{2\eta} F(s|t,u) &(s t u)^{\eta} G(s,t,u) &(s t u)^{\eta} G(s,t,u) &H(s,t,u) \\
       (s t u)^\eta G(s,t,u) &u^{2\eta} F(u|s,t) &t^{2\eta} F(t|s,u) &(s t u)^{\eta} G(s,t,u) \\
       (s t u)^\eta G(s,t,u) &t^{2\eta} F(t|s,u) &u^{2\eta} F(u|s,t) &(s t u)^{\eta} G(s,t,u) \\
       H(s,t,u) &(s t u)^{\eta} G(s,t,u) &(s t u)^\eta G(s,t,u) &s^{2\eta} F(s|t,u)
    \end{pmatrix}
    \end{gathered}
    \label{amplitudespin12}
\end{gather}
where $\eta=1$ for the spin-1 case, $G(s, t, u)$ and $H(s, t, u)$ are triple crossing symmetric and $f(s|t,u)$ is only symmetric in swapping its second and third arguments. Also, taking into account the leading terms in the photon EFT Lagrangian, in the IR, we can parameterize these independent amplitudes according to their symmetries
\begin{align}
    A_{++--} (s, t) &= F_2 s^2 + F_3 s^3 + F_{4,1} s^4 + F_{4,2} s^2 (s^2 + t^2 + u^2) + \cdots \ , \\
    A_{+++-} (s, t) &= G_3 stu + G_5 stu (s^2 + t^2 + u^2) + \cdots \ , \\
    A_{++++} (s, t) &= H_2 (s^2 + t^2 + u^2) + H_3 stu + H_4 (s^2 + t^2 + u^2)^2 + \cdots \ .
\end{align}
Expanding these expressions in terms of $v=s+\frac{t}{2}$ and $t$, we can match one of $F_i,H_i,G_i$ to multiple $g^{i,j}_{\textbf{1234}}$. 
We can choose one of them to formulate the moment problem, disregarding the ones with $g^{i< N,j}_{\textbf{1234}}$, and use the rest as null constraints. 
Let us take $A_{++++}$ for example. We can re-write it in terms of $v$ and $t$:
\begin{align}
    A_{++++} = 2 {H_2} v^2+\frac{3 {H_2}}2 t^2+\frac{H_3}4 t^3-{H_3} v^2t +4 {H_4} v^4 +6 {H_4}v^2 t^2  +\frac{9H_4}4 t^4 + \cdots
\end{align}
Matching to the $v$ and $t$ coefficients at, say, the fourth order, we get $H_4 = \frac{1}{4} g^{4,0}_{++++}= \frac{1}{6} g^{2,2}_{++++} =\frac{4}{9} g^{0,4}_{++++}$. 
In this case, we use $H_4=\frac{1}{4} g^{4,0}_{++++}$ to get the moment formulation for $H_4$ and use $\frac{1}{4} g^{4,0}_{++++}= \frac{1}{6} g^{2,2}_{++++}$ as one null constraint, discarding $\frac{1}{6} g^{2,2}=\frac{4}{9} g^{0,4}$ because it contains $g^{0,4}$.

Thanks to the $tu$ crossing symmetry and selection rules, some of the partial wave components of the spectral functions $\rho_{1} = \operatorname{Im} A_{++++}$, $\rho_{2}=\operatorname{Im} A_{+++-}$ and $\rho_{3,s}=\operatorname{Im} A_{++--}$ vanish identically, which simplifies the computations. 
Specifically, $\rho^{\ell}_{1}$ and $\rho^{\ell}_{3,s}$ are nonzero only for $\ell = 0,2,4,\cdots$; $\rho^{\ell}_2$ and $\rho^{\ell}_{3,u} +  \rho^{\ell}_{3,t}$ are nonzero only for $\ell = 2,4,6,\cdots$; $ \rho^{\ell}_{3,u} -  \rho^{\ell}_{3,t}$ is nonzero only for $\ell = 3,5,7,\cdots$.

\begin{figure}
    \centering
    \includegraphics[scale=0.395]{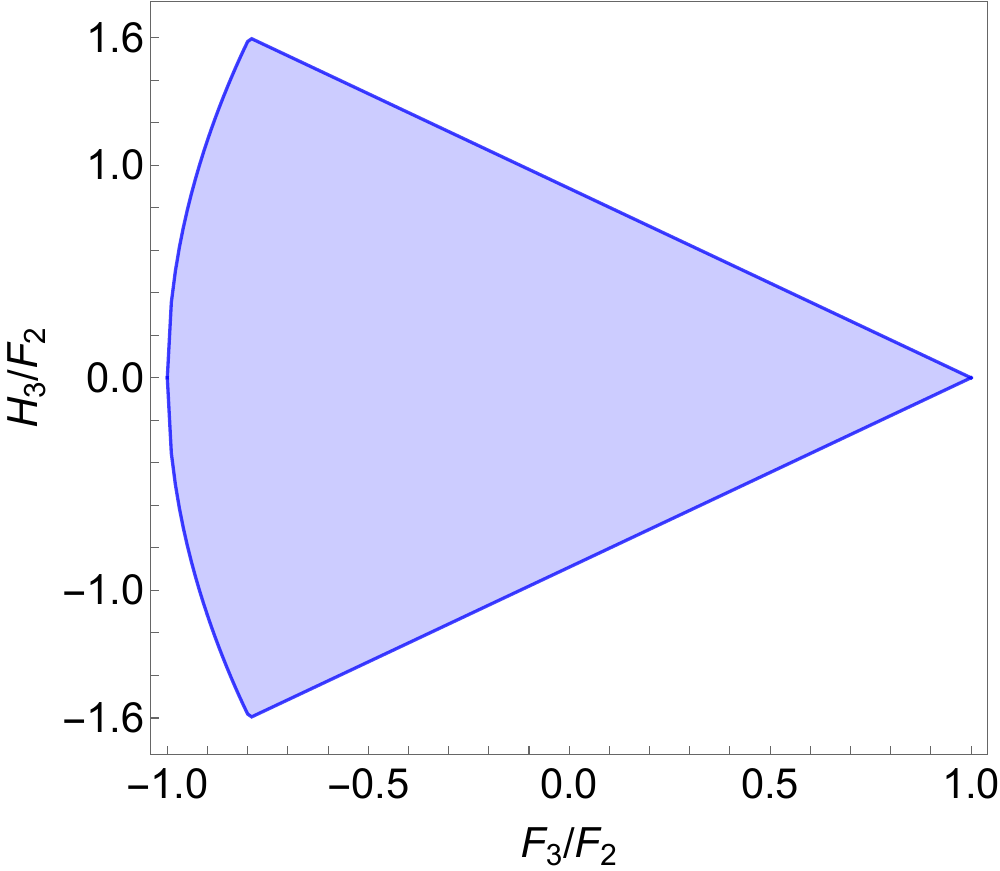}
    ~~~~~
    \includegraphics[scale=0.39]{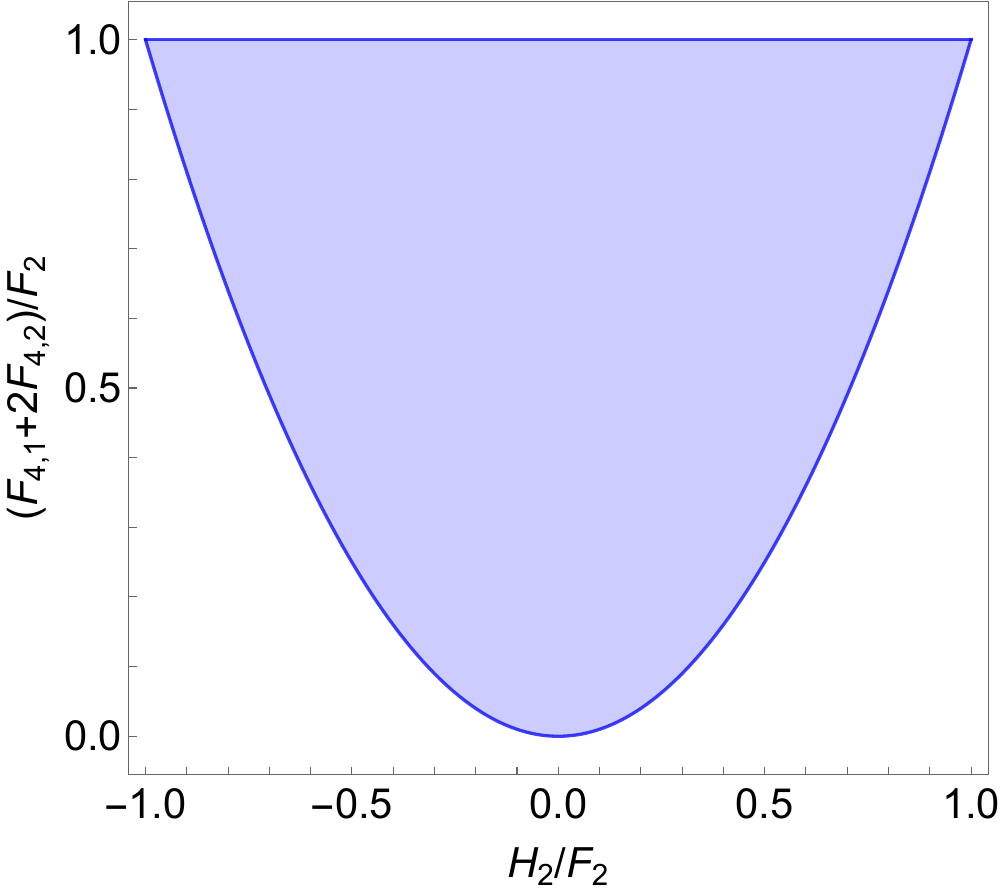}
    \caption{Positivity bounds on amplitude coefficients in the $H_3$-$F_3$ and $F_{4,1}+2F_{4,2}$-$H_2$ planes in the Photon EFT. Here we choose the truncation orders: $G=5$, $\ell_M=20$ and $i+j \leq 5$ for null constraints. }
    \label{fig:Vec_f4}
\end{figure}

As discussed in Section \ref{sec:formSpin}, for a field theory with spin, its moment formulation is slightly different from that of pure scalars. 
This is due to the appearance of non-integer powers of polynomials of moment variables within the dispersive integrals. 
A simple solution is to introduce extra moment variables. 
For the pure massless spin-1 case, we only need to introduce the extra variable $x_3=\sqrt{x_2 (x_2 - 2 x_1)}$, which means we need to define a tri-variate moment sequence
\begin{align}
     \mathfrak{b}_{\textbf{1234}}^{ \gamma_1,\gamma_2, \gamma_3} &= \int_{\mathcal{K}^{(3)}_{\mathfrak{b}}} x_1^{\gamma_2} x_2^{\gamma_2} x_3^{\gamma_3} \rho^{\prime\prime}_{\textbf{1234}} (x_1, x_2, x_3) \dif x_1 \dif x_2 \dif x_3.
\end{align}
as discussed in Section \ref{sec:formSpin}.
This will cause the Hankel matrices of $\mathfrak{b}_{\textbf{1234}}^{ \gamma_1,\gamma_2, \gamma_3}$ to become larger. In this case, the sufficient set of base polynomials to use is given by: 
\begin{align}
&x_1, 1 - x_1, x_2 - \ell_M (\ell_M + 1) x_1, x_3^2 - x_2^2 + 2 x_1 x_2, - (x_3^2 - x_2^2 + 2 x_1 x_2), x_1 (1 - x_1 ),\nn
&x_1 (x_2 - \ell_M (\ell_M + 1) x_1), (1 - x_1) (x_2 - \ell_M (\ell_M + 1) x_1), x_1 (x_3^2 - x_2^2 + 2 x_1 x_2), \nn
&(1 - x_1) (x_3^2 - x_2^2 + 2 x_1 x_2), - x_1 (x_3^2 - x_2^2 + 2 x_1 x_2), - (1 - x_1) (x_3^2 - x_2^2 + 2 x_1 x_2) .
\end{align}
Following the reasoning in Section \ref{sec:TruncatedMP}, we shall use $\mathfrak{b}_{\textbf{1234}}^{ \gamma_1,\gamma_2, \gamma_3}$ with $0 \leq \gamma_1 + \gamma_2 + \gamma_3 \leq G$ to construct the Hankel matrices. 
In Figure \ref{fig:Vec_f4}, we plot the positivity bounds for the ratios of the EFT coefficients $H_{3}/F_2$ and $F_3/F_2$ as well as $F_{4,1}+2F_{4,2}/F_2$ and $H_2/F_2$, which are in agreement with those of Ref.~\cite{Henriksson:2022oeu}.

\subsection{Gravitational EFT}
\label{sec:gravEFT}

Now, we turn to the case of a spin-2 EFT. Instead of constraining ratios of the EFT coefficients, in this case, we will apply the $L$-moment formulation for multiple degrees of freedom (see Section \ref{sec:Lmoment}) to obtain enclosed bounds for the EFT coefficients themselves. While the positivity bounds in the last subsections only use the positivity part of UV partial wave unitarity, we now also use the upper bounds on the UV partial waves.

The leading Lagrangian terms of a generic pure gravitational EFT in 4D can be written as (see, {\it e.g.}, \cite{Bern:2021ppb})
\begin{align}
    \label{eq:Grav_Lag}
    S = \frac{1}{16\pi G_N} \int \dif x^4 \sqrt{-g} \left( R + \dfrac{\lambda_3}{3!} R^{(3)} + \dfrac{\lambda_4}{2^3} (R^{(2)})^2 + \dfrac{\tilde{\lambda}_4}{2^3} (R \tilde{R})^2 + \cdots \right)
\end{align}
where $R^{(2)} = R^{\mu\nu\kappa\lambda} R_{\mu\nu\kappa\lambda}$, $R^{(3)} = R^{\mu\nu\kappa\lambda} R_{\kappa\lambda\alpha\gamma} R^{\alpha\gamma}_{\phantom{\alpha\gamma}\mu\nu}$ and $R \tilde{R} = \frac{1}{2} R^{\mu\nu\alpha\beta} \epsilon_{\alpha\beta}{}^{\gamma\delta} R_{\gamma\delta\mu\nu}$.
We consider 2-to-2 tree-level scatterings around Minkowski space for the spin-2 fields $h_{\mu\nu}$: $g_{\mu\nu} = \eta_{\mu\nu} + \kappa h_{\mu\nu}$, where $\kappa^2 = 32 \pi G_N$. 
With the explicit leading terms of the Lagrangian, it is easy to see that the independent scattering amplitudes for this case can be parameterized by \eref{amplitudespin12} with $\eta=2$ and
\begin{align}
    F(s|t,u) &= \frac{8 \pi G_N}{s t u} + \alpha_1 \frac{1}{s} + \alpha_2 \frac{t^2}{s} + \sum_{k, q} F_{k, q} s^{k-q} t^q \ , \\
    G(s,t,u) &= \frac{\alpha_3}{s t u} + \sum_{k, q} G_{k, q} s^{k-q} t^q \ , \\
    H(s,t,u) &= \sum_{k, q} H_{k, q} s^{k-q} t^q \ .
\end{align}
where $\alpha_i$, $F_{k,q}$, $G_{k,q}$ and $H_{k,q}$ are EFT coefficients. With this setup, the null constraints can be easily extracted. (Here, we choose the subtraction order $N_s = 3$ because we are considering 4D gravity.) Then, we can follow the method described in Section \ref{sec:Lmoment} to derive the enclosed bounds on the lowest order EFT coefficients themselves.

\begin{figure}[h]
    \centering
    \includegraphics[scale=0.5]{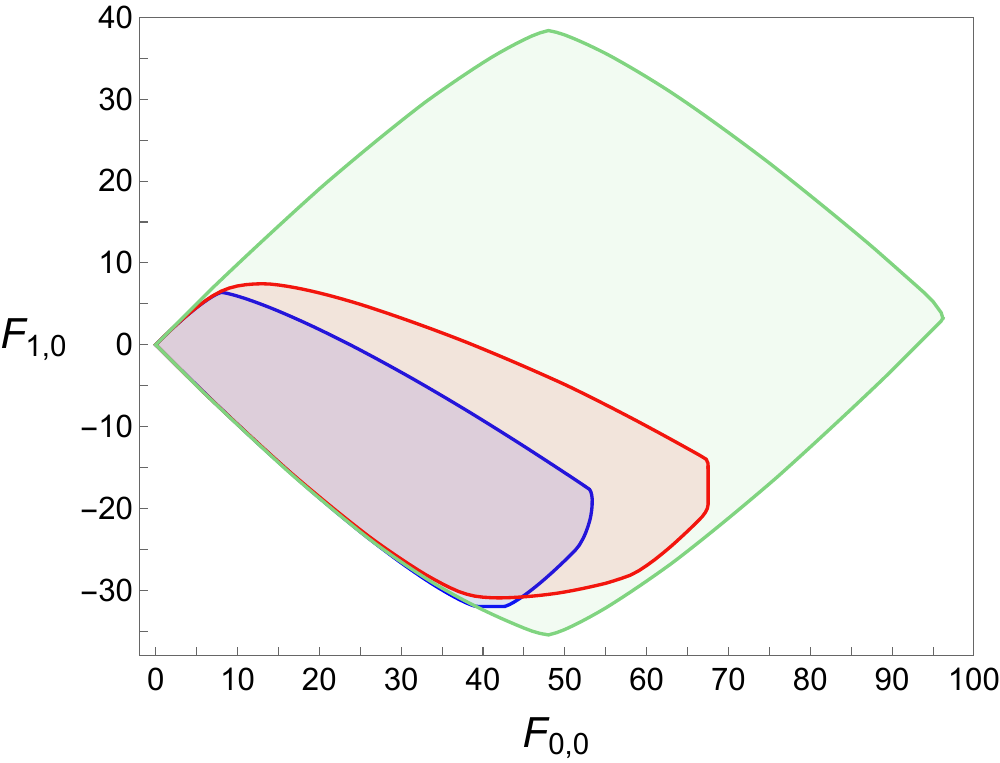}
    \caption{
    Bounds on $F_{1,0}$ and $F_{0,0}$ using both the positivity part and the upper bounds of partial wave unitarity. 
    The blue region is computed from the $L$-moment formulation for a single DoF without null constraints but with the extra assumption that $\lambda_3=0$ in \eref{eq:Grav_Lag}.
    The green region is from the $L$-moment formulation for a single DoF without the $\lambda_3 = 0$ condition but with null constraints from $A_{++--} (s, t) = A_{++--} (s, u)$ involving $g^{i,j}_{\textbf{1234}},~ i+j \leq 7$. These two regions are in agreement with Ref.~\cite{Chiang:2022jep}. The blue region is from our matrix $L$-moment approach for multi-DoFs without the $\lambda_3 = 0$ condition but with null constraints from all crossing relations involving $g^{i,j}_{\textbf{1234}}, i+j \leq 7$.
    }
    \label{fig:Grav_F00F10_Upper}
\end{figure}

\begin{figure}[h]
    \centering
    \includegraphics[scale=0.47]{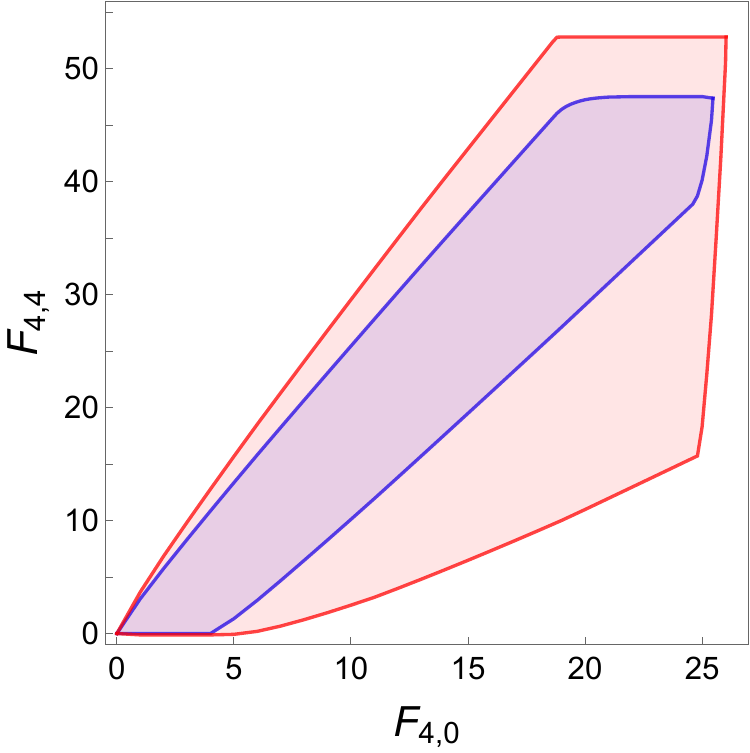}
    \caption{
    Bounds on $F_{4,0}$ and $F_{4,4}$ for $\lambda_3=0$ (see \eref{eq:Grav_Lag}) using both the positivity part and the upper bounds of partial wave unitarity. For the red region, we only impose null constraints from $A_{++--} (s, t) = A_{++--} (s, u)$ involving $g^{i,j}_{\textbf{1234}}, i+j \leq 8$. For the blue region, we impose null constraints from all available crossing symmetries involving $g^{i,j}_{\textbf{1234}}, i+j \leq 8$.
    }
    \label{fig:Grav_F40F44_Upper}
\end{figure}

The bounds on $F_{1,0}$ and $F_{0,0}$ have previously been computed with the $L$-moment formulation for a single DoF using the technique of Minkowski sums as well as numerically with linear programming \cite{Chiang:2022jep}. In Figure \ref{fig:Grav_F00F10_Upper}, we have reproduced these results in our approach: (1) the special case where the cubic curvature term $R^{(3)}$ is absent in the effective Lagrangian, which in effect gives rise to a null condition $F_{1,1}=\alpha_2=0$, without using any null constraints; (2) the case with a nonzero cubic curvature term $R^{(3)}$, using only the null constraints from the following crossing relation
\begin{align}
     A_{++--} (s, t) = A_{++--} (s, u).
\end{align}
For the (1) case, more specifically, we obtain the following upper bound on $F_{0,0}$
\begin{align}
\max\left( \frac{F_{0,0}}{(4 \pi)^2}\right) = 0.33783
\end{align}
which agrees with that of Ref.~\cite{Chiang:2022jep} to at least three decimal places. Furthermore, the matrix moment formulation allows us to use the null constraints from all the crossing relations, which, as we can see in Figure \ref{fig:Grav_F00F10_Upper}, gives rise to much stronger bounds for the case with a nonzero cubic curvature term $R^{(3)}$.
Similarly, we have also computed the bounds on $F_{4,0}$ and $F_{4,4}$ as another example to showcase the improvement due to the use of all the crossing relations; See Figure \ref{fig:Grav_F40F44_Upper}.

\section{Inverse problem: from IR coefficients to UV spectrum}
\label{sec:Spec}

As already mentioned, mathematically, a moment problem asks whether there exists a possible measure $\mu_{\mathbf{ij}}$ for a given moment sequence $\{\mathfrak{m}_{\mathbf{ij}}^{\gamma_1\gamma_2} \mid (\gamma_1, \gamma_2) \in \Theta \}$ such that an integral representation \eqref{momentpro0} exists \cite{Schmdgen2017TheMP}.
If there exists a measure for such integrals, it is said that the moment problem is {\it solvable}, and this measure is a \textit{representing measure}. 
In other words, the solvability condition discussed in Section \ref{sec:MP} is the necessary and sufficient condition for the existence of a representing measure.
Physically, solving a specific moment problem is to determine whether there exists a UV spectrum for the given values of the EFT coefficients. Additionally, we may also ask whether it is possible to reconstruct a UV spectrum by the given set of EFT coefficients and whether the inferred UV spectrum is unique or the determinacy of the moment problem.

In this section, we shall discuss how to reverse engineer the UV model from the IR information of the EFT. We will start with a toy (mathematical) example of truncated Hamburger moment problem. 
In Section \ref{sec:MassSpec}, we connect the problem of reconstructing the UV mass spectrum from the forward amplitude to a Hausdorff moment problem and demonstrate the reconstruction with the the Veneziano amplitude. 
In Section \ref{sec:AllSpec}, we generalize the problem to the case of a non-forward amplitude, which allows us to reconstruct the UV spectrum of the masses and couplings from a given IR moment sequence. 
In Section \ref{sec:EFT_gives_Moments}, we demonstrate how to derive the moment sequence from the EFT coefficients in the amplitude.
In Section \ref{sec:Determinacy}, we briefly discuss the mathematical concept of determinacy of a moment problem and its connections to uniqueness of the inferred UV spectrum. 
In Section \ref{sec:SpecExamples}, we illustrate the concepts and the methods in the UV reconstruction with explicit examples.

\subsection{Toy example: Truncated Hamburger}
\label{sec:SpecToyExam}

Let us first illustrate how to compute the representing measure or the UV spectrum with a toy example. 
Suppose there is a solvable $\Theta(3)$-truncated Hamburger (uni-variate, $1\times 1$ matrix) moment sequence $\set{\mathfrak{m}^0, \mathfrak{m}^1, \mathfrak{m}^2, \mathfrak{m}^3}$, which means they has the following integral representation with at least one possible measure $\mu$
\begin{align}
    \mathfrak{m}^{\gamma} = \int_{\mathbb{R}} x^\gamma \dif \mu, \gamma = 0,1,2,3.
\end{align}
From Section \ref{sec:TruncatedMP}, we know the following Hankel matrix shall be PSD ($\mathfrak{m}$ in $\mathcal{H}_{1}^{\mathfrak{m}}$ will be omitted if there is no ambiguity):
\begin{align}
    \mathcal{H}_{1} (2) &\equiv
    \begin{pmatrix}
        \mathfrak{m}^0 & \mathfrak{m}^1 \\
        \mathfrak{m}^1 & \mathfrak{m}^2 \\
    \end{pmatrix}.
\end{align}
Besides, the Hankel matrix shall have an eventual flat PSD extension.
For simplicity, we further assume that, given the first $4$ moments, there exists an $\mathfrak{m}^4$ such that the extended Hankel matrix
\begin{align}
    \mathcal{H}_{1} (4) &\equiv
    \begin{pmatrix}
        \mathfrak{m}^0 & \mathfrak{m}^1 & \mathfrak{m}^2 \\
        \mathfrak{m}^1 & \mathfrak{m}^2 & \mathfrak{m}^3 \\
        \mathfrak{m}^2 & \mathfrak{m}^3 & \mathfrak{m}^4 \\
    \end{pmatrix}
\end{align} 
satisfies
$\mathcal{H}_{1} (4) \succeq 0$ and $\mathcal{H}_{1} (4)$ is flat, \textit{i.e.,} $\operatorname*{rank} \mathcal{H}_{1} (4) = \operatorname*{rank} \mathcal{H}_{1} (2)$.
Moreover, we also assume that $\operatorname*{rank} \mathcal{H}_{1} (2) = 2$, which implies $\mathfrak{m}^{0} >0, ~\mathfrak{m}^{2} > 0,~ \mathfrak{m}^{0} \mathfrak{m}^{2} - (\mathfrak{m}^{1})^2 > 0$.
Then, the rank equality condition fixes this fictitious $\mathfrak{m}^4$.

Given these assumptions, it is possible to prove this moment problem is really solvable by actually finding a possible set $\set{(T^n, \omega_n)}$ such that the corresponding measure admits the following representation with delta functions
\begin{align}
    \label{eq:MomsAreDiscreteSum}
    \dif \mu = \sum_n  T^n \delta (x - \omega_n) \dif x  ~~\Rightarrow~~
    \mathfrak{m}^\gamma = \sum_{n} T^n \omega_n^\gamma,~~ \gamma =0, 1, 2, 3, 4.
\end{align}
where $T^n$ are non-negative real numbers and $\omega_n$ are real numbers. 
A measure like this, which consists of a sum of delta functions, is called an \textit{atomic measure}. 
To see this, note that, as a result of $\operatorname*{rank} (\mathcal{H}_{1} (4)) = \operatorname*{rank}  (\mathcal{H}_{1} (2))$, it will always be possible to define a real, $1\times 2$ matrix $W\equiv  (W_{x^2,1},  W_{x^2,x})^T$ via the following equation
\begin{align}
    \label{eq:DefOfWMartix1}
    \begin{pmatrix}
        \mathfrak{m}^0 & \mathfrak{m}^1 \\
        \mathfrak{m}^1 & \mathfrak{m}^2 \\
    \end{pmatrix} 
    \begin{pmatrix}
        W_{x^2,1} \\
        W_{x^2,x}
    \end{pmatrix} = 
    \begin{pmatrix}
        \mathfrak{m}^2 \\
        \mathfrak{m}^3  \\
    \end{pmatrix} .
\end{align}
Once the $W$ matrix is obtained, we can choose $\omega_n$ to be the roots of the following equation
\begin{align}
    \begin{pmatrix}
        1 & \omega
    \end{pmatrix} \cdot
    \begin{pmatrix}
        W_{x^2,1} \\
        W_{x^2,x}
    \end{pmatrix} = \omega^2.
\end{align}
Indeed, with these $\omega_n$, \eref{eq:DefOfWMartix1} is automatically satisfied if  $\mathfrak{m}^\gamma = \sum_{n} T^n \omega_n^\gamma$ is plugged into \eref{eq:DefOfWMartix1}.
Also, $W_{x^2,x}^2 + 4 W_{x^2,1}$ can be shown to be PSD by the solution of \eref{eq:DefOfWMartix1}, which means that $\omega_n$ are always real roots. 
(If the intergration region were a general semialgebraic set $\mathcal{K}$, instead of $\mathbb{R}$, this would give rise to roots $\omega_n$ living within $\mathcal{K}$ \cite{Kimsey2022OnAS}.)

Having found $\omega_n$, $T^n$ can be obtained by solving
\begin{align}
    V(\omega_1, \omega_2; \{ (0), (1) \})
    \begin{pmatrix}
        T^1 \\
        T^2
    \end{pmatrix}
    = 
    \begin{pmatrix}
        \mathfrak{m}^{0} \\
        \mathfrak{m}^{1}
    \end{pmatrix}
\end{align}
where $V(\omega_1, \omega_2; \{ (0), (1) \})$ is a Vandermonde matrix
\begin{align}
    V(\omega_1, \omega_2; \{ (0), (1) \}) = 
    \begin{pmatrix}
        (\omega_1)^0 & (\omega_2)^0 \\
        (\omega_1)^1 & (\omega_2)^1 \\
    \end{pmatrix}
\end{align}
$T^n$ can be verified to be indeed PSD
\begin{align}
    T^1 &= \dfrac{- \mathfrak{m}^{1} + \mathfrak{m}^0 \omega_2}{\omega_2 - \omega_1} \geq 0, &T^2 &= \dfrac{ \mathfrak{m}^{1} -\mathfrak{m}^0 \omega_1}{\omega_2 - \omega_1} \geq 0 .
\end{align}
With the above choice of $\set{(T^n,\omega_n)}$, as one can easily verify, $\mathfrak{m}^0$, $\mathfrak{m}^1$, $\mathfrak{m}^2$ and $\mathfrak{m}^3$ indeed admit the following integral representations
\begin{align}
    \mathfrak{m}^\gamma = \int_{\mathbb{R}} x^\gamma \dif \mu, ~~~~ \dif \mu = \Big( T^1 \delta (x - \omega_1) + T^2 \delta (x - \omega_2) \Big) \dif x  .
\end{align}
Due to the condition $\operatorname*{rank} \mathcal{H}_{1} (4) =  \operatorname*{rank} \mathcal{H}_{1} (2) = 2$, $\mathfrak{m}^{4}$ is in fact uniquely determined:
\begin{align}
\label{getm4}
    \mathfrak{m}^4 =  \dfrac{(\mathfrak{m}^2)^3 - 2\mathfrak{m}^1 \mathfrak{m}^2 \mathfrak{m}^3 + 3 \mathfrak{m}^0 (\mathfrak{m}^3)^2}{\mathfrak{m}^0 \mathfrak{m}^2 -(\mathfrak{m}^1)^2} .
\end{align}
Alternatively, we may also write it as $\mathfrak{m}^4=T^1 (\omega_1)^4 + T^2 (\omega_2)^4$, meaning that $\mathfrak{m}^4$ has the same representing measure. 

In other words, for this truncated Hamburger problem with an odd integer truncation order $G=3$ (thus $\mathcal{H}_1 (G) = \mathcal{H}_1 (G-1)$), we can find an eventual flat extension with least atoms \footnote{
    Note that we can write $\mathcal{H}_1 (\infty)= \sum_n (1,x_n, x_n^2, \cdots)^T T^n (1,x_n, x_n^2, \cdots)$, so $\operatorname*{rank} \mathcal{H}_1 (\infty)$ measures the number of delta functions.
    }
such that $\operatorname*{rank} \mathcal{H}_1 (G-1) = \operatorname*{rank} \mathcal{H}_1 (G+1)$, if $\mathcal{H}_1 (G-1) \succeq 0$ is satisfied. 
Generally speaking, with a given order of truncation, this method allows us to capture the UV spectrum to a certain order for the lowest lying modes, and as the truncation order increases, we shall see more modes to be accurately determined. 
A caveat is that there can also be some other non-minimal eventual flat extensions, and also a moment problem may not have a minimal flat extension, i.e., which means our representing measure satisfies $\operatorname*{rank} \mathcal{H}_1 (\infty) = \operatorname*{rank} \mathcal{H}_1 (G)$. 

The simplicity of this toy model is that the moment integration range is $\mathbb{R}$, so we only need to evaluate the $\mathcal{H}_1$ matrices. 
In principle, for a generic (semialgeraic) set $\mathcal{K}$, one may need to deal with more Hankel matrices. 
However, as we shall discuss later, for a moment sequence that remains determinate even if $\mathcal{K}$ is relaxed to $\mathbb{R}$, it is sufficient to only evaluate the $\mathcal{H}_1$ matrices. 

\subsection{UV mass spectrum from EFT at $t=0$}
\label{sec:MassSpec}

If the integration region $\mathcal{K}$ of the uni-variate problem is restricted to the interval $[0,1]$ (instead of $\mathbb{R}$), we arrive at the Hausdorff moment problem, which, as already mentioned, has a direct physics bearing: it arises directly from the forward limit of the 2-to-2 scattering of a scalar field theory.
To be concrete, the EFT coefficients of the forward-limit of $2$-to-$2$ scattering amplitude could be expressed as a sum of moments $\mathfrak{m}^{\gamma_1}$, which are given by
\begin{align}
\label{mgamma1Haus}
    \mathfrak{m}^{\gamma_1} = \int_{0}^{1} x_1^{\gamma_1} \dif \mu,
\end{align}
where $x_1 = {\Lambda^2}/{s^\prime}$ and $\dif \mu = x_1^{N_s - 1} \operatorname{Abs} A ( \Lambda^2 / x_1, 0) \dif x_1$.
Then, the existence of an atomic measure constructed by delta functions (see \eref{eq:MomsAreDiscreteSum}) 
means that we can find a UV completion with a number of particles with mass $M_I$:
\begin{align}
\label{eq:TIdef3}
    A(s,0) \sim \sum_I \frac{\mathcal{T}^I}{ M_I^2-s} .
\end{align}
In this sense, the solution of the moment problem or the representing measure provides a possible UV completion of the EFT. 
Thus, this method allows us to re-construct a possible UV mass spectrum and couplings from the IR data.

Additionally, every solvable Hausdorff moment sequence $\set{\mathfrak{m}^{\gamma}}$ satisfies Carleman's condition \cite{Schmdgen2017TheMP}
\begin{align}
    \sum_{\gamma=1}^{\infty} (\mathfrak{m}^{2 \gamma})^{-\frac{1}{2\gamma}} = \infty ,
\end{align}
which is easy to see by noting the fact that $\mathfrak{m}^{\gamma} \geq \mathfrak{m}^{\gamma+1}$, because of \eref{mgamma1Haus} and $x_1\in [0,1]$, leading to $\sum_{\gamma=1}^{\infty} (\mathfrak{m}^{2 \gamma})^{-\frac{1}{2\gamma}} \geq \sum_{\gamma=1}^{\infty} (\mathfrak{m}^{0})^{-\frac{1}{2\gamma}} = \infty$. This means that every solvable full Hausdorff moment problem has a unique representing measure.  
In other words, the forward-limit UV mass spectrum we can construct is unique. 
The uniqueness of the moment problem away from the forward limit will be discussed in Section \ref{sec:Determinacy}. For a truncated Hausdorff moment problem, however, multiple representing measures may exist.
Intuitively, this can be understood because the measure function contains an infinite number of degrees of freedom, so there are many ways to accommodate a finite moment sequence with different representing measures. 
For the forward-limit, it is sufficient to choose a representing measure with a minimal number of delta functions. As the truncation order increases, as we shall see in the example below, the representing measure will increasingly approach the unique measure from the full moment problem.  

\subsubsection{Veneziano amplitude}

Let us inject a UV model and show how to recover the UV model from the truncated IR data.
For concreteness, we consider a model whose full moment sequence is given by
\begin{align}
\label{eq:ForwardVeneziano}
    \mathfrak{m}^{\gamma} =  \sum_{I=1}^\infty \left(\frac{1}{I}\right)^{2} \left(\frac{1}{I}\right)^{\gamma} = \zeta (\gamma+2),
\end{align}
where $\zeta(z)$ is the Riemann zeta function. 
This injected UV model is acutally nothing but the Veneziano amplitude in the tree-level open string scattering with the color and kinematic factors stripped. To see this, note that the Veneziano amplitude with zero Regge intercept is given by
\begin{align}
    \mathcal{A}_{\text{Ven}} (s, t) = \dfrac{\Gamma (-s) \Gamma(-t)}{\Gamma(1 - s - t)} = \dfrac{1}{s t} \exp \left( \sum_{k \geq 2} \dfrac{\zeta(k)}{k} \left[ s^k + t^k - (s+t)^k \right] \right). 
\end{align}
which has a pole $1/(st)$ in the forward limit. The pole subtracted Veneziano amplitude at $t=0$ can be written as
\begin{align}
    \tilde{\mathcal{A}}_{\text{Ven}}(s,0) = \sum_{k \geq 2} \zeta(k) s^k  = \sum_{I=1}^{\infty} \dfrac{1/I}{I-s},
\end{align}
which yields the above moment sequence. 
Here we have chosen the mass units such that the Regge slope $\alpha^\prime = 1$.

Suppose only the first few EFT coefficients $\set{\mathfrak{m}^\gamma}$ are known where $\gamma \leq G$, $G$ being an odd integer, so we can precisely evaluate the Hankel matrix $\mathcal{H}_{1} (G)$. 
For this special case, it is always possible to find a fictitious $\mathfrak{m}^{G+1}$ such that the Hankel matrix $\mathcal{H}_{1} (G+1)$ is a flat extension of $\mathcal{H}_{1} (G)$:
\begin{align}
    \label{eq:Flat}
    \operatorname*{rank} \mathcal{H}_{1} (G+1) = \operatorname*{rank} \mathcal{H}_{1} (G).
\end{align}
This means that it is always possible to find a minimal representing measure. 
While the flatness condition already guarantees the PSD of $\mathcal{H}_{1} (G+1)$, it happens that this fictitious $\mathfrak{m}^{G+1}$ also makes $\mathcal{H}_{x} (G+1)$ and $\mathcal{H}_{1-x} (G+1)$ PSD. 
However, this is not true generically, as we will shortly demonstrate with a simple example.

We emphasize that the true value of the order-$(G+1)$ moment $\mathfrak{m}^{G+1} = \zeta (G + 3)$ can {\it not} satisfy the minimal flatness condition, because the flatness condition is in essence the condition used to obtain the minimal representing measure for the first several moments from infinitely many representing measures of the solvable truncated moment problem. Also, in general, it may not be possible to find $\mathfrak{m}^{G+1}$ such that $\mathcal{H}_{1} (G+1)$ is flat and $\mathcal{H}_{x} (G+1)$ and $\mathcal{H}_{1-x} (G+1)$ are both PSD simultaneously, in which case one should go to a higher order. 
If the moments are reasonable/solvable, {\it i.e.}, within the positivity bounds, it is always possible to find a larger $G'$ such that $\mathcal{H}_{1} (G')$ is flat. 
For the problem at hand, however, if all the leading (odd order) moments are known precisely, it happens that a flat extension can be found at each order.

\begin{figure}
    \centering
    \includegraphics[scale=0.47]{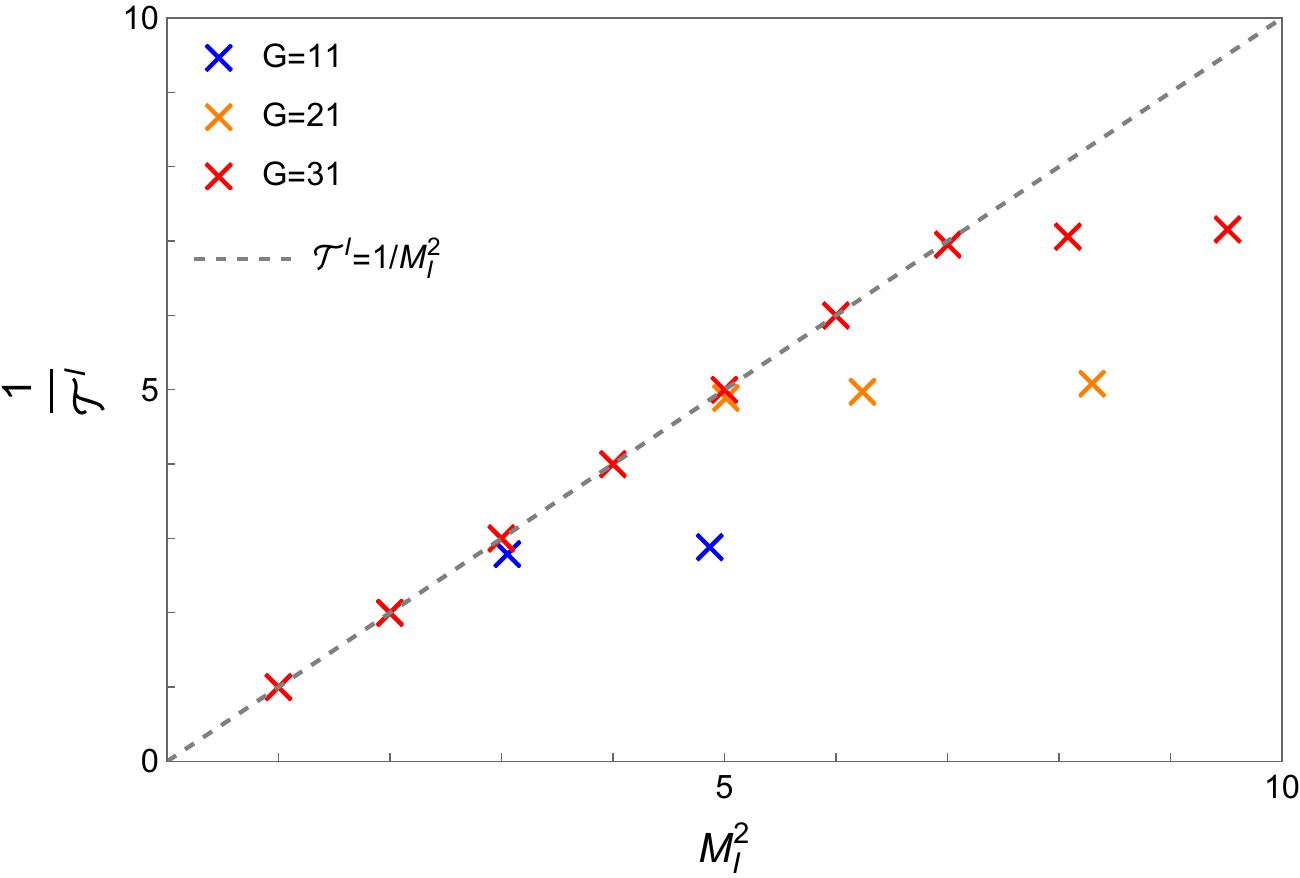}
    \caption{
    UV mass spectra reconstructed from truncated sets of EFT coefficients for the forward limit of the Veneziano amplitude; see \eref{eq:ForwardVeneziano}. 
    $M_I^2$ denotes the mass of the UV particle, and $\mathcal{T}^I$ denotes the ``coupling strength''(see \eref{eq:TIdef3} for its definition). 
    The true spectrum lies on the dashed line, \eref{eq:ForwardVeneziano}, while the crosses are the reconstructed UV spectra from EFT coefficients $\set{g^{i,0}}$ truncated up to order $i \leq G + 2$.
    }
    \label{fig:SingleFullMPExample1}
\end{figure}

Recalling $x_1=1/s^\prime$ (here $\Lambda=1$) and following the algorithm outlined in Eqs.~(\ref{eq:DefOfWMartix1}-\ref{getm4}), we can reverse-engineer the measure from a truncated set of EFT coefficients
\begin{align}
\dif \mu \sim \sum_{I=1} \mathcal{T}^I \delta (s'-M_I^2) \dif s',
\end{align}
In Figure \ref{fig:SingleFullMPExample1},  we plot how the UV mass spectrum obtained from the IR coefficients approximate the underlying true UV spectrum $\mathcal{T}^I = 1/I = 1/M_I^2,~M_I^2=1,2,\cdots$.
As the truncated order $G$ increases, the reconstructed UV mass spectrum can more accurately capture a growing number of low-lying modes in the true mass spectrum.

Note that additional UV information may not be necessary for the reserve engineering. 
For example, the Veneziano amplitude has a mass gap between the lowest lying UV modes (located at $M_I^2=1$) and the heavier modes (located above $M_I^2=2$). 
If this information is incorporated into our moment problem, rather than having the semialgebraic set $\mathcal{K}=[0,1]$, the measure is instead defined on
\begin{align}
\mathcal{K} = \set{ x_1 \mid x_1 (1-x_1) \geq 0, (x_1-1)(x_1-1/2) \geq 0}= [0, 1/2] \cup \{ 1 \}
\end{align}
For this case, the PSD of Hankel matrix $\mathcal{H}_{(x_1-1)(x_1-1/2)}$ is required, which is more difficult to solve numerically.
However, just as we have seen in Figure \ref{fig:SingleFullMPExample1}, the numerical result still matches the real spectrum, largely thanks to the determinacy property of the Hausdorff moment problem. So, when possible, it is preferable to use sufficient but loose constraints to compute the spectrum. 

\subsection{UV reconstruction for $t\neq 0$}
\label{sec:AllSpec}

Now we turn to the reverse engineering of the UV theory for the general case of a multi-field theory away from the forward limit. 
For an arbitrary set of EFT coefficients, we wish to seek a UV amplitude with countable UV particles that can realize this EFT at low energies. 
As we shall see, such a UV amplitude can always be found if the set of coefficients are within the positivity bounds; that is, if the corresponding PSD Hankel matrices admit eventual flat extensions.

We will see that for a multi-field theory, it is always possible to find a representing measure of the following form
\begin{align}
\label{eq:dmuikTo}
    \d\mu_{\mathbf{ij}} (x_1,x_2) &= \sum_{n} T^{n}_{\mathbf{ij}} \delta (x_1 - \omega_{1 n}) \delta (x_2 - \omega_{2 n}) \dif x_1 \dif x_2 \\
    \label{x1x2explicit77}
    &= \sum_{I,\ell} T^{I,\ell}_{\textbf{1234}} \delta \Big( x_1 - \dfrac{\Lambda^2}{M_I^2} \Big) \delta \Big(x_2 - \dfrac{\Lambda^2}{M_I^2} \ell (\ell + 1) \Big) \d x_1 \d x_2 \\
    &= \sum_{I,\ell} T^{I,\ell}_{\textbf{1234}} \delta (s^\prime - M_I^2) \delta ( J^2 - \ell (\ell + 1)) \d s^\prime \d  J^2
\end{align}
for any given full moment sequence $\set{\mathfrak{b}^{\gamma_1 \gamma_2}_{\textbf{ij}}}$. 
In the second line above, we have translated back to the amplitude language and separate the summation index $n$ into $I$ and $\ell$, denoting the mass and spin respectively.
However, ultimately, we are interested in reconstructing the UV spectrum for any given values of all EFT coefficients, rather than moments $\set{\mathfrak{b}^{\gamma_1 \gamma_2}_{\textbf{ij}}}$.
As we see in \eref{gij1234Exp}, because of the simultaneous presence of the $s$ and $u$ channel, the conversion from the moments to the EFT coefficients is not straightforward. 
We will address this subtlety shortly in Section \ref{sec:EFT_gives_Moments}.

Once we have obtained $(T^n_{\mathbf{ij}},\omega_{1n},\omega_{2n})$ or $(T^{I,\ell}_{\textbf{1234}}, M_I^2,\ell)$, using the definition (\eref{sigmaspe} and \eref{rhospe}),
the UV spectral functions of the theory can be written as
\begin{align}
    \label{eq:DiscResult}
    \operatorname*{Abs} A^\ell_{\textbf{1234}} (s') = \dfrac{1}{8 \pi (2 \ell + 1)} \sum_{I}  ~ 
    \mathcal{T}^{I, \ell}_{\textbf{1234}} ~ \pi \delta (s^\prime - M_I^2) ,~~~~{\rm for}~~s' \geq \Lambda^2~~,
\end{align}
where $\mathcal{T}^{I, \ell}_{\textbf{1234}} \equiv (1/2) \Lambda^2 (M_I^2/\Lambda^2)^{N_s+1} T^{I, \ell}_{\textbf{1234}}$.
This means that the UV amplitude contains heavy particles with mass $M_{I}$, spin $\ell$, and coupling strength $\mathcal{T}^{I,\ell}_{\textbf{1234}}$ respectively:
\begin{align}
        A_{\textbf{1234}}^{\rm (UV)} (s,t) & =  \text{ poles } + 
        \sum_{I,\ell} \frac{\mathcal{T}^{I,\ell}_{\textbf{1234}} d^\ell_{h_{12},h_{34}}(\cos\theta)}{M_I^2-s-i\epsilon}
        + {\rm (crossing~terms)},
\end{align}
and the dispersive relation gives the IR behaviour
\begin{align}
        &A_{\textbf{1234}}^{\rm (IR)} (s,t) = \dfrac{\lambda_{\textbf{1234}}}{-s} + \dfrac{\lambda_{\textbf{1324}}}{-t} + \dfrac{\lambda_{\textbf{1432}}}{-u} + A^{(0)}_{\textbf{1234}} (t) +  A^{(1)}_{\textbf{1234}} (t) s \\ 
         & + \sum_{I,\ell} \left( 
         \dfrac{\mathcal{T}^{I,\ell}_{\textbf{1234}}}{M_I^2 - s} \dfrac{(s+t/2)^2}{(M_I^2+t/2)^2} d^\ell_{h_{12} h_{43}} \left(1 + \dfrac{2t}{M_I^2} \right) + \dfrac{\mathcal{T}^{I,\ell}_{\textbf{1432}} }{M_I^2 - u} \dfrac{(u+t/2)^2}{(M_I^2+t/2)^2} d^\ell_{h_{14} h_{23}} \left( 1 + \dfrac{2t}{M_I^2} \right) 
         \right), \nonumber
\end{align}
In general, there are crossing symmetric terms, unless it is a dual amplitude such as those in string theory.

Let us now outline the exact procedure for finding the moment measure using the Curto-Fialkow method \cite{curto1998flat} (see \cite{Kimsey2022OnAS} for the generalization to the matrix moment problem), assuming that we have found the flat extensions of the Hankel matrices.  
It is essentially generalization of the steps illustrated in the toy example in Section \ref{sec:SpecToyExam}.

(1) Assuming Hankel matrix $\mathcal{H}_{1} (G)$ is flat at truncation order $G+2$, we define the $(\lfloor G/2 \rfloor)(\lfloor G/2 \rfloor + 1) / 2 \times (\lfloor G/2 \rfloor + 1)$ $W$-matrix as follows
\begin{align}
\label{eq:HWequation0}
    \sum_{\beta_1 \beta_2\, \mathbf{j}} \Big(\mathcal{H}_1 (G) \Big)^{\alpha_1 \alpha_2,\beta_1 \beta_2}_{\mathbf{ij}} W^{\beta_1 \beta_2, \gamma_1 \gamma_2}_{\mathbf{jm}} = B^{\alpha_1 \alpha_2,\gamma_1 \gamma_2}_{\mathbf{im}}
\end{align}
where the indices take values within the range $\beta_1 + \beta_2 \leq \lfloor G/2 \rfloor$ and $\gamma_1 + \gamma_2 = \lfloor G/2 \rfloor + 1$. The matrix $B$, which is of the same dimension as $W$, is defined by 
\begin{align}
    \mathcal{H}_{1} (G+2) =
    \begin{pmatrix}
        \mathcal{H}_{1} (G) & B \\
        B^T & C \\
    \end{pmatrix}.
\end{align}
As long as there exists a flat extension, one can prove that $W$ must have a solution, which can be expressed in terms of the moments $\mathfrak{m}^{\gamma_1\gamma_2}_{\mathbf{ij}}$. 

(2) Having found the solution for $W$, we then compute $\Omega = (\omega_1, \omega_2)$ by solving
\begin{align}
    \operatorname*{Det}_{\mathbf{ij}} \left( \sum_{\alpha_1 \alpha_2} (\omega_1)^{\alpha_1} (\omega_2)^{\alpha_2} W^{\alpha_1 \alpha_2, \beta_1 \beta_2}_{\mathbf{ij}} \right) = (\omega_1)^{\beta_1} (\omega_2)^{\beta_2}.
\end{align}
Suppose that there are $n_{\rm M}$ solutions labeled by $\Omega_{n},~n=1,2,...,n_{\rm M}$. One can show that $\Omega_n$ must live within the semialgebraic set $\mathcal{K}$ of our moment problem \cite{curto2000truncated}. 
$\Omega_n$ specifies the energy scales and spins of the UV particles.
 
(3) Then, we can compute the coupling strength $T^{n}_{\mathbf{ij}}$ by solving
\begin{align}
    V (\Omega_1, \Omega_2, \cdots, \Omega_{n_{\rm{M}}} ; \Xi ) 
    \begin{pmatrix}
        T^1_{\mathbf{jk}} \\
        T^2_{\mathbf{jk}} \\
        \vdots \\
        T^{n_{\rm{M}}}_{\mathbf{jk}}
    \end{pmatrix} = 
    \begin{pmatrix}
        \mathfrak{m}^{\gamma_{11} \gamma_{21}}_{\mathbf{ik}} \\
        \mathfrak{m}^{\gamma_{12} \gamma_{22}}_{\mathbf{ik}} \\
        \vdots \\
        \mathfrak{m}^{\gamma_{1 n_{\rm{M}}} \gamma_{2 n_{\rm{M}}}}_{\mathbf{ik}} \\
    \end{pmatrix}
\end{align}
where the generalized Vandermonde matrix is defined as 
\begin{align}
    V (\Omega_1, \Omega_2, \cdots, \Omega_{n_{{\rm{M}}}} ; \Xi ) 
    =
    \begin{pmatrix}
        \omega_{11}^{\gamma_{11}} \omega_{21}^{\gamma_{21}} I_{\mathbf{ij}} &
        \omega_{12}^{\gamma_{11}} \omega_{22}^{\gamma_{21}} I_{\mathbf{ij}} & 
        \cdots & \omega_{1n}^{\gamma_{11}}  \omega_{2n}^{\gamma_{21}} I_{\mathbf{ij}} \\ \omega_{11}^{\gamma_{12}} \omega_{21}^{\gamma_{22}} I_{\mathbf{ij}} &
        \omega_{12}^{\gamma_{12}} \omega_{22}^{\gamma_{22}} I_{\mathbf{ij}} &
        \cdots & \omega_{1n_{\rm{M}}}^{\gamma_{12}} \omega_{2n_{\rm{M}}}^{\gamma_{22}} I_{\mathbf{ij}} \\
        \vdots & \vdots & \ddots & \vdots \\
        \omega_{11}^{\gamma_{1n_{\rm{M}}}} \omega_{21}^{\gamma_{2n_{\rm{M}}}} I_{\mathbf{ij}} &
        \omega_{12}^{\gamma_{1n_{\rm{M}}}} \omega_{22}^{\gamma_{2n_{\rm{M}}}} I_{\mathbf{ij}} & 
        \cdots & \omega_{1 n_{\rm{M}}}^{\gamma_{1 n_{\rm{M}}}}  \omega_{2 n_{\rm{M}}}^{\gamma_{2 n_{\rm{M}}}} I_{\mathbf{ij}}
    \end{pmatrix}
\end{align}
where $I_{\textbf{ij}}$ is the identity matrix. 
There is some freedom in choosing the set 
\begin{align}
\label{Xilist}
    \Xi= \set{(\gamma_{11}, \gamma_{21}),(\gamma_{12}, \gamma_{22}) \cdots, (\gamma_{1n}, \gamma_{2n})}.
\end{align}
However, as long as it is chosen in a way to make sure that $V (\Omega_1, \Omega_2, \cdots, \Omega_n ; \Lambda )$ is invertible, it amounts to the same $T^{n}_{\mathbf{ij}}$. 
In fact, such a set always exists and $T^{n}_{\mathbf{ij}}$ are guaranteed to be positive. The positivity of $T^{n}_{\mathbf{ij}}$ ensures that the UV particles are not plagued by ghost instabilities.

\subsubsection{From EFT coefficients to moments}
\label{sec:EFT_gives_Moments}

In the preceding discussion, we glossed over the subtlety of how to uniquely determine the bi-variate moment sequence from the EFT coefficients. Let us now specify the explicit procedures to do so.

\begin{table}[h]
    \centering
    \begin{tabular}{|c||c|c|c|c|}
        \hline
        $\Lambda^{2i+2j} g^{i,j}$ & $j=0$ & $j=1$ & $j=2$ & $j=\cdots$\\
        \hline
        \hline
        $i+j=2$ & $\mathfrak{b}^{0,0}$  & $\times$ & $\times$ & $\times$ \\
        \hline
        $i+j=3$ & $0$ & $\mathfrak{b}^{0,1} - \frac{3}{2} \mathfrak{b}^{1,0}$ & $\times$ & $\times$ \\
        \hline
        $i+j=4$ & $\mathfrak{b}^{2,0}$    & $0$                                                    & $\frac{1}{4} \mathfrak{b}^{0,2} - 2 \mathfrak{b}^{1,1} + \frac{3}{2} \mathfrak{b}^{2,0}$ & $\times$ \\
        \hline
        $i+j=5$ & $0$ & $\mathfrak{b}^{2,1} - \frac{5}{2} \mathfrak{b}^{3,0}$ & $0$ & $\cdots$ \\
        \hline
        $i+j=6$ & $\mathfrak{b}^{4,0}$ & $0$ & $\frac{1}{4} \mathfrak{b}^{2,2} -  3 \mathfrak{b}^{3,1} + \frac{15}{4} \mathfrak{b}^{4,0}$ & $\cdots$ \\
        \hline
        $i+j=7$ & $0$ & $\mathfrak{b}^{4,1} - \frac{7}{2} \mathfrak{b}^{5,0}$ & $0$ & $\cdots$ \\
        \hline
        $i+j=8$ & $\mathfrak{b}^{6,0}$ & $0$ & $\frac{1}{4} \mathfrak{b}^{4,2} -  4 \mathfrak{b}^{5,1} + 7 \mathfrak{b}^{6,0}$ & $\cdots$ \\
        \hline
        $i+j=9$ & $0$ & $\mathfrak{b}^{6,1} - \frac{11}{2} \mathfrak{b}^{7,0}$ & $0$ & $\cdots$ \\
        \hline
        $i+j=10$ & $\mathfrak{b}^{8,0}$ & $0$ & $\frac{1}{4} \mathfrak{b}^{6,2} -  5 \mathfrak{b}^{7,1} + \frac{45}{4} \mathfrak{b}^{8,0}$ & $\cdots$ \\
        \hline
        $i+j=11$ & $0$ & $\mathfrak{b}^{8,1} - \frac{11}{2} \mathfrak{b}^{9,0}$ & $0$ & $\cdots$ \\
        \hline
        $i+j=12$ & $\mathfrak{b}^{10,0}$ & $0$ & $\frac{1}{4} \mathfrak{b}^{8,2} -  6 \mathfrak{b}^{9,1} + \frac{33}{2} \mathfrak{b}^{10,0}$ & $\cdots$ \\
        \hline
    \end{tabular}
    \caption{Relations between $g^{i,j}$ and $\mathfrak{b}^{\gamma_1,\gamma_2}$ for the single scalar case in the pure bi-variate formulation. For example, $g^{3,0}=0$ and $\Lambda^6 g^{2,1}=\mathfrak{b}^{0,1}-\frac32\mathfrak{b}^{1,0}$.}
    \label{tab:EFTCoeffs_And_Moments}
\end{table}

At first glance, since a sum rule contains both the $s$ and $u$ channel part, the map from the EFT coefficients to the moment sequence does not seem to be invertible. 
For concreteness, let us take the single scalar case for example and work with a pure bi-variate formulation (setting $\ell_M=0$ in \eref{gij1234Exp} and without $\mathfrak{a}^{\gamma_1}$). For this case, the sum rules \eqref{gij1234Exp} become\,\footnote{
Remember that the sum rules for the single scalar case can be written explicitly as \begin{align}
    g^{i,j} \equiv \dfrac{1 + (-1)^i}{2 (\Lambda^2)^{i+j}} \int \sum_{k=0}^{j} x_1^{i+j-k-2} \dfrac{\prod_{l=1}^{k} ( x_2 - l (l-1) x_1)}{(k!)^2} \dfrac{(-2)^{j-k} (j-k+i)!}{(j-k)!i!} \rho (x_1, x_2) \dif x_1 \dif x_2.
\end{align}
}
\begin{align}
    g^{i,j} =   \sum_{\gamma_2^\prime} V^{(\gamma)}_{\gamma_2,\gamma_2^\prime} \mathfrak{b}^{\gamma_1,\gamma_2^\prime} + (\text{$u$-channel}) ,
\end{align}
where now $\mathfrak{b}^{\gamma_1 \gamma_2}$ are defined by
 \begin{align}
 \label{bg1g2delta56}
     \mathfrak{b}^{\gamma_1 \gamma_2} \equiv 
     \int x_1^{\gamma_1} x_2^{\gamma_2} \sum_{\ell = 0}^{\infty} \sigma^{\ell} (x_1) \delta \Big(x_2 - \ell (\ell + 1) x_1 \Big) \dif x_1 \dif x_2 .
\end{align}
The first few explicit relations between EFT coefficients $g^{i,j}$ and the moments $\mathfrak{b}^{\gamma_1, \gamma_2}$ are shown in Table \ref{tab:EFTCoeffs_And_Moments}. 
The general expression for the first three columns are given by 
\begin{align}
    &\Lambda^{4i} g^{2i,0} = \mathfrak{b}^{2i-2,0},~~~~~~ \Lambda^{4i+2} g^{2i,1} = \mathfrak{b}^{2i-2,1} - \frac{2i+1}{2} \mathfrak{b}^{2i-1,0},
    \nonumber   \\ 
    \label{gtobgeneral3}
    &\Lambda^{4i+4} g^{2i,2} = \frac{1}{4} \mathfrak{b}^{2i-2,2} - (i+1) \mathfrak{b}^{2i-1,1} + \frac{(2i+1)(i+1)}{4} \mathfrak{b}^{2i,0}.
\end{align}
We see that half of the entries are zero: $g^{2i+1,j}=0$, making it impossible to express $\mathfrak{b}^{\gamma_1 \gamma_2}$ in term of $g^{i,j}$ from this table alone. 
For instance, as we can see in the table, for $i + j \leq 3$, we have $\Lambda^4 g^{2,0} = \mathfrak{b}^{0,0}$, $g^{3,0} = 0$ and $\Lambda^6 g^{2,1} = \mathfrak{b}^{0,1} - \frac{3}{2} \mathfrak{b}^{1,0}$. Therefore, without further input, it is impossible to express $\mathfrak{b}^{0,0},~ \mathfrak{b}^{1,0},~ \mathfrak{b}^{0,1}$ as linear combinations of $g^{2,0},~g^{3,0},~g^{2,1}$. 
More generally, we can not express $\mathfrak{b}^{2\gamma+1,0}$ in terms of $g^{i,j}$ only, as $g^{2i,1}$ is equal to a linear combination of $\mathfrak{b}^{2\gamma+1,0}$ and $\mathfrak{b}^{2\gamma,1}$, and similarly for some other $\mathfrak{b}^{\gamma_1,\gamma_2}$.

However, by further taking into account the spectral information or the positivity of the Hankel matrices, we can determine $\mathfrak{b}^{\gamma_1,\gamma_2}$ from $g^{i,j}$, as we will demonstrate below.

\bigskip
\noindent{\bf Method 1:}

Let us first specify the procedure to determine all the moments with further spectral information. 
This relies on the fact that the first moment variable $x_1 = \Lambda^2/ s^\prime$ is defined on $[0,1]$. 
First, note that 
by defining a new moment variable $x_1^\prime = (x_1)^2$, we can express the moments $\mathfrak{b}^{2\gamma,0}$ as:
\begin{align}
    \mathfrak{b}^{2\gamma, 0} = \int_0^1 (x_1^\prime)^{\gamma} \rho^{\prime}_0 (x_1^\prime) \dif x_1^\prime ~\text{ with }~ \rho^{\prime}_0 (x_1^\prime) \equiv \sum_{\ell = 0}^{\infty} \sigma_{}^{\ell} \left( \sqrt{\smash[b]{\mathstrut x_1^{\prime}}} \right)  / \left(2\sqrt{\smash[b]{\mathstrut x_1^{\prime}}}\right) 
\end{align}
where $\sigma^\ell$ is defined in \eref{sigmaspe}.
This formulation leads to a Hausdorf moment problem, which, as discussed in Section \ref{sec:MassSpec}, allows us to write the modified spectral function as a sum of delta functions of $x_1^\prime$:
\begin{align}
     \rho^{\prime}_0 (x_1^\prime) = \sum_{I} T_0^I \; \delta (x_1^\prime - \Lambda^4 /M_I^4) 
\end{align}
with some calculable $(T_0^I,M_I)$. 
That is, we can solve this Hausdorf moment problem explicitly to get the spectrum $(T_0^I,M_I)$, which will be used to determine $\mathfrak{b}^{2\gamma+1, 0}$. 
Since $x_1$ is positive, we can change the delta functions of $x_1^\prime$ into delta functions of $x_1$, and get the following relation
\begin{align}
    \sum_{\ell = 0}^{\infty} \sigma_{}^{\ell} \left( x_1 \right) = \sum_{I} T_0^I \; \delta (x_1 - \Lambda^2 /M_I^2)
\end{align}
As mentioned in Section \ref{sec:MassSpec}, the UV spectrum obtained in this way is unique. 
Thus, the key observation is that we can determine the moment sequence $\set{\mathfrak{b}^{2\gamma+1,0}}$ using this UV spectrum:
\begin{align}
\label{b2gp1EQ}
     \mathfrak{b}^{2\gamma+1, 0} = \int_0^1 x_1^{2\gamma+1} \left( \sum_{\ell = 0}^{\infty} \sigma_{}^{\ell} \left( x_1 \right) \right) \dif x_1 = \sum_I T_0^I (\Lambda^2/M_I^2)^{2\gamma+1}
\end{align}
With $\mathfrak{b}^{2\gamma+1, 0}$ determined, one can now infer the value of $\mathfrak{b}^{2\gamma,1}$ from the relation $\mathfrak{b}^{2n-2,1} = \Lambda^{4n+2} g^{2n,1} + \frac{2n+1}{2} \mathfrak{b}^{2n-1,0}$ (see Table \ref{tab:EFTCoeffs_And_Moments} or \eref{gtobgeneral3}). Now, the $\mathfrak{b}^{2\gamma, 1}$ moment itself has the following dispersive relation:
\begin{align}
    \mathfrak{b}^{2\gamma, 1} = \int_0^1 (x_1^\prime)^{\gamma} \rho_1^{\prime} (x_1^\prime) \dif x_1^\prime ~ \text{ with }~ \rho_1^{\prime} (x_1^\prime) \equiv \sum_{\ell = 0}^{\infty} \ell (\ell+1) \sigma_{}^{\ell} \left( \sqrt{\smash[b]{\mathstrut x_1^{\prime}}} \right)  / (2\sqrt{\smash[b]{\mathstrut x_1^{\prime}}})
\end{align}
(Note that although $\rho_1^{\prime} (x_1^\prime)$ is similar to $\rho_0^{\prime} (x_1^\prime)$, in general, one can not infer $\rho_1^{\prime} (x_1^\prime)$ from $\rho_0^{\prime} (x_1^\prime)$.) In this formulation, $\mathfrak{b}^{2\gamma, 1}$ again is a Hausdorff problem, so we can write $ \rho_1^{\prime} (x_1^\prime)$ as 
\begin{equation}
    \rho^{\prime}_1 (x_1^\prime) = \sum_{I} T_1^I \; \delta (x_1^\prime - \Lambda^4/M_I^4)
\end{equation}
Using the spectrum $(T_1^I,M_I)$ and repeating the previous procedure, we can determine the moments $\mathfrak{b}^{2\gamma+1,1}$, thus the entire $\mathfrak{b}^{\gamma,1}$ sequence.
Similarly, the other moments $\mathfrak{b}^{\gamma_1,\gamma_2}$ with $\gamma_2\geq 2$ can also be determined analogously.

In practice, we usually only know a few EFT coefficients, so the Hausdorff spectra mentioned above are truncated ones. 
In this case, as mentioned, it is suffice to choose a spectrum with the least number of delta functions. 
With more EFT coefficients included, the obtained moments will increasingly approach the true moments, as we will see in Section \ref{sec:Vis}.

\bigskip
\noindent{\bf Method 2:}

An alternative approach to effectively determine $\mathfrak{b}^{\gamma_1,\gamma_2}$ from $g^{i,j}$, which is often easier to implement numerically, is to constrain the extra undetermined moments using the PSD of the Hankel matrices. 
Specifically, by numerically running SDPs, we can constrain a given $\mathfrak{b}^{\gamma_1,\gamma_2}$ from both above and below, and as we include more EFT coefficients, the gap between the upper and lower bounds on this $\mathfrak{b}^{\gamma_1,\gamma_2}$ shrinks. 
As we will see in Section \ref{sec:Vis}, this method can often determine the moments to a good precision, closely approaching the first method.
Concretely, if we want to determine a given $\mathfrak{b}^{\gamma_1,\gamma_2}$ from a set of EFT coefficients, we can compute its maximum and minimum, subject to the relevant relations in Table \ref{tab:EFTCoeffs_And_Moments} along with the PSD conditions of the Hankel matrices truncated to include all the $\mathfrak{b}^{\gamma_1,\gamma_2}$ involved in the relevant $g^{ij}$ and $\mathfrak{b}^{\gamma_1,\gamma_2}$ relations. 
That is, we can compute the minimum of $\mathfrak{b}^{\gamma_1,\gamma_2}$ by the following SDP
\begin{align}
\label{numSDPmethod2}
    \text {\bf minimize } & \mathfrak{b}^{\gamma_1,\gamma_2} \\
    \text {\bf such that } & g^{2,0} = \mathfrak{b}^{0,0}, g^{2,1} = \mathfrak{b}^{0,1} - \frac{3}{2} \mathfrak{b}^{1,0}, \cdots \nn
    & \begin{pmatrix}
        \mathfrak{b}^{1,0} & \mathfrak{b}^{2,0} & \mathfrak{b}^{1,1} & \cdots\\
        \mathfrak{b}^{2,0} & \mathfrak{b}^{3,0} & \mathfrak{b}^{2,1} & \cdots\\
        \mathfrak{b}^{1,1} & \mathfrak{b}^{2,1} & \mathfrak{b}^{1,2} & \cdots \\
        \vdots & \vdots & \vdots & \ddots
    \end{pmatrix} \succeq 0, ~~~
     \begin{pmatrix}
        \mathfrak{b}^{0,0} - \mathfrak{b}^{1,0} & \mathfrak{b}^{1,0} - \mathfrak{b}^{2,0} & \mathfrak{b}^{0,1} - \mathfrak{b}^{1,1} & \cdots\\
        \mathfrak{b}^{1,0} - \mathfrak{b}^{2,0} & \mathfrak{b}^{2,0} - \mathfrak{b}^{3,0}  & \mathfrak{b}^{1,1} - \mathfrak{b}^{2,1}  & \cdots\\
        \mathfrak{b}^{0,1} - \mathfrak{b}^{1,1}  & \mathfrak{b}^{1,1} -  \mathfrak{b}^{2,1}  & \mathfrak{b}^{0,2} -  \mathfrak{b}^{1,2} & \cdots \\
        \vdots & \vdots & \vdots & \ddots
    \end{pmatrix} \succeq 0,\cdots\nonumber
\end{align}
and similarly for the maximum. The maximum and minimum quickly approach each other as we include more EFT coefficients. 

\subsection{Determinacy/Uniqueness}
\label{sec:Determinacy}

We have specified how to construct a UV spectrum (or representing measure) for a given set of EFT coefficients (or a given moment sequence). One may wonder whether the obtained spectrum (or measure) is unique. This is often called the determinacy problem of the moments. Let us see how the existing results in this problem can be used for the case of a single scalar field.

To this end, we shall separate the bi-variate moment sequence into a set of uni-variate moment sequences for different UV masses, and we will be concerned with the determinacy of the spin moment sequences. This is because, as we saw in Section \ref{sec:MassSpec} in the forward limit, the mass spectrum is always unique (in the full moment problem).

Let us define these uni-variant spin moments and express them in terms of the bi-variant moments. Suppose the UV spectrum is given by the masses $M_I$ and couplings $T^{I,\ell}$, which leads to the following representing measure
\begin{align}
\label{cmomentdef}
        \rho (x_1, x_2) = \sum_{I,\ell} T^{I,\ell} \, \delta \left( x_1 - \frac{\Lambda^2}{M_I^2} \right) \delta \left(x_2 - \frac{\Lambda^2}{M_I^2} \ell (\ell + 1) \right).
\end{align}
For each UV mass, we can define the following uni-variate moment sequence
\begin{align}
    \mathfrak{c}_I^{\gamma_2} \equiv \sum_{\ell=0}^{\infty} T^{I,\ell} \Big( \ell (\ell + 1) \Big)^{\gamma_2} \equiv \int_{0}^{\infty} ( J^2)^{\gamma_2} \hat{\rho}_I ( J^2) \dif  J^2,
\end{align}
where $\hat{\rho}_I ( J^2) \equiv \sum_{\ell=0}^{\infty} T^{I,\ell} \delta (  J^2 - \ell (\ell + 1) )$. 
Then, the bi-variant moments are related to $ \mathfrak{c}_I^{ \gamma_2}$ by 
\begin{equation}
\label{momentsbtoc}
 \mathfrak{b}^{\gamma_1, \gamma_2} =  \sum_{I} (\Lambda^2/M_I^2 )^{\gamma_1+\gamma_2} \mathfrak{c}_I^{\gamma_2}
\end{equation}
Since the Vandermonde matrix is invertible, we can write $\mathfrak{c}^{\gamma_2}_I$ in terms of $\mathfrak{b}^{\gamma_1,\gamma_2}$
\begin{align}
    \begin{pmatrix}
        \mathfrak{c}_1^{\gamma_2} \\
        \mathfrak{c}_2^{\gamma_2} \\
        \mathfrak{c}_3^{\gamma_2} \\
        \vdots \\
    \end{pmatrix} = 
    \operatorname*{diag} \left( \frac{\Lambda^2}{M_1^2}, \frac{\Lambda^2}{M_2^2}, \frac{\Lambda^2}{M_3^2}, \cdots \right)^{-\gamma_2}
    \begin{pmatrix}
        1 & 1 & 1 & \cdots \\[5pt]
        \dfrac{\Lambda^2}{M_1^2} & \dfrac{\Lambda^2}{M_2^2} & \dfrac{\Lambda^2}{M_3^2} & \cdots \\[10pt]
        \dfrac{\Lambda^4}{M_1^4} & \dfrac{\Lambda^4}{M_2^4} & \dfrac{\Lambda^4}{M_3^4} & \cdots \\[10pt]
       \vdots & \vdots & \vdots & \ddots \\[10pt]
    \end{pmatrix}^{-1}
        \begin{pmatrix}
        \mathfrak{b}^{0, \gamma_2} \\
        \mathfrak{b}^{1, \gamma_2} \\
        \mathfrak{b}^{2, \gamma_2} \\
        \vdots \\
    \end{pmatrix}
\end{align}
We shall then try to determine whether the moment sequences $\mathfrak{c}^{\gamma_2}_I$ have unique UV measures in the physical problems. 

In general, the determinacy of a moment problem with a more restricted range for the moment variable is guaranteed by determinacy under a more relaxed range.
For example, a (uni-variate) Stieltjes moment problem with range $[0,\infty)$ can be thought as a special Hamburger moment problem, which in general has range $(-\infty,\infty)$. Thus, the condition for a solvable Stieltjes moment sequence to be determinate is weaker than the condition for the corresponding Hamburger moment sequence to be determinate. In other words, a Stieltjes determinate moment sequence can be Hamburger indeterminate, while a Hamburger determinate moment sequence must be Stieltjes determinate.
For our current problem (for a scalar), the $\mathfrak{c}_I^{\gamma_2}$ moment sequence is defined on the set 
\begin{equation}
\label{setellscalar}
\set{ \ell (\ell + 1) \mid \ell = 0, 2, 4, \cdots} .
\end{equation}
So the determinacy conditions for this problem are weaker than those for the corresponding Stieltjes or Hamburger moment problem. We can often use the corresponding Stieltjes or Hamburger moment problem to check the determinacy of $\mathfrak{c}_I^{\gamma_2}$. 

For a Hamburger moment sequence, Carleman's condition suggests that if $\mathfrak{c}^{\gamma+1} / \mathfrak{c}^{\gamma} \lesssim O ( \gamma ) \text{ when } \gamma \rightarrow \infty$, the moment sequence is determinate on $\mathbb{R}$. On the other hand, if $\mathfrak{c}^{\gamma+1} / \mathfrak{c}^{\gamma} \succsim O ( \gamma ) \text{ when } \gamma \rightarrow \infty$, the moment sequence is not necessarily indeterminate on $\mathbb{R}$. 

However, the case of the Stieltjes moment problem is different.
A concise summary for the determinacy and indeterminacy criteria for the Stieltjes moment problem can be found in Ref.~\cite{Lin_2017}. 
In particular, we have the following results:
\begin{itemize}
    \item A Stieltjes moment problem is determinate if its sequence satisfies $\mathfrak{c}^{\gamma+1} / \mathfrak{c}^{\gamma} \sim O ( \gamma^2 )$ when $\gamma \rightarrow \infty$.  
    \item A solvable Stieltjes moment sequence $\mathfrak{c}^{\gamma}$ is indeterminate if it has a continuous representing measure $\hat{\rho}(x) \dif x$ satisfying $-{\dif \ln \hat{\rho} (x)}/\dif (\ln x) \rightarrow + \infty \text{ when } x \rightarrow + \infty$ and grows faster than $c \gamma^{(2 + \epsilon) \gamma}$ for some positive $c$ and $\epsilon$.
\end{itemize}
Therefore, the boundary between an indeterminate and a determinate Stieltjes moment problem is
 \begin{align}
 \label{cdeterminacyboundary}
    \mathfrak{c}^{\gamma+1} / \mathfrak{c}^{\gamma} \sim c \gamma^2 ~\text{ when }~ \gamma \rightarrow \infty,~~~~ c>0.
\end{align}
For our current problem with the set (\ref{setellscalar}), which is a subset of $[0,\infty)$, this criterion acts as a sufficient condition for each $\mathfrak{c}_I^{\gamma}$.

A tantalizing aspect of this determinacy boundary is that it seems to be aligned with the Froissart bound for the amplitude. To see this, note that a spectral density of the form $\hat{\rho}_I( J^2) \sim  \exp(- c_I ( J^2)^{1/2})$ with $c_I>0$ lies at the boundary of condition \eqref{cdeterminacyboundary}, which can be checked by substituting it into \eref{cmomentdef}. On the other hand, the Froissart bound implies that at large $\ell$ and small $x_1$ we have \cite{Gribov_2003}
\begin{align}
   \hat{\rho}_I \sim \operatorname{Im} a_\ell (x_1) \sim \exp{(- 2 m \, \ell M_I / \Lambda^2)} ,
\end{align}
where $m$ is the mass of the scattering particle. This matches the above determinacy boundary for the fastest varying factor. However, it is possible that discrepancies arise for the small $\ell$ or $x_1 \sim 1$ spectral measures.

\subsection{Some examples}
\label{sec:SpecExamples}

In this subsection, we will illustrate the connections between the UV spectra and the EFT coefficients, as well as the reverse engineering from the IR through the moment formulation, using a few specific examples

\subsubsection{Multiple UV scalars}

Let us begin with a simple case where the UV theory only contains a number of scalars with mass $M_I$.
That is, we consider a renormalizable theory with a massless scalar field $\phi$ and heavy scalars $\Phi_I$ with mass $M_I$, whose interacting part of the Lagrangian is schematically given by
\begin{align}
     \mathcal{L}_{\text{int}} &= \dfrac{1}{24} \lambda \phi^4 + \dfrac{1}{6} \eta \Lambda \phi^3 + \dfrac{1}{2}\eta_I \Lambda \phi^2 \Phi_I + \left( \phi \Phi \Phi \right) + \left( \Phi \Phi \Phi \right)  \\
    &~~~~~+ \left( \phi \phi \phi \Phi \right) + \left( \phi \phi \Phi \Phi \right) + \left( \phi \Phi \Phi \Phi \right) + \left( \Phi \Phi \Phi \Phi \right)
\end{align}
where $\Lambda$ is the mass of the lightest $\Phi_I$. 
From the UV point of view, the tree-level amplitude for the $\phi\phi\to\phi\phi$ scattering is given by
\begin{align}
    A (s, t) &= \lambda + \eta^2 \dfrac{\Lambda^2}{-s} + \sum_I \eta_I^2 \dfrac{\Lambda^2}{M_I^2 - s} + ( s \leftrightarrow t) + ( s \leftrightarrow u).
\end{align}
From the IR point of view, it may be instructive to rewrite this amplitude as follows
\begin{align}
    &\phantom{+} A(s, t) = \eta^2 \left( \dfrac{\Lambda^2}{-s} + \dfrac{\Lambda^2}{-t} + \dfrac{\Lambda^2}{-u} \right) + A_{(0)} (t) +  s A_{(1)} (t) \nonumber \\
    &~~~~+ \sum_I \eta_I^2 \left[ \dfrac{\Lambda^2}{M_I^2 - s} \dfrac{(s+\frac{t}{2})^2}{(M_I^2+\frac{t}{2})^2} P_{\ell = 0} \left(1 + \frac{2t}{M_I^2}\right) + \dfrac{\Lambda^2}{M_I^2 - u} \dfrac{(u+\frac{t}{2})^2}{(M_I^2+\frac{t}{2})^2} P_{\ell = 0} \left( 1 + \frac{2t}{M_I^2} \right) \right] \nonumber
\end{align}
This is essentially the dispersion relation where the integration over $s'$ becomes a sum over spinless particles with masses $M^2_I$.

To see how the UV reconstruction from the EFT coefficients works, let us match the UV amplitude with the EFT coefficients $g^{i,j}$, which are defined by
\begin{align}
    A (s, t) &= \cdots + \sum_{i \geq 2,j \geq 0} g^{i,j} \left(s + \frac{t}{2}\right)^i t^j,
\end{align}
where for $i \geq 2$ we have 
\begin{align}
    g^{i,0} &= \dfrac{1 + (-1)^i}{(\Lambda^2)^{i}} \sum_I \eta_I^2 \left( \dfrac{\Lambda^2}{M_I^2} \right)^{i+1}, ~~~~~~
    g^{i,1} = - \dfrac{i+1}{2} \dfrac{ 1 + (-1)^i}{(\Lambda^2)^{i+1}} \sum_I \eta_I^2 \left( \dfrac{\Lambda^2}{M_I^2} \right)^{i+2}, \nonumber\\
    g^{i,2} &= \dfrac{(i+1)(i+2)}{8} \dfrac{1 + (-1)^i}{(\Lambda^2)^{i+2}} \sum_I \eta_I^2 \left( \dfrac{\Lambda^2}{M_I^2} \right)^{i+3},~~~~~~~~~~~~~ \cdots 
\end{align}
Our goal is that we start from a purely low energy point of view, only knowing these EFT coefficients, and we wish to infer the UV theory from them. 

To proceed in a purely bi-variant formulation ($\ell_M = 0$), we need to first convert these EFT coefficients to the bi-variant moments $\mathfrak{b}^{\gamma_1,\gamma_2}$. From the dispersive sum rules, we know that they are related by
\begin{align}
    &\Lambda^{4i} g^{2i,0} = \mathfrak{b}^{2i-2,0},~~~~~~~~ \Lambda^{4i+2} g^{2i,1} = \mathfrak{b}^{2i-2,1} - \frac{2i+1}{2} \mathfrak{b}^{2i-1,0},
    \nonumber\\ 
    &\Lambda^{4i+4} g^{2i,2} = \frac{1}{4} \mathfrak{b}^{2i-2,2} - \dfrac{2i+2}{2} \mathfrak{b}^{2i-1,1} + \frac{(2i+1)(2i+2)}{8} \mathfrak{b}^{2i,0}, ~~~~~
    \cdots  .
\end{align}
 For the forward-limit moments where $\gamma_2=0$, we can easily solve the relevant equations above to get $\mathfrak{b}^{\gamma_1, 0}$ in terms of $g^{i,0}$, which results in
\begin{align}
    \mathfrak{b}^{\gamma_1, 0} &=  \sum_I 2\eta_I^2 \left( \dfrac{\Lambda^2}{M_I^2} \right)^{3} \left( \dfrac{\Lambda^2}{M_I^2} \right)^{\gamma_1},~~~ \gamma_1 = 0,2,4, \cdots.
\end{align}
From these moments, we can infer that the UV masses are $M_I$ and the UV couplings $\mathcal{T}^I = \eta_I^2 \Lambda^2$. 
(With the hindsight of the UV theory along with its analytical expressions, we can determine this spectrum precisely. If we did it numerically with some truncation, we would only approximate this spectrum numerically, following the procedure of Eqs.~\eqref{eq:HWequation0} to \eqref{Xilist}, as we shall see in later examples.)
Due to the uniqueness of the UV spectrum in the forward limit, we can infer that for $\gamma_1 = 1,3,5, \cdots$, we must have the same spectrum, and therefore we have
\begin{align}
    \mathfrak{b}^{\gamma_1, 0} &= \sum_I 2\eta_I^2 \left( \dfrac{\Lambda^2}{M_I^2} \right)^{3} \left( \dfrac{\Lambda^2}{M_I^2} \right)^{\gamma_1}, ~~~ \gamma_1 = 0,1,2, \cdots;
\end{align}
Then using the relations between $g^{i,j}$ and $\mathfrak{b}^{\gamma_1,\gamma_2}$, we can infer that
$\mathfrak{b}^{\gamma_1, 1} = 0$ for $\gamma_1 = 0,2,4, \cdots$.
By the determinacy of the Hausdorff moment problem again, we can infer $\mathfrak{b}^{\gamma_1, 1} = 0$ for $\gamma_1 = 0,1,2, \cdots$. Then using the relations between $g^{i,j}$ and $\mathfrak{b}^{\gamma_1,\gamma_2}$, we can infer $\mathfrak{b}^{\gamma_1, 2} =0$ for $\gamma_1 = 0,2,4,\cdots$. Repeating this procedure, we get
\begin{align}
    \mathfrak{b}^{\gamma_1, 1} = \mathfrak{b}^{\gamma_1, 2} = \cdots = 0, ~~ \gamma_1 = 0,1,2, \cdots .
\end{align}
and thus all the bi-variant moments are given by
\begin{align}
    \mathfrak{b}^{\gamma_1, \gamma_2} &=  \sum_{\ell=0} \sum_I 2\eta_I^2 \left( \dfrac{\Lambda^2}{M_I^2} \right)^{3} \left( \dfrac{\Lambda^2}{M_I^2} \right)^{\gamma_1} \left( \dfrac{\Lambda^2 (\ell^2 + \ell)}{M_I^2} \right)^{\gamma_2}.
    \\
    &= \int_{\mathcal{K}_\mathfrak{b}} x_1^{\gamma_1} x_2^{\gamma_2}  \sum_{\ell=0} \sum_{I} 2\eta_I^2 \left( \dfrac{\Lambda^2}{M_I^2} \right)^{3}\delta \Big(x_1 - \dfrac{\Lambda^2}{M_I^2} \Big) \delta \Big( x_2 - \dfrac{\Lambda^2}{M_I^2} \ell (\ell + 1) \Big)  \dif x_1 \dif x_2, 
\end{align}
where we have defined $0^0=1$ for convenience. Comparing with \eref{x1x2explicit77}, we see that this gives a spectrum of scalars with masses $M_I$ and couplings $\mathcal{T}^{I,\ell} = \eta_I^2 \Lambda^2$.

\subsubsection{UV vector}

Now, let us consider two massless/IR scalars $\phi_1$ and $\phi_2$, with the injection of one UV vector $A^\mu$ with mass $M$, and see how the IR coefficients can be used infer the heavy vector.
The interacting part of the UV Lagrangian for this setup is
\begin{align}
    \mathcal{L}_{\text{int}} 
    &= \dfrac{1}{4} \lambda \phi_1^2 \phi_2^2 + \dfrac{1}{6} \eta_{111} M \phi_1^3  + \dfrac{1}{6} \eta_{222} M \phi_2^3 + \dfrac{1}{2} \eta_{112} M \phi_1^2 \phi_2 + \dfrac{1}{2} \eta_{122} M \phi_1 \phi_2^2 \\
    &~~~ + \dfrac{1}{2} \lambda_{11} \phi_1^2 \partial_\mu A^\mu + \dfrac{1}{2} \lambda_{22} \phi_2^2 \partial_\mu A^\mu + \lambda_{12} \phi_1 \phi_2 \partial_\mu A^\mu + 2 \lambda_{12}^{\prime} \phi_1 \partial_\mu \phi_2 A^\mu \\
    &~~~ + \left( \phi \phi \phi \phi \right) + \left( A^2 \partial A \right) + \left( A A A A \right) \nonumber
\end{align}
For such a simple setup, it is sufficient to focus on the tree-level amplitude for the $\phi_1 \phi_2 \rightarrow \phi_1 \phi_2$ scattering, which is given by
\begin{align}
     A_{{1212}} (s, t) 
    &= \lambda + \left( \eta_{112} \eta_{222} +\eta_{111} \eta_{122} \right)  \dfrac{M^2}{-t} + \left( \eta_{112}^2 + \eta_{122}^2 \right) \left( \dfrac{M^2}{-s} + \dfrac{M^2}{-u} \right) \nonumber \\
    &~~~ - 2 \lambda_{12}^{\prime \, 2} + ( \lambda_{11} \lambda_{22} - \lambda_{12}^2 - \lambda_{12}^{\prime \, 2} ) \dfrac{-t}{M^2} + \lambda_{12}^{\prime \, 2} \left( \dfrac{M^2 + 2 t}{M^2 - s} + \dfrac{M^2 + 2 t}{M^2 - u} \right) 
\end{align}
Without any prior knowledge of the UV information, one may need to include more scattering processes.
This amplitude can be re-written in the dispersion relation form as
\begin{align}
     &A_{{1212}} (s, t)
    = \left( \eta_{112} \eta_{222} +\eta_{111} \eta_{122} \right)  \dfrac{M^2}{-t} + \left( \eta_{112}^2 + \eta_{122}^2 \right) \left( \dfrac{M^2}{-s} + \dfrac{M^2}{-u} \right) + A_{(0)} (t) + s A_{(1)} (t) \nonumber \\
    & ~~~~~~+ \dfrac{\lambda_{12}^{\prime \, 2} M^2}{M^2 - s} \dfrac{(s+\frac{t}2)^2}{(M_k+\frac{t}2)^2} P_{\ell = 1} \left(1 + \frac{2t}{M^2}\right) + \dfrac{\lambda_{12}^{\prime \, 2} M^2}{M^2 - u} \dfrac{(s+\frac{t}2)^2}{(M^2+\frac{t}2)^2} P_{\ell = 1} \left( 1 + \frac{2t}{M^2} \right)  .
\end{align}
We see that the UV spectrum contains only one vector with mass $M$, which is also the EFT cutoff $\Lambda=M$. Matching this amplitude to the EFT expansion $ A_{{1212}} (s, t) = \cdots + \sum_{i \geq 2,j \geq 0} g_{{1212}}^{i,j}  \left(s+\frac{t}2\right)^i t^j$, we get the EFT coefficients
\begin{align}
    g_{{1212}}^{i,j}  &= g^{i,j}_{\text{S}} + \dfrac{2}{M^2} g^{i,j-1}_{\text{S}},~~~~~~~ i \geq 2, j \geq 1,
\end{align}
where the ``scalar'' coefficients $g^{i,j}_{\text{S}}$ are defined via 
\begin{equation}
\frac{M^2}{M^2 - s} + \frac{M^2}{M^2 - u} = \cdots + \sum_{i \geq 2,j \geq 0} g^{i,j}_{\text{S}} \left(s+\frac{t}2\right)^i t^j .
\end{equation}
In the scalar theory, we can find that (with the convention $0^0 = 1$)
\begin{align}
    \mathfrak{b}^{\gamma_1, \gamma_2}_{\text{S}} = 2 \, \lambda^{\prime \, 2}_{12} \left( \dfrac{M^2}{M^2} \right)^{\gamma_1} \left( \dfrac{0 \times (0+1) M^2}{M^2} \right)^{\gamma_2}.
\end{align}

To reconstruct the UV spectrum from these low energy coefficients, we first extract the bi-variant moments $\mathfrak{b}^{\gamma_1,\gamma_2}$ from them. Again, this is done by utilizing the sum rules 
\begin{align}
    &M^{4i} g^{2i,0} = \mathfrak{b}^{2i-2,0},~~~~ M^{4i+2} g^{2i,1} = \mathfrak{b}^{2i-2,1} - \frac{2i+1}{2} \mathfrak{b}^{2i-1,0},
    \nonumber \\ 
    &M^{4i+4} g^{2i,2} = \frac{1}{4} \mathfrak{b}^{2i-2,2} - \dfrac{2i+2}{2} \mathfrak{b}^{2i-1,1} + \frac{(2i+1)(2i+2)}{8} \mathfrak{b}^{2i,0}, ~~~~~
    \cdots 
\end{align}
and the spectral representations of $\mathfrak{b}^{\gamma_1,\gamma_2}$ for fixed $\gamma_2$, as specified in Section \ref{sec:EFT_gives_Moments} or more explicitly in the previous subsection. By an explicit computation, we can easily find that for the first few $\gamma_2$ the moments are given by
\begin{align}
\label{bvectorresults}
    \mathfrak{b}^{\gamma_1, \gamma_2} & =  2\lambda^{\prime \, 2}_{12} \left( \dfrac{M^2}{M^2} \right)^{\gamma_1} \left( \dfrac{1 \times (1 + 1) M^2}{M^2} \right)^{\gamma_2}  
    \\
    &= \int_{\mathcal{K}_\mathfrak{b}} x_1^{\gamma_1} x_2^{\gamma_2} \sum_{\ell=1} 2\lambda^{\prime \, 2}_{12} \delta \Big(x_1 - \dfrac{M^2}{M^2} \Big) \delta \Big( x_2 - \dfrac{M^2}{M^2} \ell (\ell + 1) \Big) \dif x_1 \dif x_2 ,
\end{align}
from which we easily can read the mass and the spin of the UV particle, which are $M$ and $1$ respectively.

Incidentally, another simple way to infer the UV spectrum is to use the relation $g^{i,j}_{{1212}}  = g^{i,j}_{\text{S}} + \frac{2}{M^2} g^{i,j-1}_{\text{S}}, ~i \geq 2, j \geq 1$, along with the following generic scalar sum rule
\begin{align}
    g_{\textbf{1234}}^{i,j} &\equiv \dfrac{1 + (-1)^i}{2 (M^2)^{i+j}} \int \sum_{k=0}^{j} x_1^{i+j-k-2}  \dfrac{\prod_{l=1}^{k} ( x_2 - l (l-1) x_1)}{(k!)^2}
    \nn
    &~~~~~~~~~~~~~~~~~~~~~~~~~~\times  \dfrac{(-2)^{j-k} (j-k+i)!}{(j-k)!i!} \rho_{\textbf{1234}} (x_1, x_2) \dif x_1 \dif x_2.
\end{align}
To see this, note that for $g^{i,j}_{\text{S}}$, we have $\rho_{\text{S}} (x_1, x_2) = 2 \lambda^{\prime \, 2}_{12} \delta (x_1 - 1) \delta (x_2 - 0)$. Substituting the sum rule for $g^{i,j}_{\text{S}}$ into $g^{i,j}_{{1212}}  = g^{i,j}_{\text{S}} + \frac{2}{M^2} g^{i,j-1}_{\text{S}}$, we easily get
\begin{align}
    g^{i,j+1}_{{1212}}  
    &= 2 \lambda^{\prime \, 2}_{12} \dfrac{1 + (-1)^i}{2 (M^2)^{i+j+1}} \sum_{k=0}^{j} \dfrac{\prod_{l=1}^{k} ( 2 - l (l-1))}{(k!)^2} \dfrac{(-2)^{j-k} (j-k+i)!}{(j-k)!i!}  \\
    &= \dfrac{1 + (-1)^i}{2 (M^2)^{i+j}} \int \sum_{k=0}^{j} x_1^{i+j-k-2} \dfrac{\prod_{l=1}^{k} ( x_2 - l (l-1) x_1)}{(k!)^2} \dfrac{(-2)^{j-k} (j-k+i)!}{(j-k)!i!} \rho_{{1212}}  \dif x_1 \dif x_2. \nonumber
\end{align}
where we have defined $\rho_{{1212}} \equiv 2 \lambda^{\prime \, 2}_{12} \delta (x_1 - 1) \delta (x_2 - 2 x_1)$ to arrive at the last equality. From this spectral function, we can read that the UV spectrum only contains a UV vector particle with mass $M$.

\subsubsection{$stu$ kink theory}

The $stu$ kink theory \cite{Caron-Huot:2020cmc} lies at the intersection between the two smooth boundaries of the positivity bounds for the scalar EFT. Let us now reconstruct the spectrum of this theory from a truncated series of the (exactly determined) EFT coefficients. 

The $stu$ kink theory is defined by the following 2-to-2 scattering amplitude
\begin{align}
    \label{Astu1stline}
            A(s,t) &= \dfrac{\Lambda^6}{(\Lambda^2 - s)(\Lambda^2 - t)(\Lambda^2 - u)} - \gamma_{stu} \left( \dfrac{\Lambda^2}{\Lambda^2 - s} + \dfrac{\Lambda^2}{\Lambda^2 - t} + \dfrac{\Lambda^2}{\Lambda^2 - u} \right)
            \\
            & = 1 - 3 \gamma_{stu} + (1 - 2 \gamma_{stu}) \dfrac{s^2 + s t + t^2}{\Lambda^4} + (-1 + 3 \gamma_{stu}) \dfrac{s^2 t + s t^2}{\Lambda^6} +...
            \label{Astu2ndline}
    \end{align}
where $\gamma_{stu} = {4 \ln 2}/{9}$ in 4D. The $\gamma_{stu}$ term in \eref{Astu1stline} subtracts the UV massive spin-0 component. We are considering massless scalar scattering in the IR, so $s + t + u = 0$, and the $g^{i,j}$ coefficients can be extracted from \eref{Astu2ndline}. Despite being simple, this UV theory consists of an infinite tower of higher spin-$\ell$ particles with mass $M_\ell=\Lambda$, $\ell=2,4,6,...$, and coupling strength
\begin{align}
\label{Tstuspectrum}
    \mathcal{T}^\ell &= \dfrac{\pi}{2} (2 \ell + 1) \int_{-1}^{1} P_{\ell} (z) \left( \dfrac{4}{9 - z^2} - \dfrac{4\ln 2}{9}  \right) \dif z .
\end{align}
The odd $\ell$ are absent due to the $tu$ crossing symmetry. The coupling strength can be extracted by replacing $s$ with $s+i\epsilon$ in the amplitude and making use of the identity ${\rm Im}(1/(s-s_*-i\epsilon))= -\pi \delta(s-s_*)$.
Here, $\mathcal{T}^\ell$ is not normalized and $\sum_{\ell=0}^{\infty} \mathcal{T}^\ell = 1 - ({8 \operatorname{ln} 2})/{9}$.

Given the explicit form of the UV theory, it is easy to get the bi-variant moments $\mathfrak{b}^{\gamma_1,\gamma_2}$ for the theory via the dispersive sum rules:
\begin{align}
\label{mstuexp}
     \mathfrak{b}^{\gamma_1,\gamma_2} &= \dfrac{\pi}{2} \operatorname*{lim}_{\xi \rightarrow 1^-} \left[ \left( 1 + 2 \xi \dfrac{\dif}{\dif \xi} \right)^{2\gamma_2+1} F(\xi) \right] - \dfrac{4 \pi \operatorname{ln} 2}{9} \delta_{\gamma_2, 0}.
\end{align}
where
\begin{align}
    \begin{gathered}
        F(\xi) \equiv \sum_{\ell=0}^{\infty} \int_{-1}^{1} P_{\ell} (z) \xi^{\ell} \dfrac{4}{9 - z^2}~\dif z = \int_{-1}^{1} \dfrac{4}{\sqrt{1 - 2 z \xi + \xi^2}(9 - z^2)}  ~\dif z
    \end{gathered}
\end{align}
The last integration in the $F(\xi)$ definition is elementary and can be carried out analytically. In this case, it is also straightforward to extract the moments $\mathfrak{b}^{\gamma_1,\gamma_2}$ from the EFT coefficients, which proceeds as follows.
First, from the sum rules, we directly get
\begin{align}
    g^{2,0} = \Lambda^4 g^{4,0} = \Lambda^8 g^{6,0} = \cdots ~~\Longrightarrow~~
    \mathfrak{b}^{0, 0} = \mathfrak{b}^{2, 0} = \mathfrak{b}^{4, 0} = \cdots.
\end{align}
Recall that the PSD of Hankel matrices implies $(\mathfrak{b}^{1,0})^2 \leq \mathfrak{b}^{0,0} \mathfrak{b}^{2,0}$ and $\mathfrak{b}^{0,0} \geq \mathfrak{b}^{1,0} \geq \mathfrak{b}^{2,0}$, which, combined with $\mathfrak{b}^{0,0} = \mathfrak{b}^{2,0}$ and the sum rules, leads to $\mathfrak{b}^{0,0} = \mathfrak{b}^{1,0} = \mathfrak{b}^{2,0}$.
Similarly, we can further infer that $\mathfrak{b}^{\gamma_1, 0} = \mathfrak{b}^{0,0}$. Then, using the spectral reconstruction method of \ref{sec:EFT_gives_Moments}, we see that there is only one UV mass scale $\Lambda$, which in turn means that  
\begin{align}
    \mathfrak{b}^{\gamma_1, \gamma_2} = \mathfrak{b}^{0, \gamma_2}.
\end{align}
Again, from the sum rules, we can write $\mathfrak{b}^{0, \gamma_2}$ as a linear combination of $g^{2,j}$.
This indeed gives the same moment sequence as in \eref{mstuexp}.

\begin{figure}[h]
    \centering
    \includegraphics[width=0.5\textwidth]{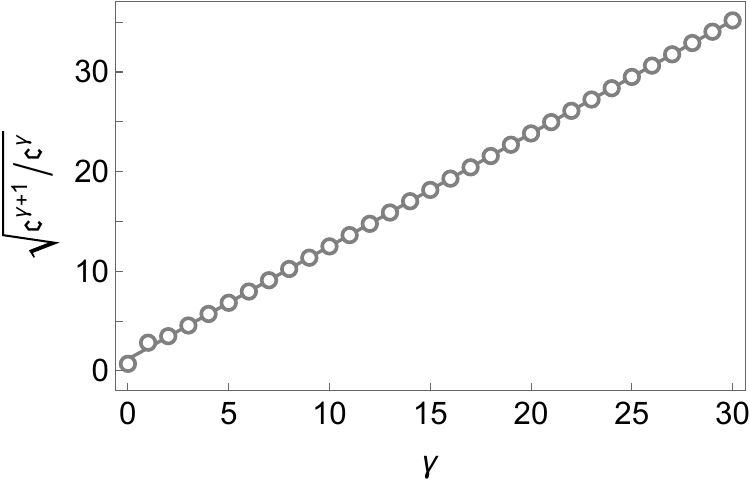}
    \caption{$\mathfrak{c}^{\gamma+1}/ \mathfrak{c}^{\gamma}$ dependence on $\gamma$. This shows that the $stu$ kink theory is Stieltjes determinate, and  thus the obtained UV spectrum is unique.} 
    \label{fig:RatiosOfMoments_stupole}
\end{figure}

Since the UV particles have the same mass $M_\ell = \Lambda=1$, when converting the bi-variant moments $\mathfrak{b}^{\gamma_1,\gamma_2}$ to uni-variant moments $\mathfrak{c}^{i,\gamma_2}$ as in \eref{momentsbtoc}, there will be only one uni-variate moment sequence:
\begin{align}
   \mathfrak{c}^{\gamma_2} &= \dfrac{\pi}{2} \operatorname*{lim}_{\xi \rightarrow 1^-} \left[ \left( 1 + 2 \xi \dfrac{\dif}{\dif \xi} \right)^{2\gamma_2+1} F(\xi) \right] - \dfrac{4 \pi \operatorname{ln} 2}{9} \delta_{\gamma_2, 0}.
\end{align}
It is easy to numerically verify that the ratio $\mathfrak{c}^{\gamma+1}/ \mathfrak{c}^{\gamma}$ quickly converges a power-law behavior
\begin{equation}
 \mathfrak{c}^{\gamma+1}/ \mathfrak{c}^{\gamma} \sim \mathcal{O} (\gamma^2),
\end{equation}
one can see from Figure \ref{fig:RatiosOfMoments_stupole}. This suggests that the $\mathfrak{c}^{\gamma}$ moment sequence for the $stu$ kink theory is Stieltjes determinate. (In fact, since there is no UV scalar particle, the $\mathfrak{c}^{\gamma}$ moment sequence is actually also Hamburger determinate. Also, since all marginal moment sequences ($\mathfrak{b}^{\gamma_1,0}$ and $\mathfrak{b}^{0,\gamma_2}$) of $\mathfrak{b}^{\gamma_1,\gamma_2}$ is determinate on $\mathbb{R}$, $\mathfrak{b}^{\gamma_1,\gamma_2}$ is determinate on $\mathbb{R}^2$ \cite{Schmdgen2017TheMP}.) Therefore, the representing measure or the UV spectrum reconstructed from the EFT coefficients originated in the $stu$ kink theory is unique. Of course, numerically, we will see that the reverse engineered UV spectrum indeed approaches the underlying theory quickly.

\begin{figure}
    \centering
    \includegraphics[width=0.45\textwidth]{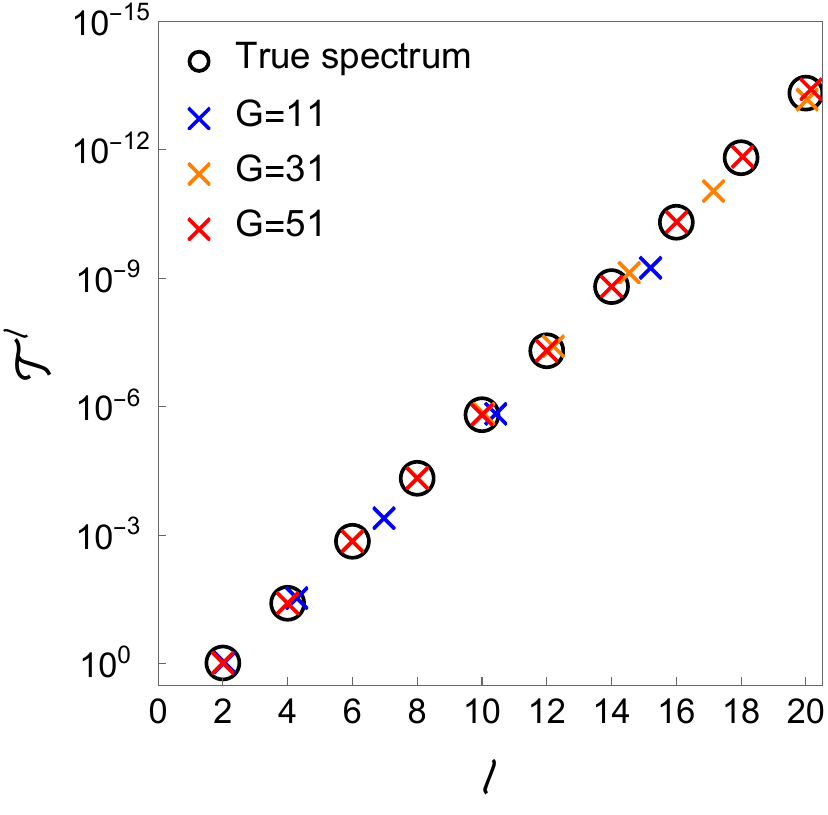}
    ~~~~~~
    \includegraphics[width=0.45\textwidth]{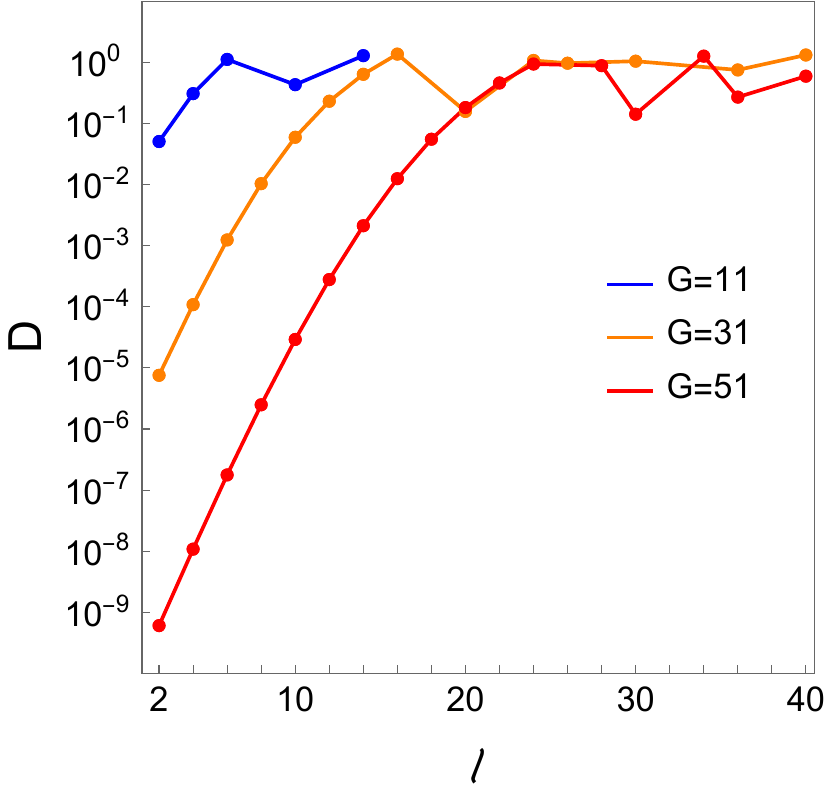}
    \caption{Coupling $\mathcal{T}^\ell$ (see \eref{eq:DiscResult} and \eref{Tstuspectrum}) of the $stu$ kink theory reverse engineered from the truncated EFT coefficients with truncation order $G$. 
    $D$ in the right subplot is the distance of a particular data point to the corresponding true spectral pole shown in the left subplot.} 
    \label{fig:DoubleFullMPExample1}
\end{figure}

Now, let us reconstruct the UV spectrum from the moment sequence $\mathfrak{b}^{\gamma_1,\gamma_2}$, following the procedure of Eqs.~\eqref{eq:HWequation0} to \eqref{Xilist}. As usually happens, suppose we only know the first few moments $\mathfrak{b}^{\gamma_1,\gamma_2}$ with $\gamma_1+\gamma_2\leq 2G+1$ from the low energy EFT: 
\begin{align}
    \mathfrak{b}^{0,0},~\mathfrak{b}^{1,0},~\mathfrak{b}^{0,1},~\mathfrak{b}^{2,0},~\mathfrak{b}^{1,1},~\mathfrak{b}^{0,2},~...,~\mathfrak{b}^{2G+1,0},~...,~\mathfrak{b}^{0,2G+1}, 
\end{align}
and for our purposes let us assume that our experiments are so precise that we know them almost exactly as prescribed by \eref{mstuexp}. Knowing only the numerical values of these $\mathfrak{b}^{\gamma_1,\gamma_2}$,
we hope to numerically reconstruct the UV spectrum \eqref{Tstuspectrum}. As a first step, to optimize the results, we should seek a flat extension for the truncated bi-variate moment problem with some extra moments $\mathfrak{b}^{\gamma_1,\gamma_2},\;\gamma_1+\gamma_2=2G+2$, satisfying the flat extension condition 
\begin{align}
    \operatorname{rank} \mathcal{H}_1 (2G+1) = \operatorname{rank} \mathcal{H}_1 (2G+2) .
\end{align}
Due to $\mathfrak{b}^{\gamma_1,\gamma_2}$ being determinate, even without the positivity of other Hankel matrices, we can still get the correct spectrum. 
With the moment sequence flatly extended, we can follow the algorithm outlined in Section \ref{sec:AllSpec} to reconstruct the couplings $T^\ell$, as shown in Figure \ref{fig:DoubleFullMPExample1}. 
To measure the accuracy of the reconstructed spectrum, we define the distance, $D$, of a particular data point to the corresponding spectral pole, and the numerical results are shown in the right subplot of Figure \ref{fig:DoubleFullMPExample1}. We see that as the truncation order increases, more UV poles can be accurately reconstructed.

\subsubsection{Virasoro amplitude}
\label{sec:Vis}

Now, we will use the Virasoro amplitude with zero Regge intercept as an example to illustrate the two methods mentioned in Section \ref{sec:EFT_gives_Moments} to infer the $\mathfrak{b}^{\gamma_1,\gamma_2}$ moments from the EFT coefficients $g^{i,j}$. Then, we will use the numerically obtained moments to reconstruct the spectrum of the Visrasoro amplitude. The Virasoro amplitude describes 2-to-2 scattering between tree-level closed strings, which with the kinematic prefactor stripped is given by:
\begin{gather}
    A_{\text{Vis}} (s,t,u) \equiv \dfrac{\Gamma(-s)\Gamma(-t)\Gamma(-u)}{\Gamma(1+s) \Gamma(1+t) \Gamma(1+u)}
\end{gather}
subject to the on-shell constraint $s + t + u = 0$. The coefficients $g^{i,j}$ in the IR behavior can be extracted from
\begin{align}
    &A_{\text{Vis}} (s,t,u) = \dfrac{1}{s t u} + \sum_{i,j \geq 0} g^{i,j} v^{i} t^{j}. \\
    &~~~= \dfrac{1}{s t u} \operatorname*{Exp} \left( \sum_{k=1}^{\infty} \dfrac{2 \zeta (2k+1)}{2k+1} (s^{2k+1} + t^{2k+1} + u^{2k+1}) \right) 
\end{align}
In the last example, as there is only one UV mass scale, we were able to analytically infer the moments $\mathfrak{b}^{\gamma_1,\gamma_2}$ from the EFT coefficients. Generically, however, with many UV mass scales, this step needs to be performed numerically. So it is instructive to explicitly demonstrate the numerical accuracy of the whole UV reconstruction chain including the conversion from the EFT coefficients to the moments.

Let us first convert the EFT coefficients to the moments. To this end, note that we directly get 
\begin{align}
\mathfrak{b}^{2\gamma,0} = g^{2 \gamma+2, 0}
\end{align}
from Table \ref{tab:EFTCoeffs_And_Moments}. As mentioned in Section \ref{sec:EFT_gives_Moments}, it is not straightforward to get the $\mathfrak{b}^{2\gamma+1,0}$ moments from the EFT coefficients. To obtain them, we can use the extra spectral information from $\mathfrak{b}^{2\gamma,0}$, and find the measure consisting of the least number of delta functions from the known $g^{2i,0}$.
For example, suppose we can extract $g^{2,0},g^{4,0},g^{6,0},g^{8,0}$ from the EFT, that is, we know $\mathfrak{b}^{0,0}, \mathfrak{b}^{2,0} ,\mathfrak{b}^{4,0},\mathfrak{b}^{6,0}$. To find the flat extension for the Hausdorff moment problem, we can solve the following equation to get ${\mathfrak{b}}^{8,0}$
\begin{align}
    \operatorname*{rank} 
    \begin{pmatrix}
        \mathfrak{b}^{0,0} & \mathfrak{b}^{2,0} & \mathfrak{b}^{4,0} \\
        \mathfrak{b}^{2,0} & \mathfrak{b}^{4,0} & \mathfrak{b}^{6,0} \\
        \mathfrak{b}^{4,0} & \mathfrak{b}^{6,0} & \mathfrak{b}^{8,0}
    \end{pmatrix} = 
    \operatorname*{rank} 
    \begin{pmatrix}
        \mathfrak{b}^{0,0} & \mathfrak{b}^{2,0} \\
        \mathfrak{b}^{2,0} & \mathfrak{b}^{4,0}
    \end{pmatrix} = 2
\end{align}
With this set of $\mathfrak{b}^{0,0}, \mathfrak{b}^{2,0} ,\mathfrak{b}^{4,0},\mathfrak{b}^{6,0}, {\mathfrak{b}}^{8,0}$, we can find an atomic measure by the method introduced in Section \ref{sec:MassSpec}. 
Then, we can use this spectrum to get ${\mathfrak{b}}^{2\gamma+1,0}$, as prescribed in \eref{b2gp1EQ}, truncated to an order similar to that in the even order moments. With all the leading orders of ${\mathfrak{b}}^{\gamma,0}$ known, we can then proceed to compute ${\mathfrak{b}}^{\gamma,1}$. From the known ${\mathfrak{b}}^{\gamma,0}$ and the $g^{ij}$-${\mathfrak{b}}^{\gamma_1,\gamma_2}$ relations in Table \ref{tab:EFTCoeffs_And_Moments}, we can get ${\mathfrak{b}}^{2\gamma,1}$, and the highest order of ${\mathfrak{b}}^{2\gamma,1}$ is chosen to satisfy the flatness condition. This allows us to construct another atomic measure, from which we can extract ${\mathfrak{b}}^{2\gamma+1,1}$. These steps can be repeated for higher order ${\mathfrak{b}}^{\gamma_1,\gamma_2}$. 

\begin{figure}[h]
    \centering
    \includegraphics[scale=0.52]{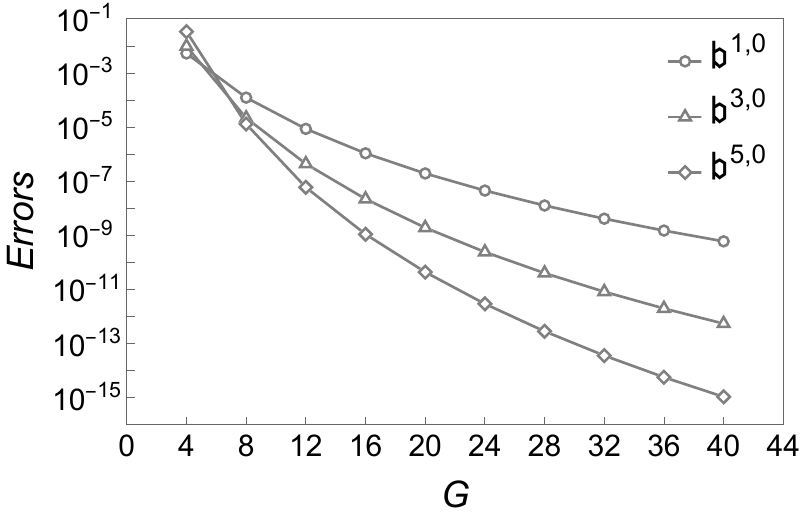}
    ~~
    \includegraphics[scale=0.52]{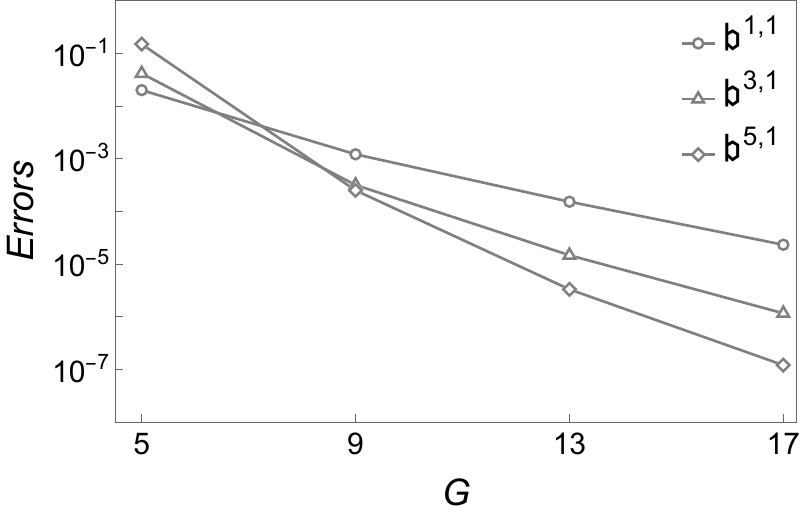}
    \caption{Relative errors for the reconstructed $\mathfrak{b}^{2\gamma+1,0}$ and $\mathfrak{b}^{2\gamma+1,1}$ from EFT coefficients $g^{i,j}, i+j \leq G$ using the spectral method. For $\mathfrak{b}^{2\gamma+1,0}$, the true values are the ones from the known UV theory, while the ``true'' values for $\mathfrak{b}^{2\gamma+1,1}$ are chosen to be the most accurately reconstructed ones.}
    \label{fig:Vis_err}
\end{figure}

Let us compare these reconstructed ${\mathfrak{b}}^{\gamma_1,\gamma_2}$ via this spectral method with the ``true'' values in Figure \ref{fig:Vis_err}. With the benefits of knowing the UV theory, we know that  $\mathfrak{b}^{2\gamma+1,0} = 2 \zeta(2\gamma+6)$, which are chosen as the true values in Figure \ref{fig:Vis_err}. For ${\mathfrak{b}}^{\gamma_1,1}$, we choose the ``true'' values to be the ones constructed with the highest order truncation of the $g^{i,j}$ coefficients. We see that this method works very well as the truncation order of the EFT coefficients increases.

\begin{figure}[h]
    \centering
    \includegraphics[scale=0.3]{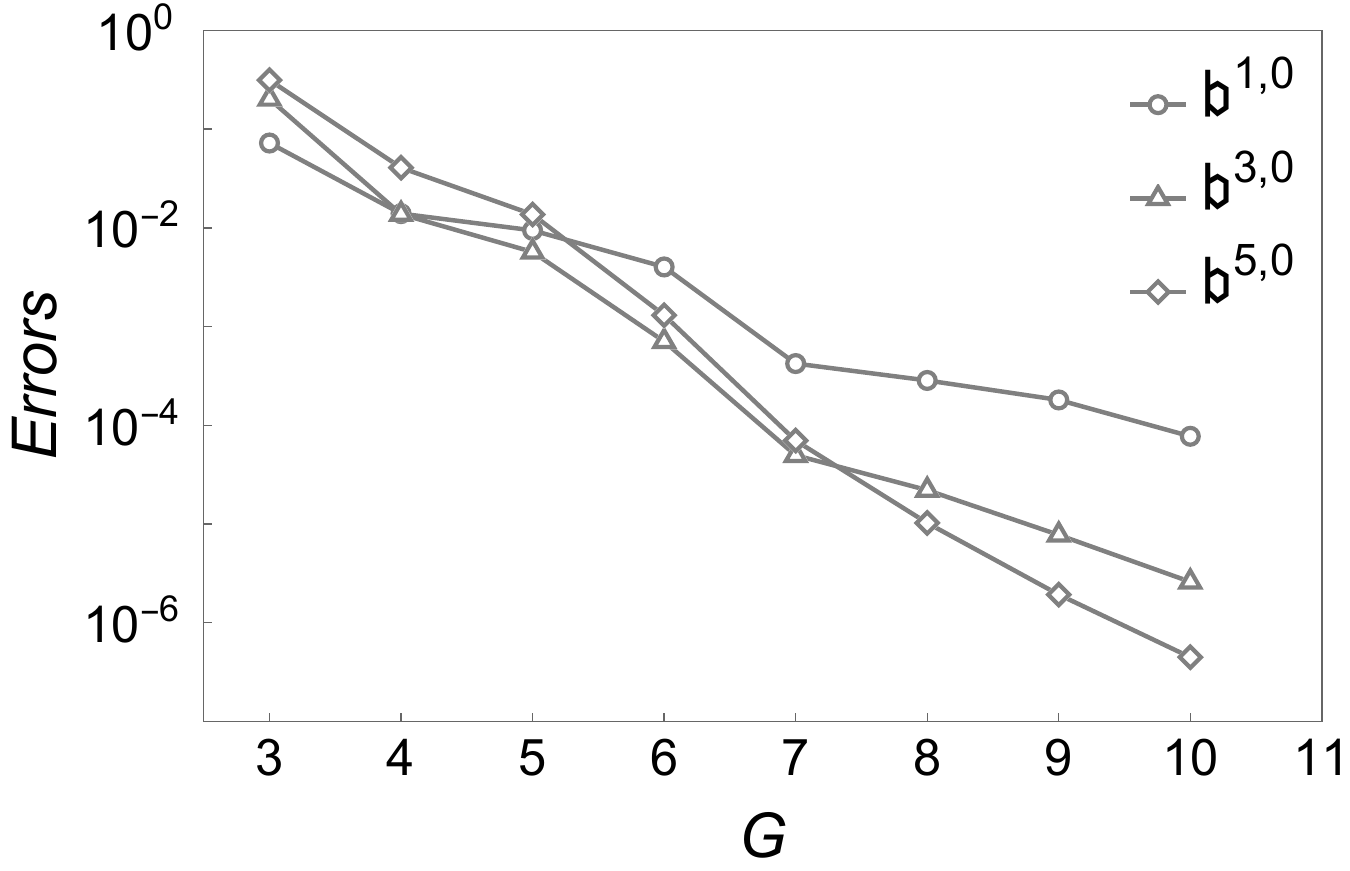}
    ~~
    \includegraphics[scale=0.3]{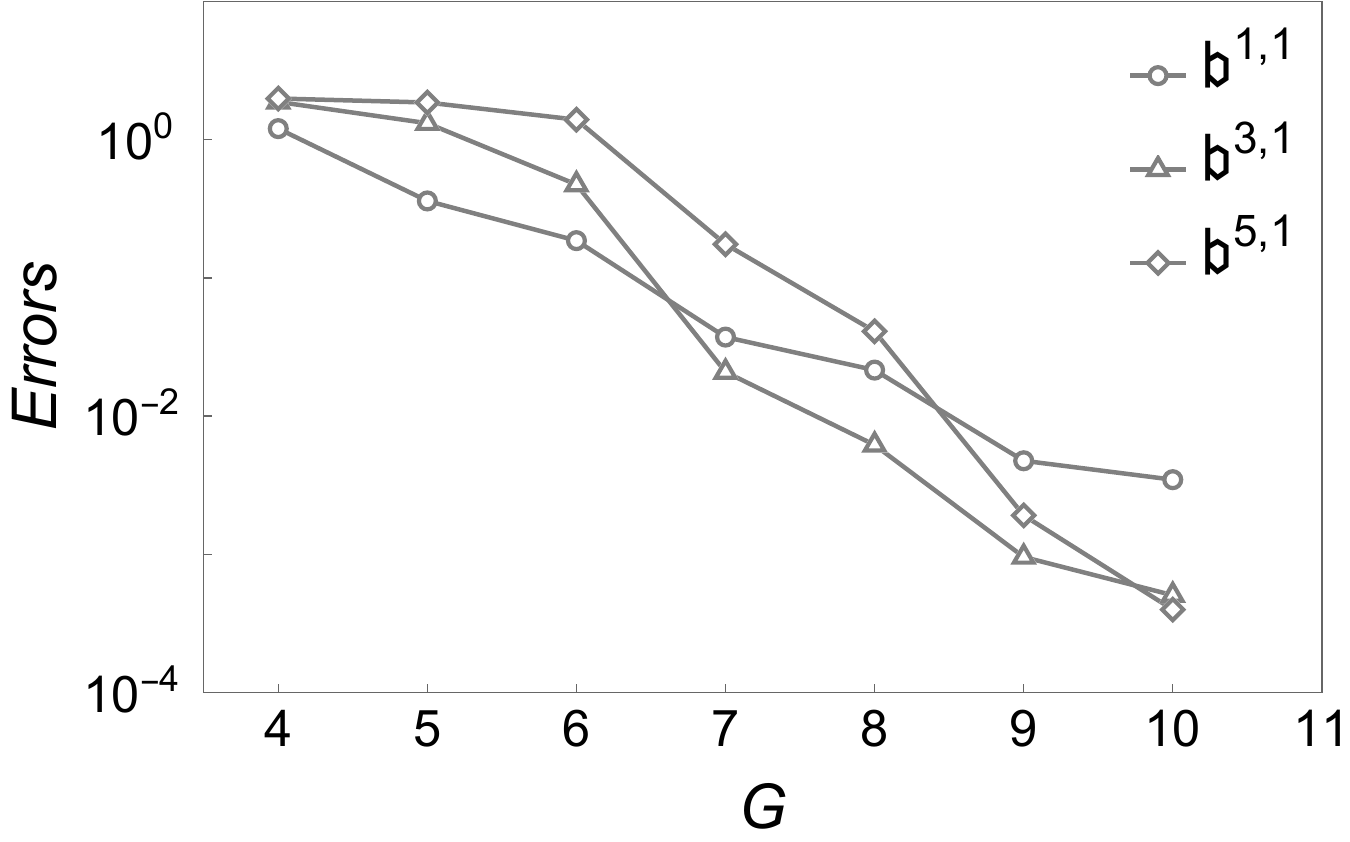}
    \caption{
    Relative errors for $\mathfrak{b}^{2\gamma+1,0}$ and $\mathfrak{b}^{2\gamma+1,1}$ from EFT coefficients $g^{i,j}, i+j \leq G$ using the SDP method. For $\mathfrak{b}^{2\gamma+1,0}$, the true values are the ones from the known UV theory, while the ``true'' values for $\mathfrak{b}^{2\gamma+1,1}$ are chosen to be the average of the maximum and minimum values.
    }
    \label{fig:Vis_SDP}
\end{figure}

Another method mentioned in Section \ref{tab:EFTCoeffs_And_Moments} is a more numerical approach. With the second method, we constrain each of the $\mathfrak{b}^{\gamma_1,\gamma_2}$ coefficients from both sides by directly implementing SDPs, such as that of \eref{numSDPmethod2}, with the positivity of the Hankel matrices and the $g^{i,j}$-${\mathfrak{b}}^{\gamma_1,\gamma_2}$ relations in Table \ref{tab:EFTCoeffs_And_Moments}.
As an illustration, in Figure \ref{fig:Vis_SDP}, we use this method to compute the upper and lower bounds on the first few $\mathfrak{b}^{\gamma_1,\gamma_2}$. We see that as the truncation order $G=\max(i+j)$ of the EFT coefficients $g^{i,j}$ increases, both the upper and lower bounds quickly converge to the true values. 

\begin{figure}[h]
    \centering
    \includegraphics[scale=0.6]{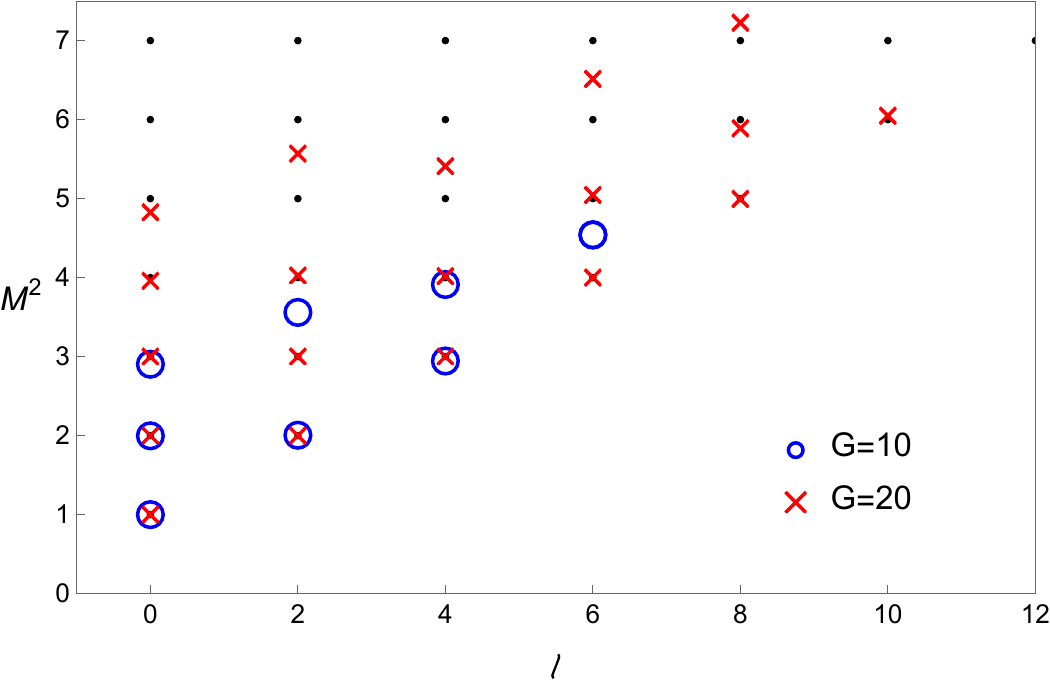}
    \caption{Reconstructed UV spectrum for the Visrasoro amplitude using EFT coefficients $g^{i,j}$, with $i+j \leq G$. The black dots are the true spectrum, while the blue circles and red crosses are the reconstructed spectra with different EFT truncation.}
    \label{fig:Vis_Spec}
\end{figure}

Now, let us use the obtained moments $\mathfrak{b}^{\gamma_1,\gamma_2}$ to reconstruct the UV spectrum of the Visrasoro amplitude with the Curto-Fialkow method (see Section \ref{sec:AllSpec}). For this problem, it is easier to implement the second method above, and the results are shown in Figure \ref{fig:Vis_Spec}. We find that an increasing number of UV masses and spins can be accurately captured as more EFT coefficients are used. In particular, those with $\ell=0$ and the lowest masses for higher $\ell$ values are the easiest to capture. 

Assuming the spectrum can determined to high orders, we can verify {\it A posteriori} that the UV spectrum is unique. To see this, simply note that, due to $\ell \leq 2 I - 2$, for every UV mass $M_I^2 = I$, the moment sequence $\mathfrak{c}_I^{\gamma}$ satisfies $\mathfrak{c}_I^{\gamma+1}/\mathfrak{c}_I^{\gamma} \sim const$. This growth is slower than the determinancy upper bound: $\mathfrak{c}^{\gamma+1} / \mathfrak{c}^{\gamma} \sim O ( \gamma^2 )$ for large $\gamma$.

\section*{Acknowledgments}

We would like to thank Dong-Yu Hong, Yu-tin Huang, Andrew Tolley and Zhuo-Hui Wang for helpful discussions.
SYZ acknowledges support from the National Natural Science Foundation of China under grant No.~12075233, No.~12475074 and No.~12247103.

\bibliographystyle{JHEP}

\bibliography{refs.bib}

\end{document}